\documentclass[useAMS,usenatbib]{mn2e}

\usepackage{psfig}
\usepackage[dvips]{graphicx}
\usepackage{lscape}

\def\lesssim{\mathrel{\hbox{\rlap{\hbox{\lower4pt\hbox{$\sim$}}}\hbox{$<$}}}}
\def\gtrsim{\mathrel{\hbox{\rlap{\hbox{\lower4pt\hbox{$\sim$}}}\hbox{$>$}}}}

\title[Extended MILES models photometric constraints]{MIUSCAT: extended MILES spectral coverage. II. Constraints from optical photometry}
\author[Ricciardelli et al.]{E. Ricciardelli$^{1}$\thanks{E-mail:
    elena.ricciardelli@uv.es},A. Vazdekis$^{2,3}$, A.J. Cenarro$^4$, J. Falc\'on-Barroso$^{2,3}$\\
$^{1}$Departament d'Astronomia i Astrofisica, Universitat de Valencia, c/ Dr. Moliner 50, E-46100 - Burjassot, Valencia, Spain\\
$^{2}$Instituto de Astrof\'isica de Canarias, V\'ia Lactea s/n, E-38200 La Laguna, Tenerife, Spain\\
$^{3}$Departamento de Astrof\'isica, Universidad de La Laguna,
E-38205, Tenerife, Spain\\
$^{4}$Centro de Estudios de F{\'{\i}}sica del Cosmos de Arag\'on, Plaza de San Juan 1, Planta 2, E-44001, Teruel, Spain\\
}

\begin{document}
\date{Accepted ...  Received ...; in original form ...
}
\pagerange{\pageref{firstpage}--\pageref{lastpage}} \pubyear{...}
\maketitle
\label{firstpage}
\begin{abstract}

In the present work we show a comprehensive comparison of our new
stellar population synthesis MIUSCAT models with photometric data of globular clusters and early-type galaxies.
The models compare remarkably well with the colours of Milky Way
globular clusters in the optical range. Likewise, the colours of M31 globular
clusters can also be explained by the models by assuming younger ages
then their Galactic counterparts. 
When compared with quiescent galaxies we reproduce the
colour evolution at intermediate redshift. 
On the other hand we find that the colour relations of nearby early-type galaxies are
still a challenge for present-day stellar population
synthesis models. We investigate a number of possible explanations and
establish the importance of $\alpha-$enhanced models to bring down the
discrepancy with observations.

\end{abstract}


\section{Introduction}\label{intro}

Stellar population synthesis (SPS) models are commonly used  
for solving a great variety of studies, from age/metallicity estimates in
globular clusters to the reconstruction of the star formation history (SFH)
in galaxies. 
The modeling of the spectral energy distribution (SED) emitted by
evolving stellar populations requires at least three main
ingredients: stellar evolutionary tracks, a spectral or photometric
library 
with well-established atmospheric parameters for each library star, 
and
an initial mass function (IMF).
A variety of models exists in the literature that differ in their treatment
of each of these ingredients, e.g., \citep{Worthey94, Vazdekis99, BC03, Maraston05, Conroy10}. 
In particular, the adoption of the spectral
library used to convert physical quantities in observed flux turned
out to be critical in the comparison with observations. 
Although theoretical libraries (e.g., \citealt{Lejeune98}) are widely used among the most popular
SPS models \citep{BC03, Maraston05} because they are easier to
handle, given the wide stellar parameter and spectral coverages and the arbitrary wavelength
resolution, empirical-based libraries represent  
an advantage as the stars are real.  The adoption of empirical spectra allows to overcome the
uncertainties in the underlying model atmosphere calculations,
giving more reliable predictions when compared with 
both spectroscopic \citep{Vazdekis03, Vazdekis10} and 
photometric \citep{Maraston09, Peacock11} data.

In Vazdekis et al (2012, hereafter Paper I) a library of single-burst stellar
population model SEDs covering the spectral range 3500 - 9500\,\AA\ at moderately
high resolution is presented. For building-up these models we
combine three extensive empirical stellar spectral libraries, namely MILES
\citep{Sanchez06}, Indo-US \citep{Valdes04} and CaT (\citealt{Cenarro01},
  \citealt{Cenarro01b}, \citealt{Cenarro07}). Here we focus
on calibrating these models by comparing these new predictions to a variety of
photometric data ranging from globular clusters (GC) to quiescent galaxies.
Broadband colours are among the simplest predictions of population
synthesis models, thus comparing the model predictions to
observed colours is the natural zeroth-order test of compatibility
between models and observations. 

Since GC are almost coeval and mono-metallic population of stars that
closely resemble a single stellar population (SSP), they are the ideal benchmark to test the
accuracy of the models. On the other hand, 
quiescent galaxies can reach larger metallicities than the metallicity distribution of most globular cluster systems
and allow us, with some caveats, to calibrate the
models even in the high metallicity regime.
Although they can be modelled as an SSP as a first
approximation,  pieces of evidence of recent  star formation in early-type
galaxies have been found by different authors (e.g.,
\citealt{Bressan06, Sarzi08}), which indicate the need
of more complex stellar populations to obtain better fits.
Moreover, galaxies may contain a certain amount of internal reddening
that can further complicate the analysis. 

The paper is organized as follows. In Section 2 we present a brief summary of 
the stellar population synthesis models presented in Paper I. In Section 3 we describe the
webtool developed to download and handle these models. In Sections 4 and 5 we
present the comparison of these models with globular cluster and galaxy data,
respectively.
Section 7 summarizes our results. 
Throughout the paper, we adopt the concordance cosmology: $H_0$=70 \mbox{km s$^{-1}$} \mbox{Mpc$^{-1}$}, 
$\Omega_m=0.3$ and $\Omega_{\Lambda}=0.7$. 

\section{MIUSCAT models}

The stellar population models we use throughout the paper have been
presented in Paper I, here we briefly summarize them. These models represent an
extension of the \citet{Vazdekis03, Vazdekis10}  models,  based on the CaT  and
MILES empirical stellar spectral libraries. MILES includes 985 stars covering
the range $\lambda\lambda$ 3525-7500\,\AA\ \citep{Sanchez06}  with a spectral
resolution of 2.5\,\AA\,FWHM, (Falc\'on-Barroso et al. 2011).
The CaT library \citep{Cenarro01} consists of 706 stars in the range
$\lambda\lambda$ 8350-9020 \,\AA\ at resolution 1.5 \,\AA. The Indo-U.S. stellar
library \citep{Valdes04} has also been added to fill-in the spectral range
around 8000\,\AA\ that is not covered by the MILES and CaT libraries and to
extend blueward and redward the wavelength coverage of the models. The Indo-U.S.
includes $\approx$ 1200 stars with spectra that typically cover from 3465 to
9469\,\AA\ at resolution $\sim$1.36\,\AA\, FWHM (Falc\'on-Barroso et
al. 2011).  

Although all the stars of the Indo-U.S. library have been used to obtain an
homogenized set of stellar parameters matching those of MILES and CaT libraries,
only a subsample composed of 432 stars have been included to build-up the final
MIUSCAT catalogue to feed the models. Only stars without gaps in the stellar
spectra or significant telluric absorption residuals, among other reasons, were
considered (see paper I for details). As the flux-calibration quality of the
Indo-U.S. is not as accurate as in the other two libraries, a procedure was
applied to correct the spectrum shape in the region $\lambda\lambda$ $\approx$
7400-8350\,\AA\ to match the MILES and CaT stellar spectra. Finally the
Indo-U.S. spectrum was used to extend the spectral range blueward the MILES and
redward the CaT spectral ranges without applying any correction to the Indo-U.S.
spectrum. Finally, the CaT and Indo-U.S. spectra were smoothed to match the
resolution of MILES, i.e. 2.5\,\AA\,(FWHM). 

The MIUSCAT stellar spectra were implemented in the models as described in
\citet{Vazdekis03, Vazdekis10} and updated in Falc\'on-Barroso et al. (2011).
Finally the relevant spectral ranges of the resulting SSP SEDs were extracted
and plugged to the original model SEDs computed on the basis of the  MILES and
CaT libraries. For the latter step a further correction, similar to that
described above for the individual stellar spectra, was applied to the model
SEDs to obtain a unique spectrum, which is identical to the MILES and CaT based
models in the corresponding spectral ranges, respectively.
 
\subsection{U magnitude}\label{sect_u}

Unlike MILES SSP spectra, the new MIUSCAT models allow us to measure from B to I
broad-band filters and related colours.
It is also possible to derive the U filter, which starts at $\approx$
3050 \,\AA\  \citep{Buser78}, by correcting for the missing flux blueward 3464.9 \,\AA. 
For this purpose we make use of the
fractions provided in paper I, which are obtained with the aid of the library of
stellar spectra of \citealt{Pickles98} ($\lambda\lambda$ 1150-25000\,\AA). For
each SSP the fractions obtained for the different stellar spectral types are
integrated along the isochrone to obtain the corresponding missing flux for the
SSP SED. Finally the total U magnitude is derived by measuring the flux of the
MIUSCAT SSP spectrum through the U filter redward 3464.9 \,\AA\ and correcting
by the missing flux. To compare our model predictions to the observational data
discussed in this paper we follow the same method to compute the SDSS {\it u}
filter at two redshift values (i.e. 0 and 0.04; see Section  6).

\section{The webtool}

The model predictions used in this work have been obtained with the aid of the
webtool facilities that we provide in the MILES website: http://miles.iac.es. 
The MILES and CaT stellar libraries and the stellar population synthesis models
can be retrieved and handled according to the user requirements.  The webtools
include the transformation of the spectra to match the instrumental set-up of
the observations (spectral resolution and sampling), measurements of
line-strength indices and synthetic magnitudes derived from the spectrum.  

As in this work we apply the MIUSCAT models to a large sample of galaxy
data we describe in more detail another webtool facility that will be
extensively used in Section 6. The tool returns the model spectrum
corresponding to a given Star Formation History (SFH), which can be either
user-defined or follow some parametric description. The first case is devoted to handle with complex
SFHs, with an arbitrary number of bursts. For instance, this
case will be particularly useful for the SFHs retrieved from cosmological
simulations.  Given the mass fraction and the metallicity at different ages, the
tool computes the corresponding SED according to: 

\begin{equation}
F_{\lambda}=\sum_{i=1}^{N_b}m_iF_{\lambda}(t_i,Z_i) 
\end{equation} 

\noindent where $N_b$ is the number of input bursts, $m_i$ is the fraction of
mass formed in the i-th bin and $F_{\lambda}(t_i,Z_i)$ is the i-th SSP spectrum
for a chosen IMF. In the case that $t_i$ and $Z_i$ are not included in the 
age/metallicity grid of the models, the closest SSP model to the desired
parameters is selected. Thus no interpolation is performed on the SSP spectra.

For the parametric SFHs four different cases are allowed:

\begin{itemize}
\item{Multiple bursts: up to five bursts occurring at different ages and
    with different metallicities are summed up}
\item{Truncated: the star formation rate (SFR) is assumed constant 
    between the time of formation, $t_F$, and the truncation time,
    $t_{T}$
    (all the times are referred to the present time) }
\item{Exponential: the SFH is modelled as a power-law rising term at early
    times plus an exponentially declining SFR:
\begin{equation}
SFR(t)=(\frac{t_{F}-t}{t_{F} } ) ^n e^{\frac{t_{F}-t}{\tau}}
\end{equation} }
where $t_F$ is the time when star formation starts, $\tau$ is the
e-folding time and $n$ parametrizes the power-law rising term at early times.
\item{Exponential plus bursts: up to five bursts can be added on top on
    the exponential SFR}
\end{itemize}

As we will show in the following sections, the SFH webtool is
particularly useful to construct galaxy SEDs, for which the
use of complex stellar populations provide better fits than the SSP
approach. In Section 6 several applications of the parametric
SFH will be explored.

\section{Comparison with Globular clusters}

\subsection{Milky Way globular clusters}
\begin{figure*}
\includegraphics[angle=0,width=0.69\columnwidth]{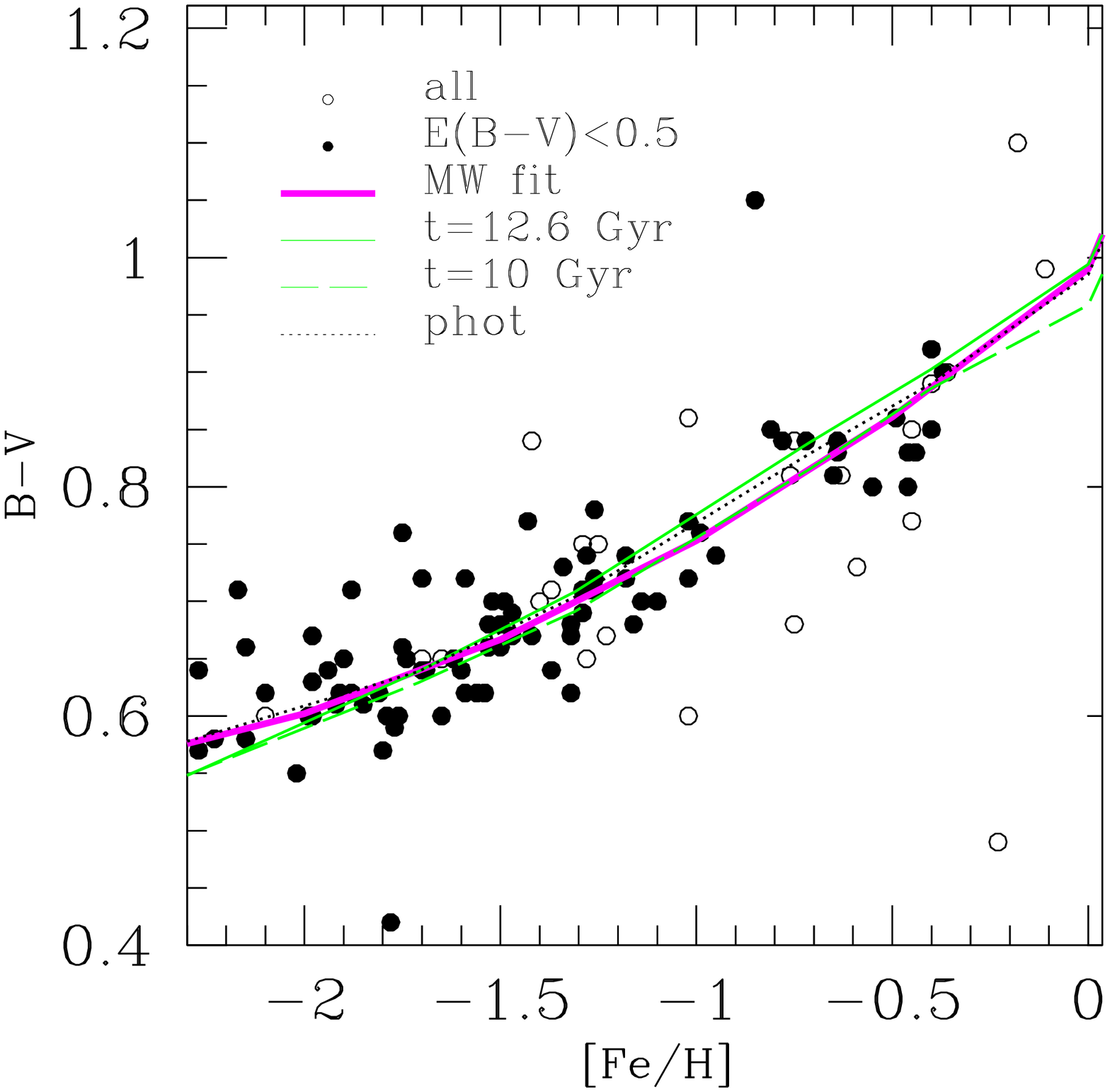}
\includegraphics[angle=0,width=0.69\columnwidth]{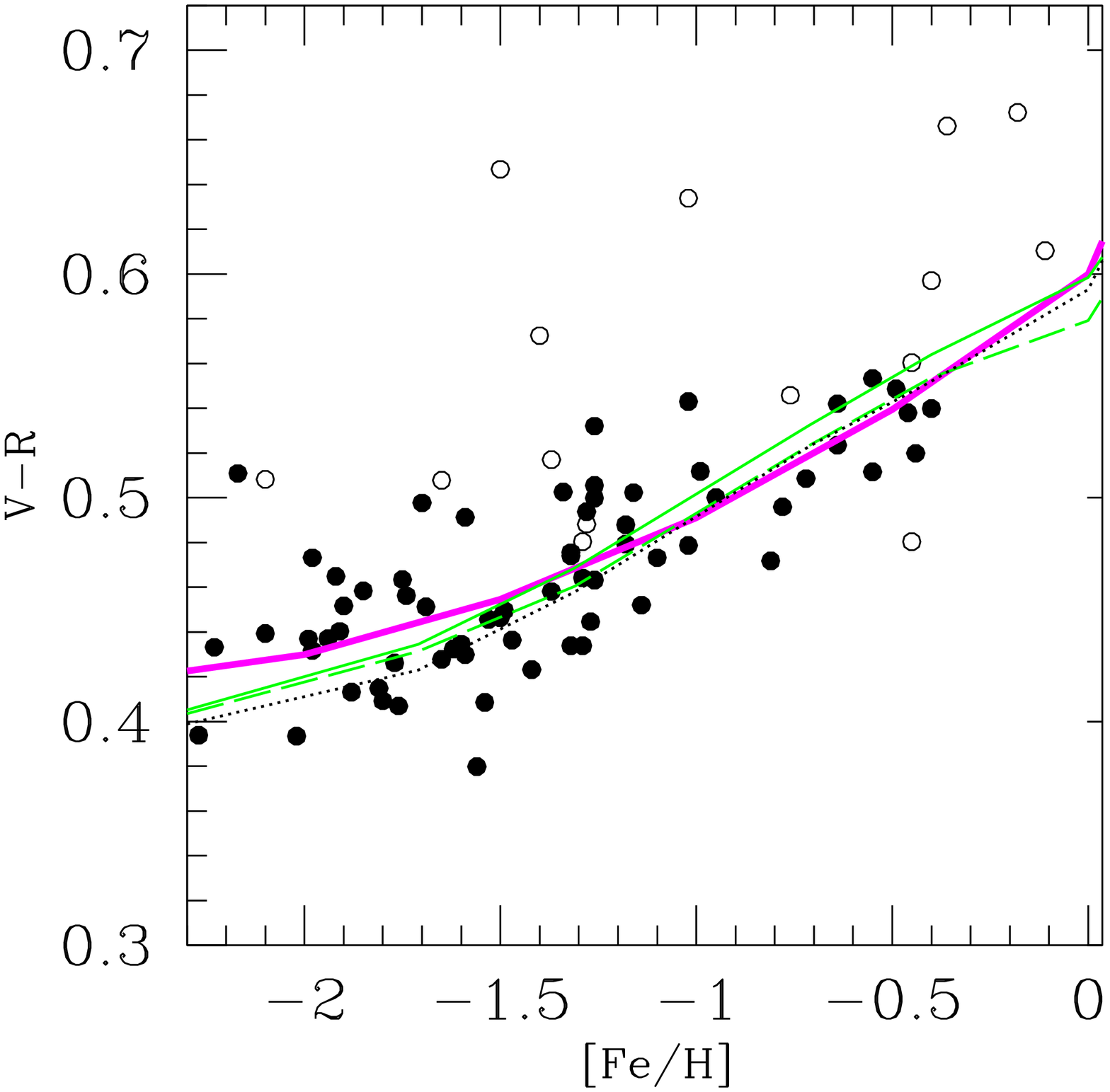}
\includegraphics[angle=0,width=0.69\columnwidth]{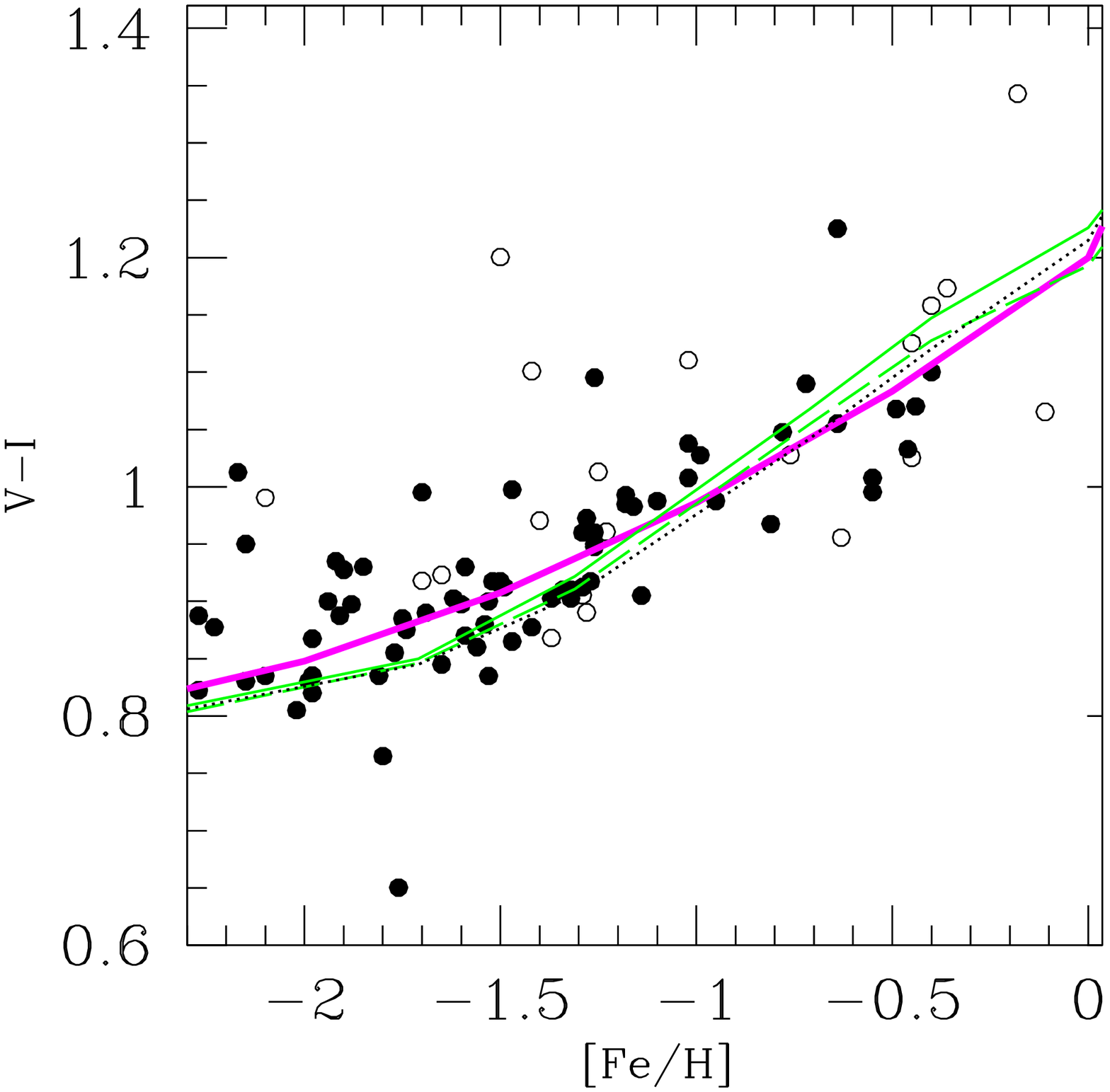}
\caption{ 
  Optical colours in several bands as a function of metallicity for the MW
  globular clusters (Harris 1996 catalog; 2010 edition). Filled (open) circles
  represent globular clusters with reddening E(B$-$V)$\le$0.5 ($>$0.5),
  de-reddened according to the extinction law of \citet{Cardelli89}. The magenta
  line shows a second order polynomial fit to the data with E(B$-$V)$\le$0.5. The black
  dotted lines show the photometric predictions of the MILES models. The
  synthetic colours derived from the MIUSCAT SSP spectra for models with age of 12.6
  and 10\,Gyr, for different metallicities and Kroupa Universal IMF, are shown
  in solid and dashed green lines, respectively.
}
\label{GC_BVRI}
\end{figure*}

\begin{figure*}
\includegraphics[angle=0,width=0.69\columnwidth]{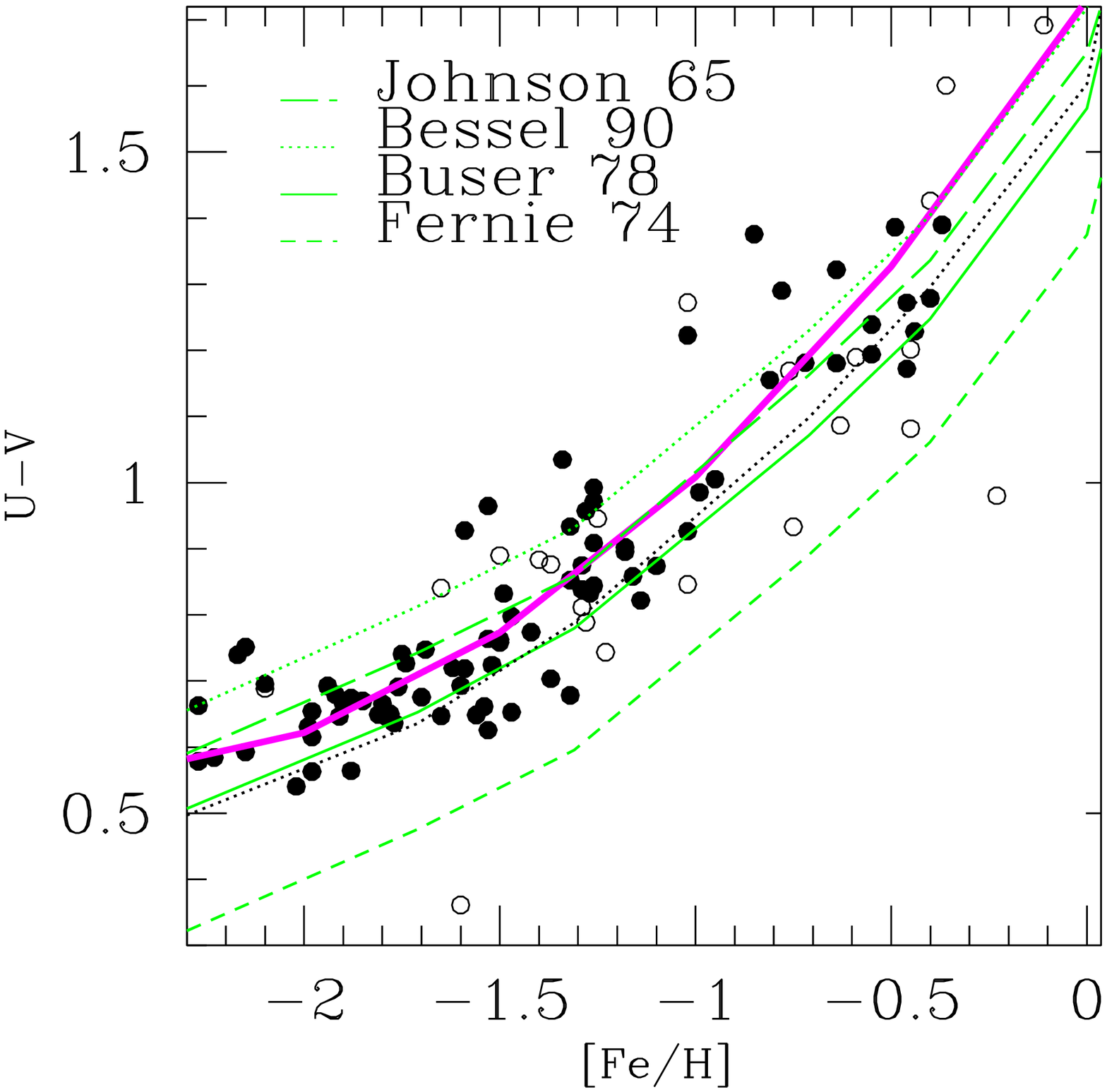}
\includegraphics[angle=0,width=0.69\columnwidth]{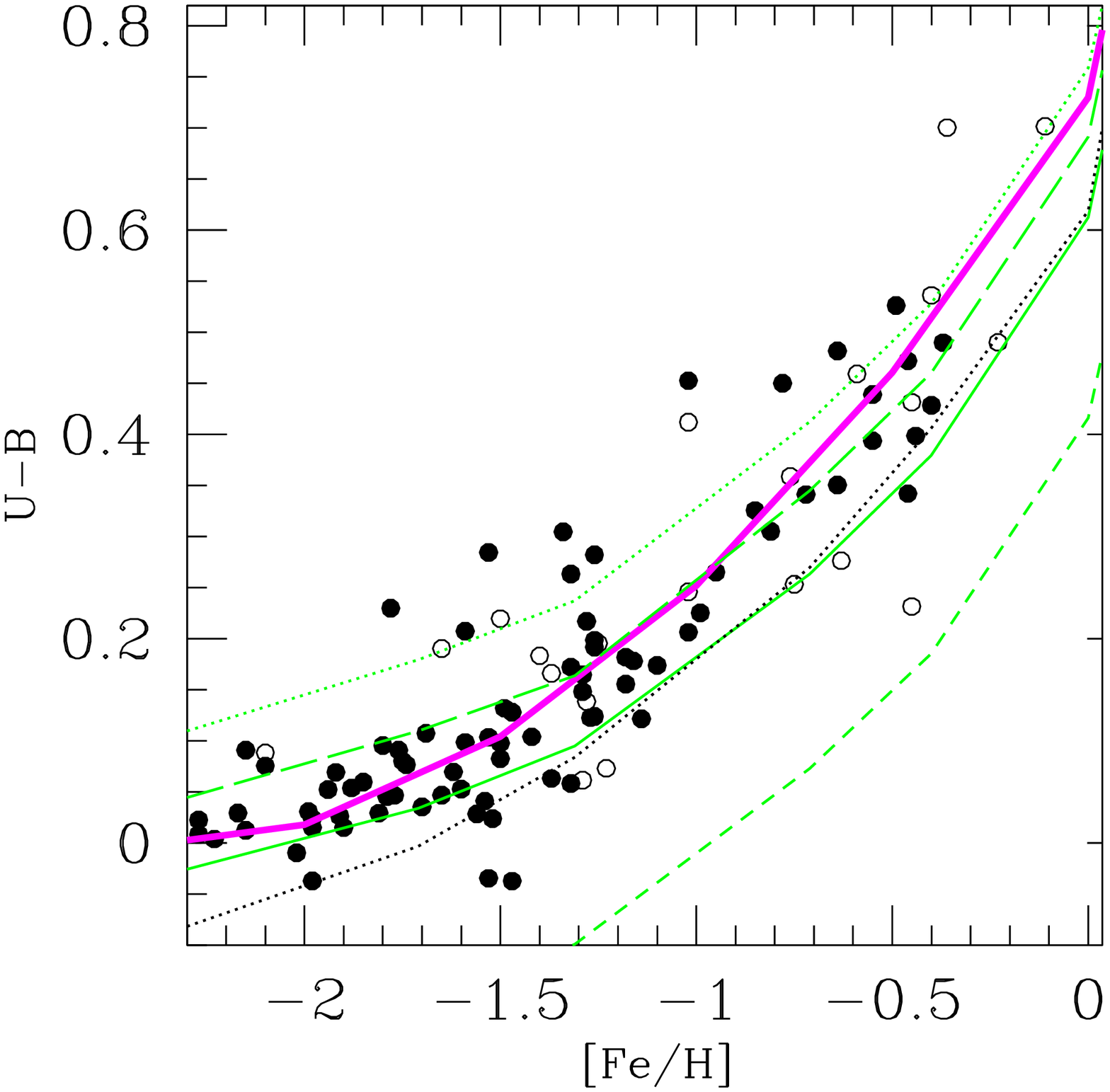}
\caption{ 
The same as figure \ref{GC_BVRI} but for U$-$V and U$-$B
  colours. The green lines show the synthetic colours derived by using
  different definitions of the U filter: solid for the Buser
\& Kurucz (1978) filter, long-dashed for the Johnson (1965) filter, 
dotted for Bessel (1990) and short-dashed for Fernie (1974).}
\label{GC_UBV}
\end{figure*}

Here we use the catalogue of \citealt{Harris96} (2010 edition) as a reference
for the Milky Way (MW) globular clusters. The catalogue is a compilation of 150
clusters including the Johnson-Cousins UBVRI photometry, [Fe/H] metallicity and
reddening E(B$-$V) among other data. Colours, except B$-$V, have been de-reddened
using the extinction law of \citet{Cardelli89}.

Figure \ref{GC_BVRI} shows the comparison of the optical broad-band colours of
the GCs with the ones derived from the models. The model metallicity indicates
the total abundance ratio [Z/H]. A fit to the data points with a second order polynomial
is indicated to guide the eye. In order to avoid clusters with
uncertain photometry due to high extinction, we excluded the most reddened
clusters with E(B$-$V) $\geq$0.5 (open circles). The MIUSCAT models are shown
for ages of 10.0 and 12.6\,Gyr for a Kroupa Universal IMF. The use of a Salpeter
IMF would leave the colours almost unaffected (with differences smaller than 
0.01\, mag). To derive the synthetic colours we adopt the \citet{Buser78} filter
responses and Vega magnitudes. The zero point  has been set by using
the \citet{Hayes85} 
spectrum of Vega with a flux of $3.44\times10^{-9}$\,erg
cm$^{-2}$\,s$^{-1}$\,\AA$^{-1}$ at 5556\,\AA  (see Falcon-Barroso et al. 2011 for
further details on the computation of synthetic magnitudes). The colours of Vega
in the  Johnson-Cousins filters are assumed to be U$-$B=U$-$V=B$-$V=V$-$R=0 
mag,  V$-$I=-0.04 mag.  The black dotted line represents the photometric SSP
model predictions computed by adopting temperature-gravity-metallicity relations
from extensive photometric stellar libraries \citep{Vazdekis96, 
  Vazdekis10}. 
The colour predictions resulting from these two approaches give very
similar results (with differences of the order of 0.01-0.02 mag) showing the
reliability of the flux-calibration of the model SEDs (see Paper I for more
details).  

Figure \ref{GC_BVRI} also  shows that the models reproduce the observed colours fairly
well. Note that the V$-$I colour is particularly relevant to test the
reliability of our models. This colour is heavily sensitive to the
details of the method used to join the three libraries (MILES,
Indo-U.S. and CaT). 
 Indeed, the V magnitude is entirely measured within the MILES
range, whereas the I filter response falls within the Indo-U.S./CaT range. The
good match to the observed colours shows the robustness of the method and the
flux-calibration of the MIUSCAT SEDs. It is also remarkable the ability of our
models to reproduce the B$-$V colour. This has been for a long time a problem for
stellar population synthesis models. As quoted in previous works
(e.g. \citealt{Worthey94, Maraston05}), 
SSP model SEDs tend to provide systematically
redder B$-$V colour than the observed values ($\approx$0.05 mag). One of the
reasons of these discrepancies may be ascribed to difficulties of the
theoretically-based models in modeling the spectral energy distribution around
$\simeq$ 5000-6000 \,\AA, where the theoretical spectra predict a flux excess. 
The use of empirical libraries \citep{Pickles98} has proven to be effective at
improving the match with colours of luminous red galaxies \citep{Maraston09}.
Likely, the use of empirical libraries with good flux-calibration quality and
good parameter coverage, including the  metal-poor regime, has allowed us to
improve the matching of the GC colours in comparison to previous
works. 
The improved match of the B$-$V vs [Z/H] relation when using
MILES-based models instead of the STELIB-Kurucz  library 
has been also shown in \citet{Maraston11}.

In Figure \ref{GC_UBV} we show the results obtained for the U$-$V and U$-$B
colours. We found that colours involving the U band are significantly more
difficult to reproduce. Indeed, our models predict bluer colours ($\approx$0.1
mag in the high metallicity regime). As explained in Sect. \ref{sect_u}, the U
magnitude is not fully derived from the MIUSCAT SEDs, which do not extend to the
blue end of the U filter response. The missing flux is accounted for with the
aid of the \citep{Pickles98} stellar library. However we rule out this
calibration as the a
main source for the obtained discrepancy, as the obtained magnitudes following
two different approaches (the second one employing SSP model SEDs from other
authors) do agree within $\approx$0.02 mag. 
Moreover, even the photometric predictions show a similar level of
disagreement with the data.
A possible explanation may be linked to the various U filter response curves that
can be adopted to reproduce the observations. Figure \ref{GC_UBV} shows the
obtained results for various U filter responses, which differ from our reference
filter (\citealt{Buser78}). Whereas for other Johnson broad-band filters the
adoption of different references for a given filter usually implies minor
changes, in the case of the U filter, the exact definition of the curve makes a
significant impact on the obtained colours. For instance, if we use the Johnson
(1965) response curve, we obtain a much better match. Since it is difficult to
trace back the compilation of data to recover which filter has been used for
every GC, the U$-$V and U$-$B colours are subject to this source of
uncertainty. 

\begin{figure*}
\includegraphics[angle=0,width=0.69\columnwidth]{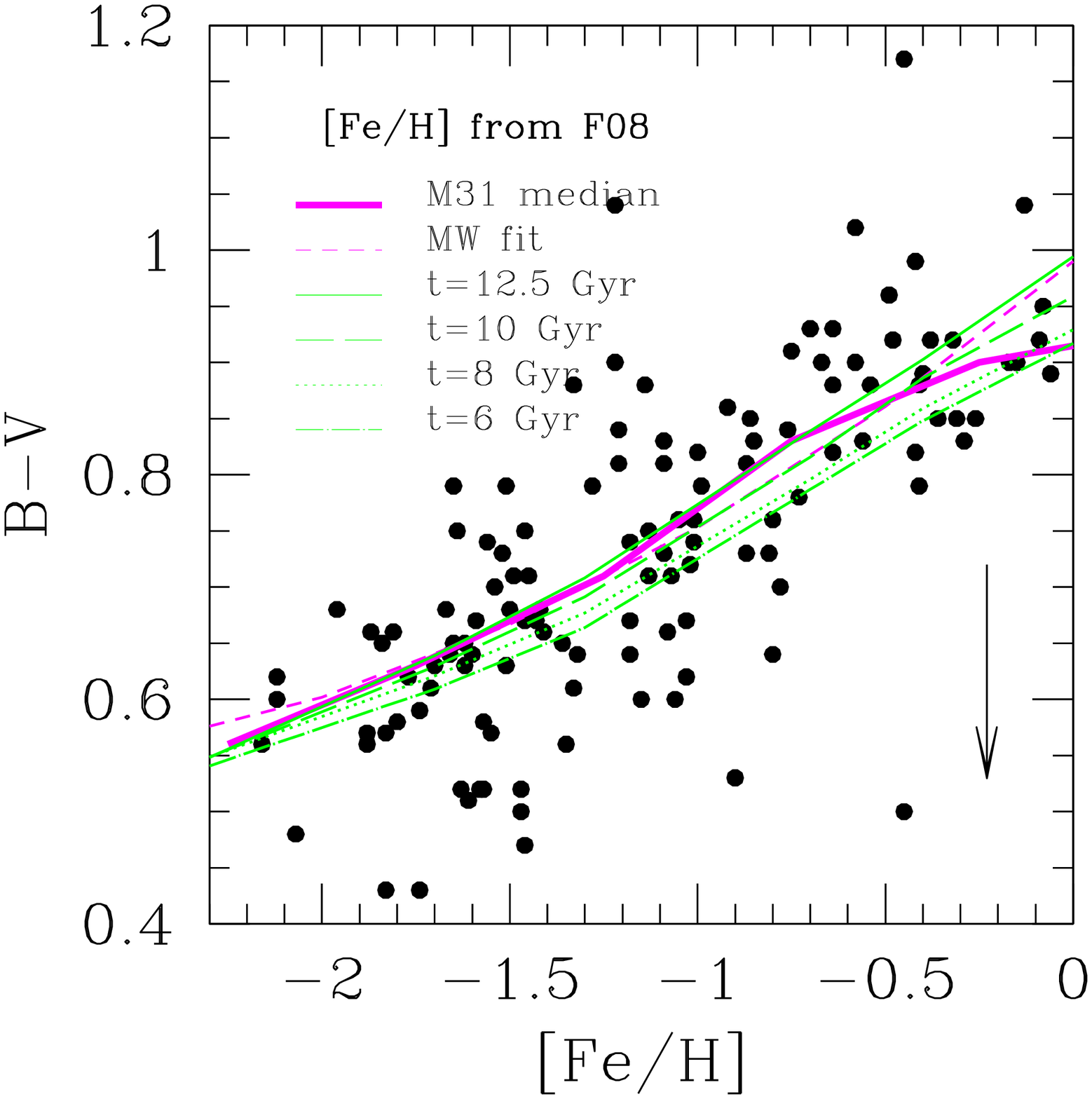}
\includegraphics[angle=0,width=0.69\columnwidth]{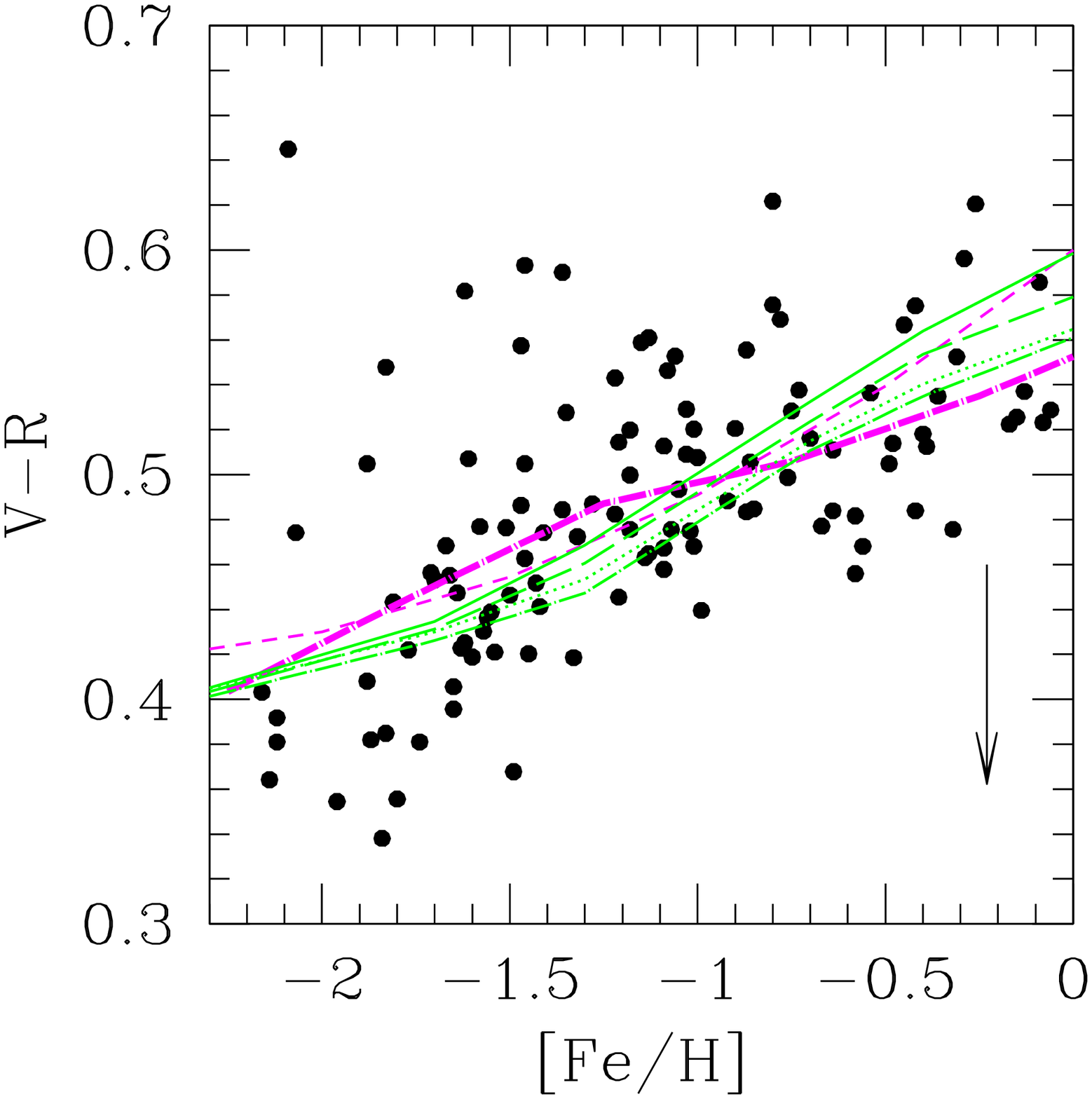}
\includegraphics[angle=0,width=0.69\columnwidth]{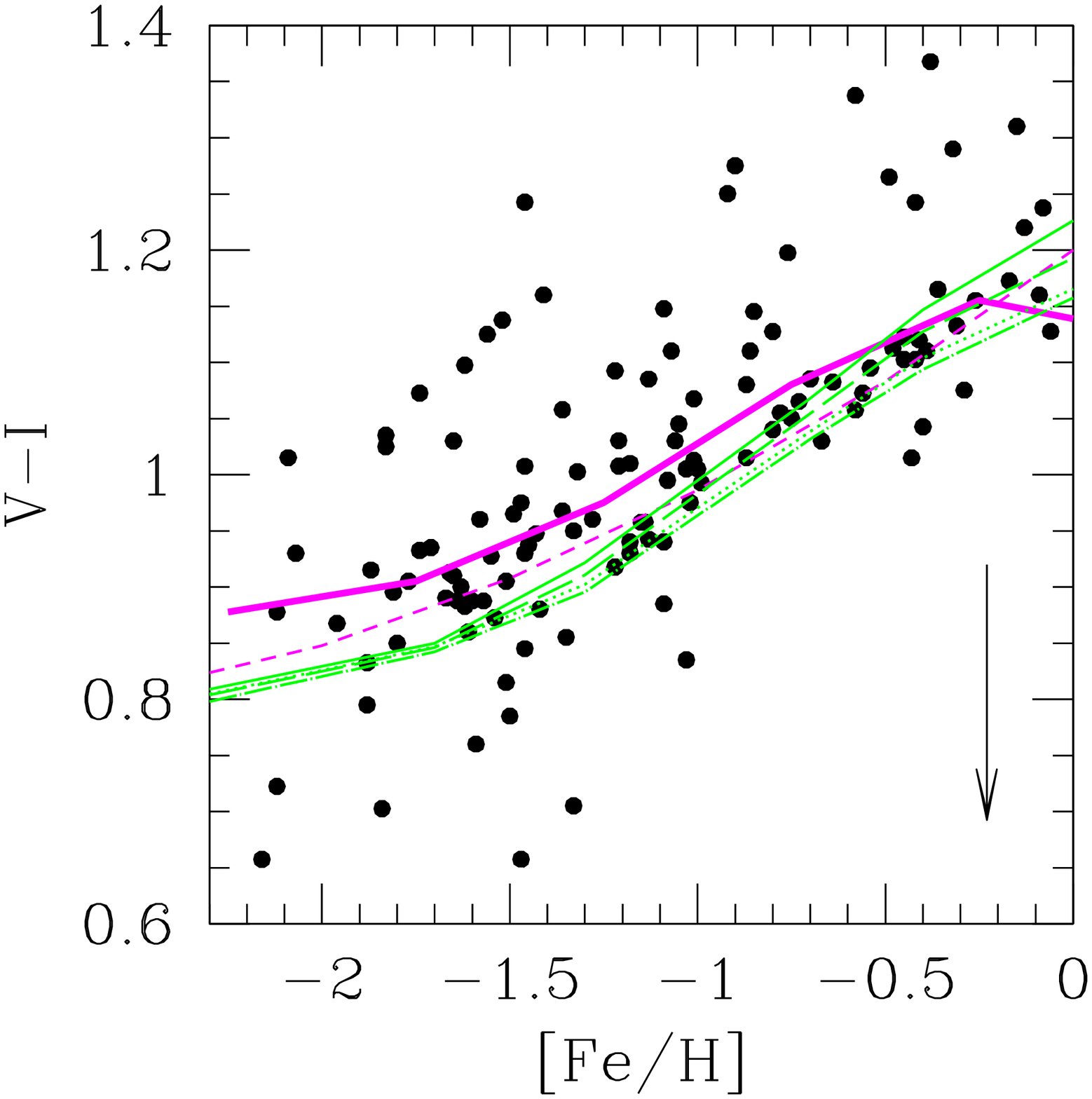}\\
\includegraphics[angle=0,width=0.69\columnwidth]{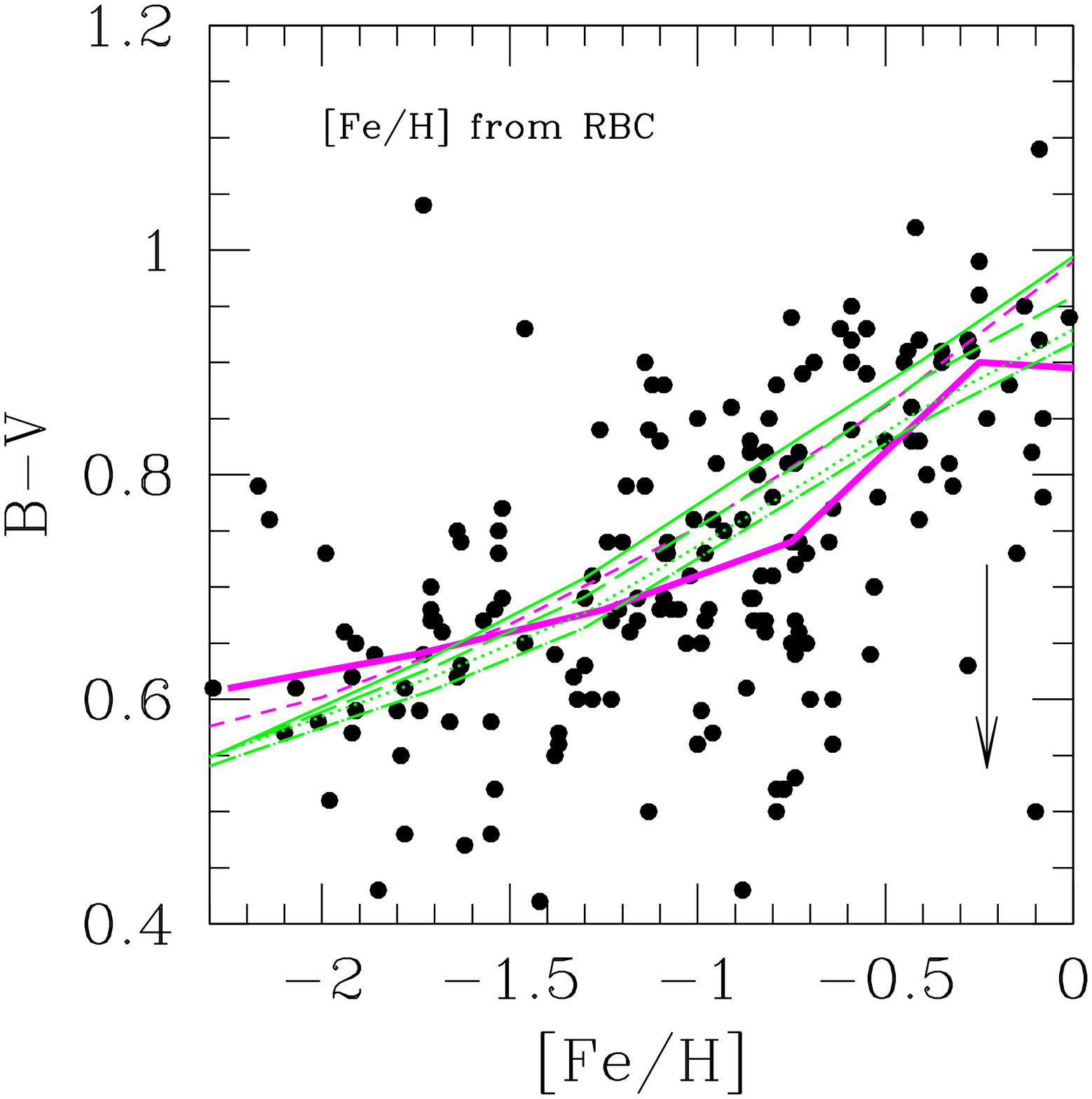}
\includegraphics[angle=0,width=0.69\columnwidth]{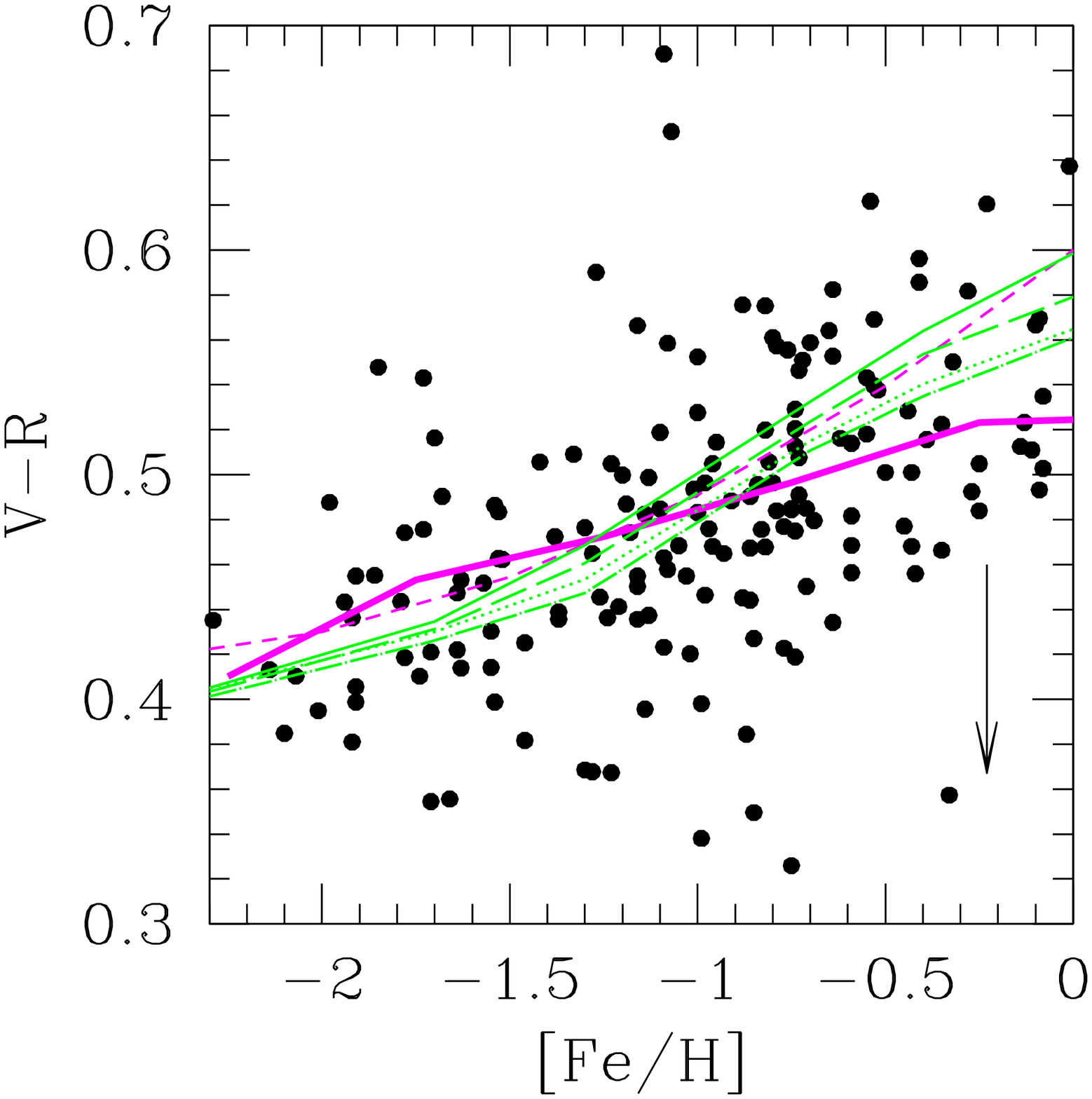}
\includegraphics[angle=0,width=0.69\columnwidth]{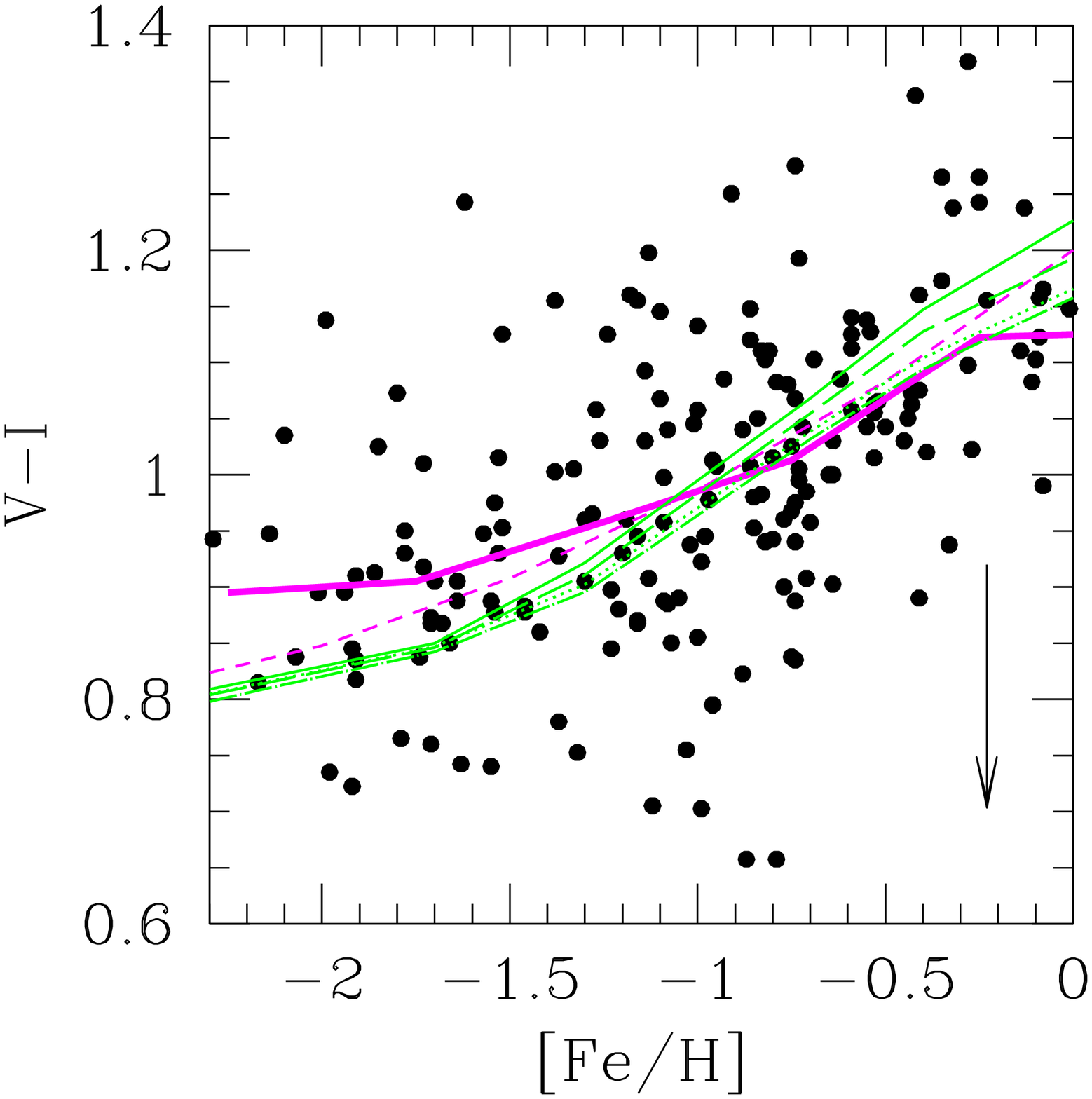}\\
\includegraphics[angle=0,width=0.69\columnwidth]{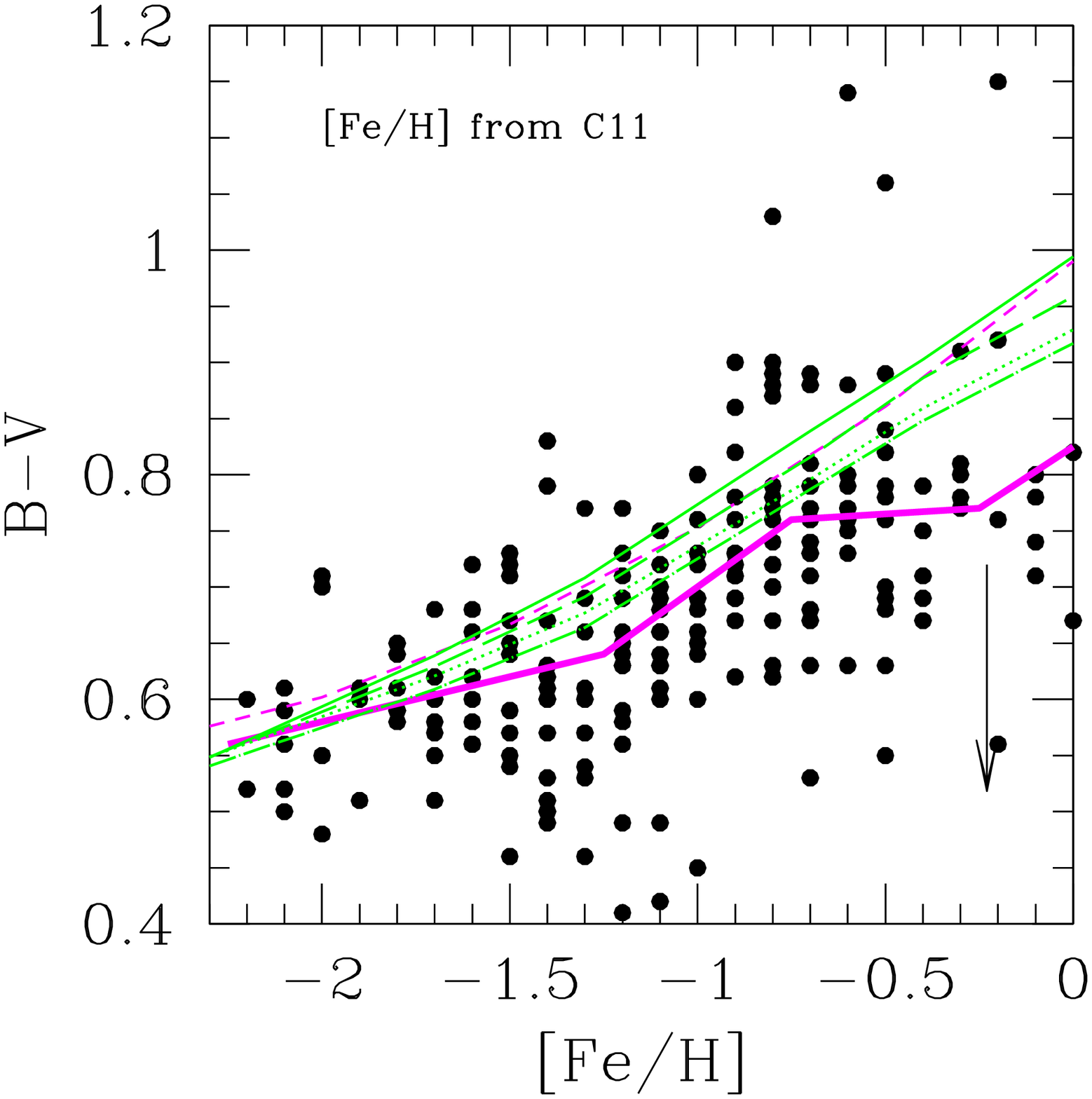}
\includegraphics[angle=0,width=0.69\columnwidth]{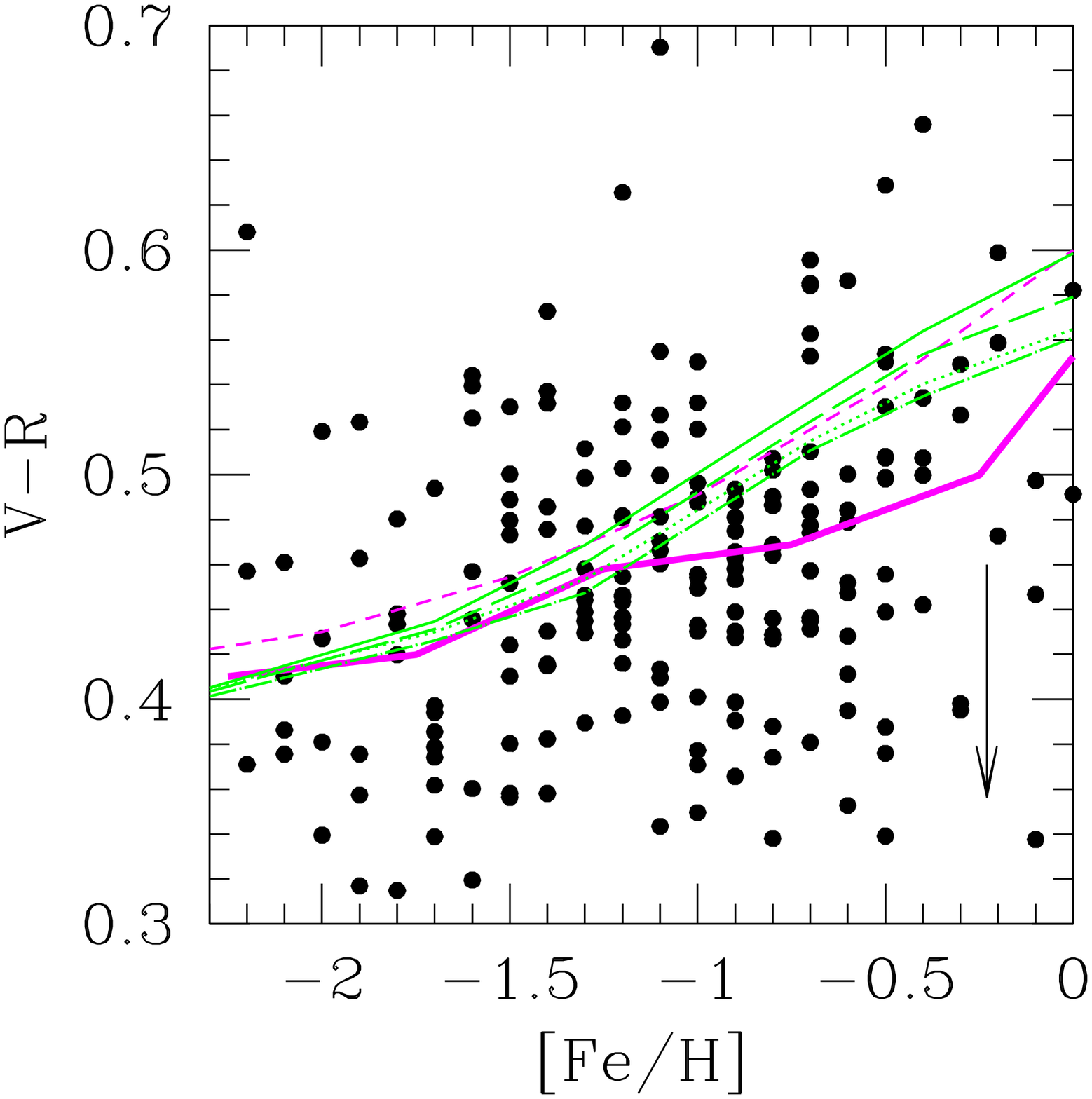}
\includegraphics[angle=0,width=0.69\columnwidth]{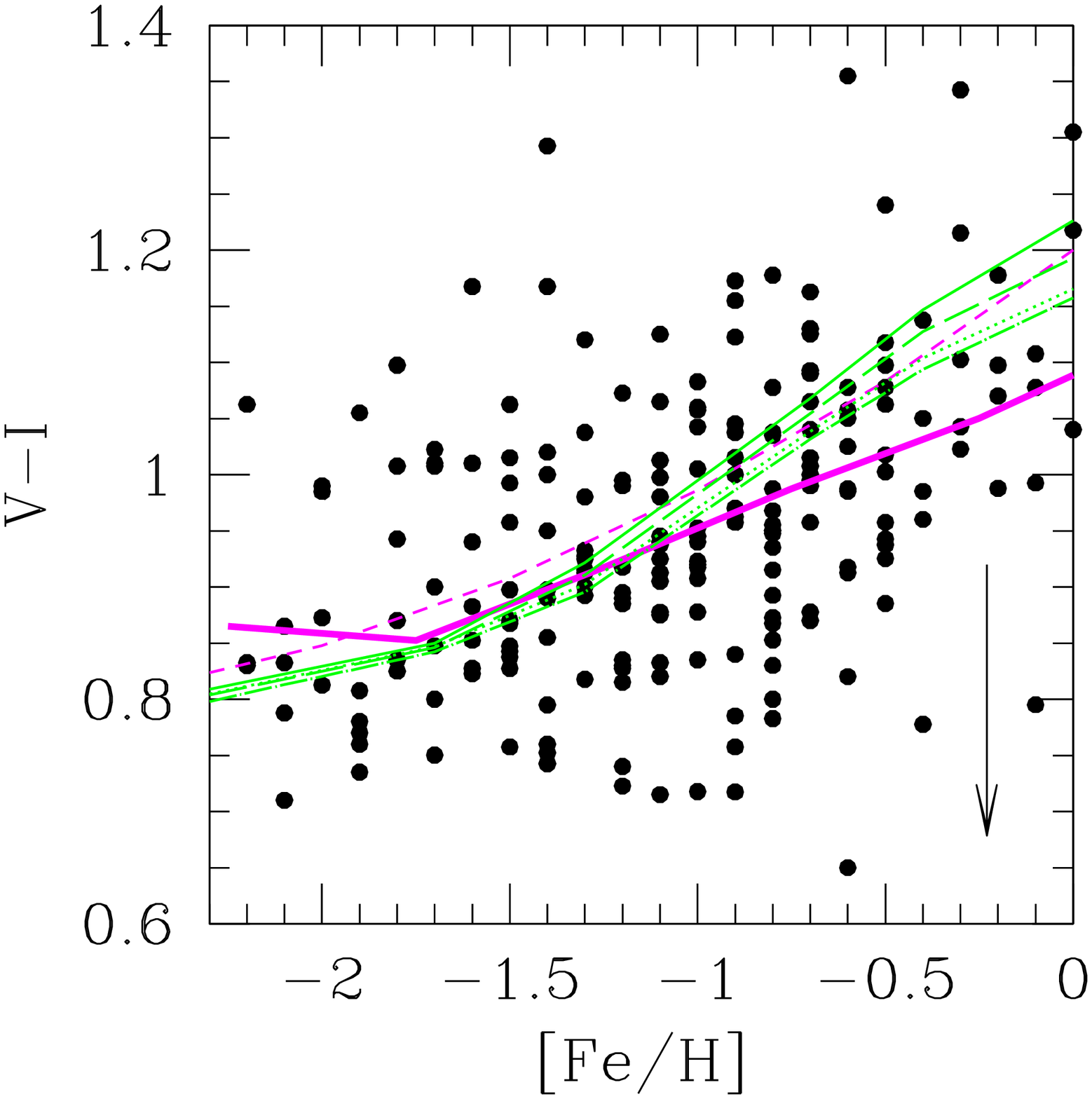}
\caption{Optical Johnson-Cousins colours plotted against the metallicity for M31
  globular clusters. In the first row of panels the metallicity and reddening measurements are
  taken from \citealt{Fan08}, in the second row they come from the
  RBC, whereas in the last row we consider the catalog of \citet{Caldwell11}. 
  Only clusters with reddening E(B$-$V)$\le$0.5 are
  shown. The vertical arrow indicates the mean amount of de-reddening applied.The magenta thick solid line indicates the median to the data
  calculated in metallicity bins of $-2, -1.5, -1, -0.5, 0 $ and $ 0.5 $   dex.
  The green lines are the synthetic colours derived
  from the MIUSCAT SSP SEDs for Kroupa Universal IMF and different ages: 12.6 (solid), 10
  (dashed), 8 (dotted) and 6\,Gyr (dashed-dotted). 
}
\label{M31_J}
\end{figure*}

\subsection{M31 globular clusters}

In this section we compare our models with the globular clusters of M31. For
this purpose we use different photometric catalogs and metallicity
estimates. 
We consider the spectroscopic metallicity determinations of \citet{Fan08}, which
is a compilation of various sources \citep{Huchra91, Barmby00,
Perrett02},  the Revised Bologna
Catalog\footnote{http://www.bo.astro.it/M31} (RBC),  
which are determined in a homogeneous
manner \citep{Galleti09} and the spectroscopic catalog of
\citet{Caldwell11}, hereafter C11. 
Integrated colours in the  Johnson-Cousins system
have been taken from the RBC, while photometry in the SDSS bands is from \citet{Peacock10}.
The extinction in each band has been derived from the reddening
values given in the corresponding catalog  and the \citet{Cardelli89}
extinction law. 
Given the higher uncertainty in the M31 measurements compared to the MW GCs, we
only consider clusters with low extinction values (E(B$-$V) $\le$0.5) to avoid
unreliable colour estimates. Only globular clusters classified as old in the
catalogs are considered.

\begin{figure*}
\includegraphics[angle=0,width=0.69\columnwidth]{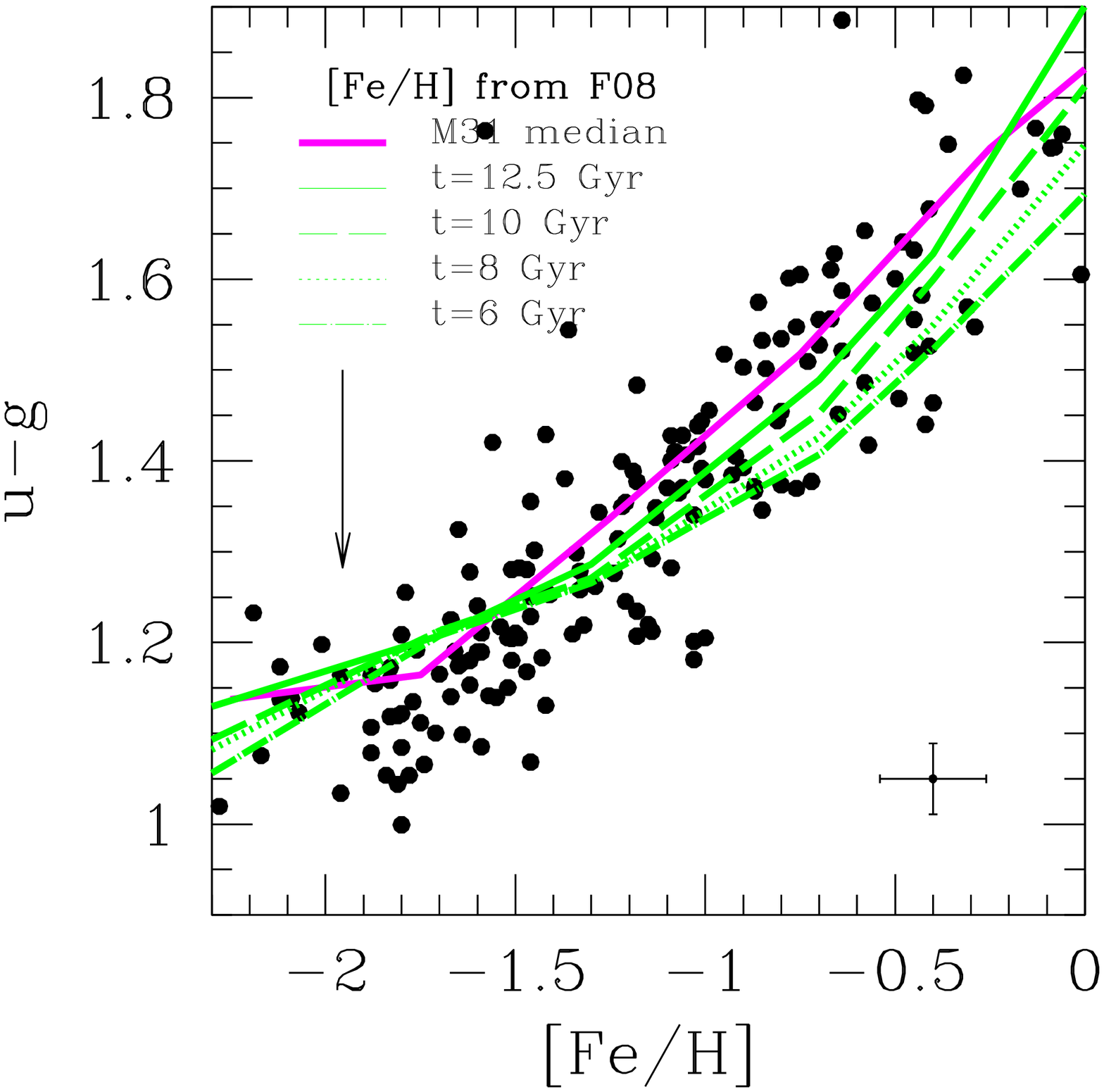}
\includegraphics[angle=0,width=0.69\columnwidth]{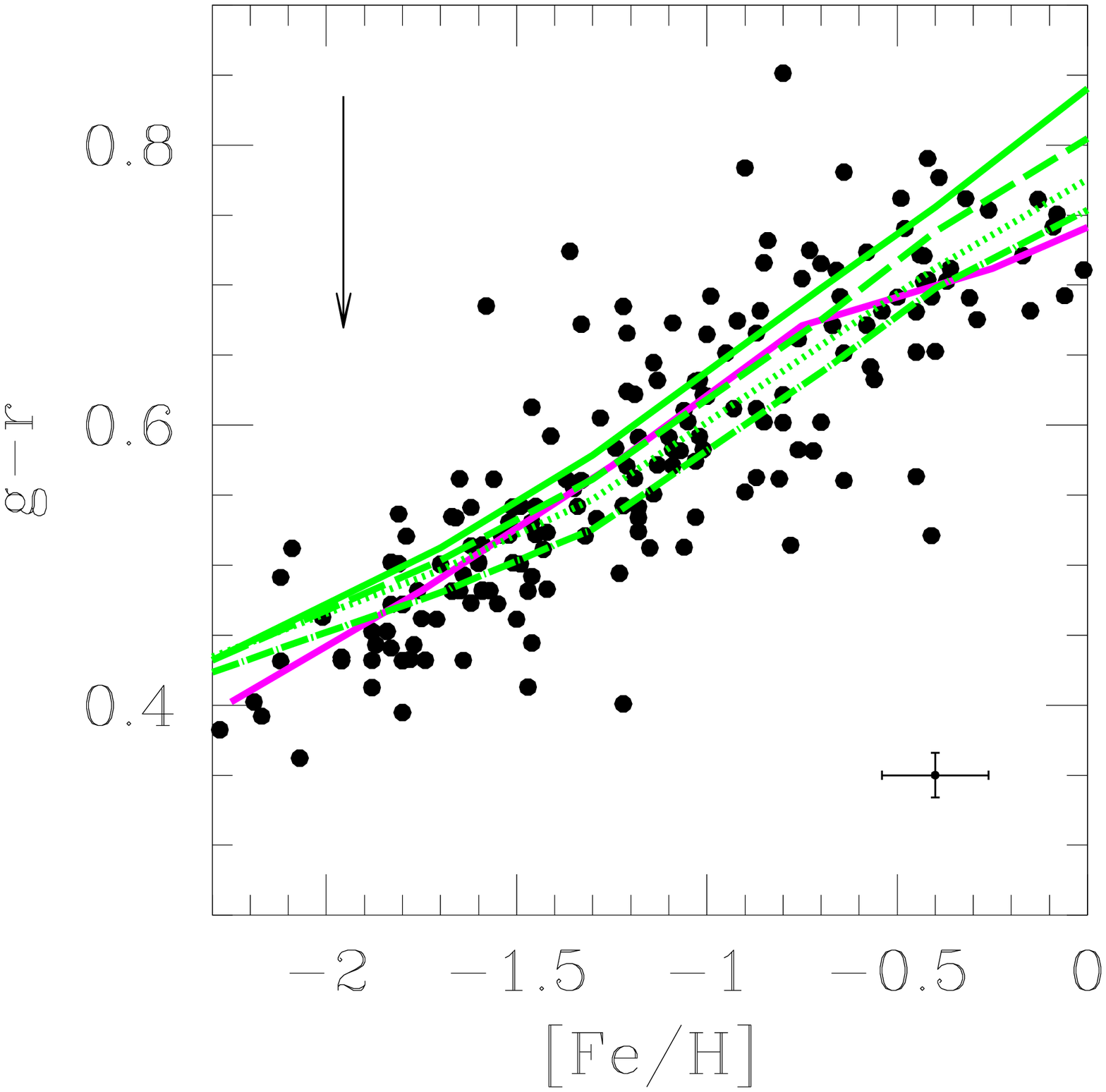}
\includegraphics[angle=0,width=0.69\columnwidth]{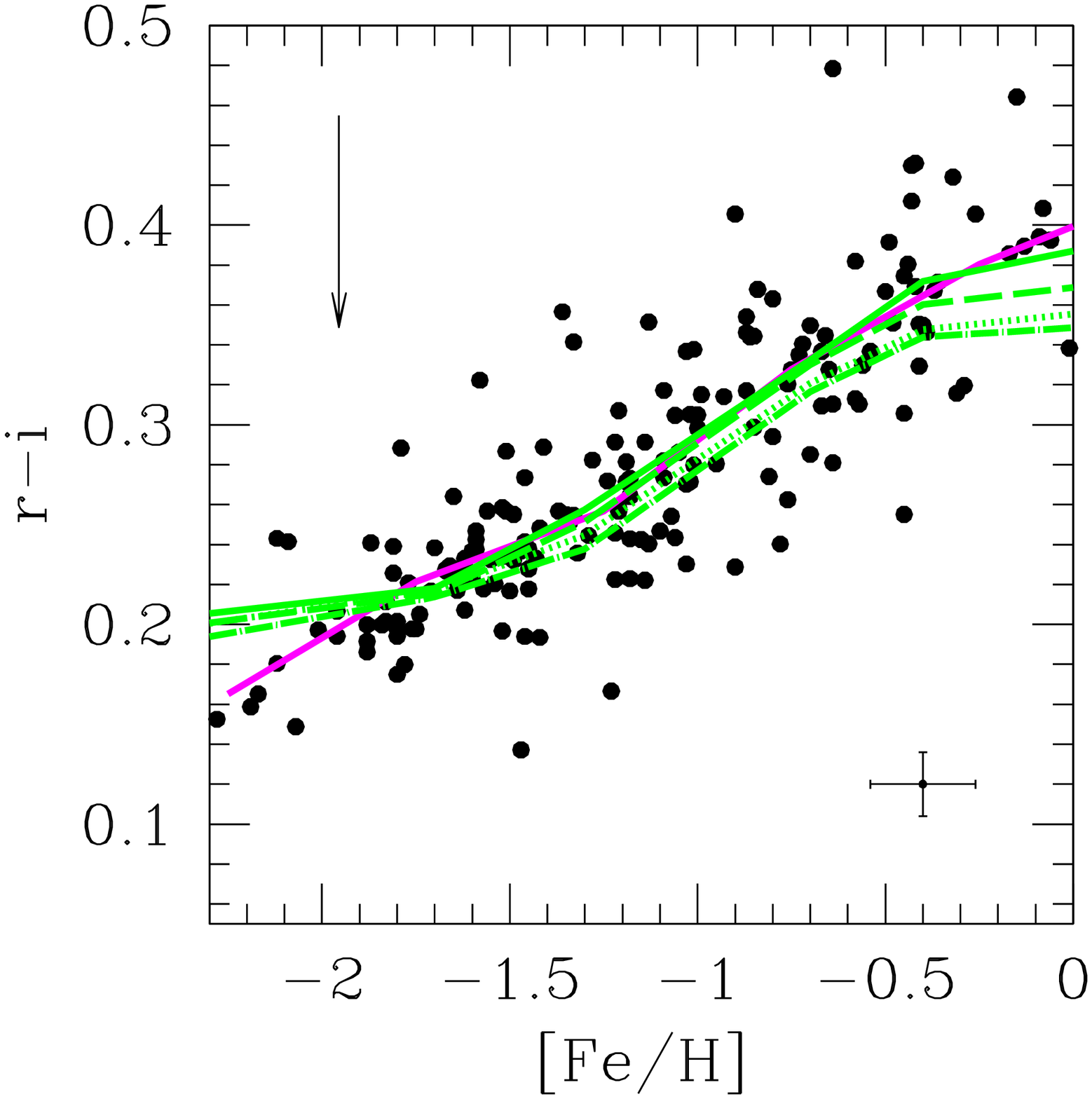}\\
\includegraphics[angle=0,width=0.69\columnwidth]{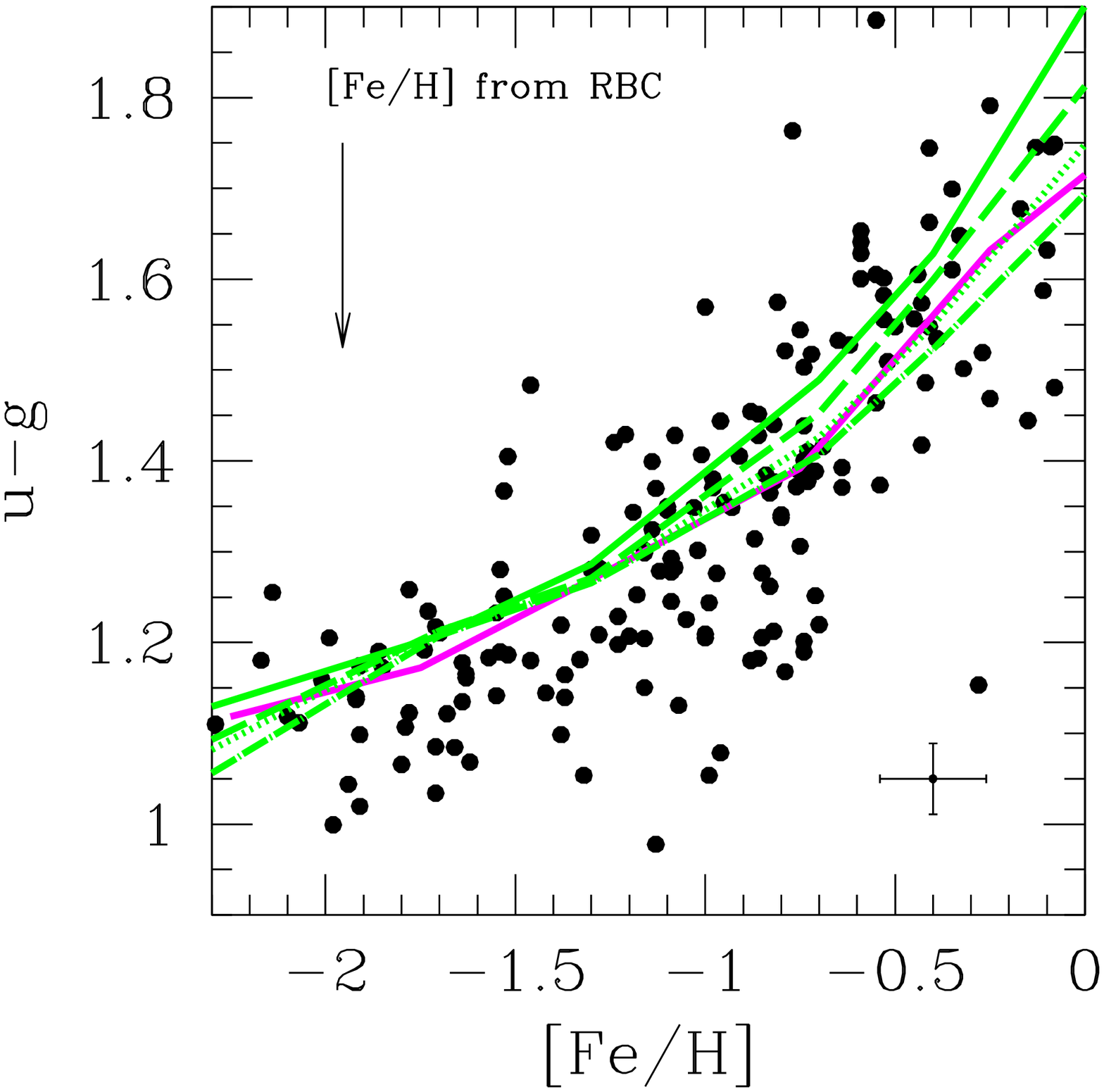}
\includegraphics[angle=0,width=0.69\columnwidth]{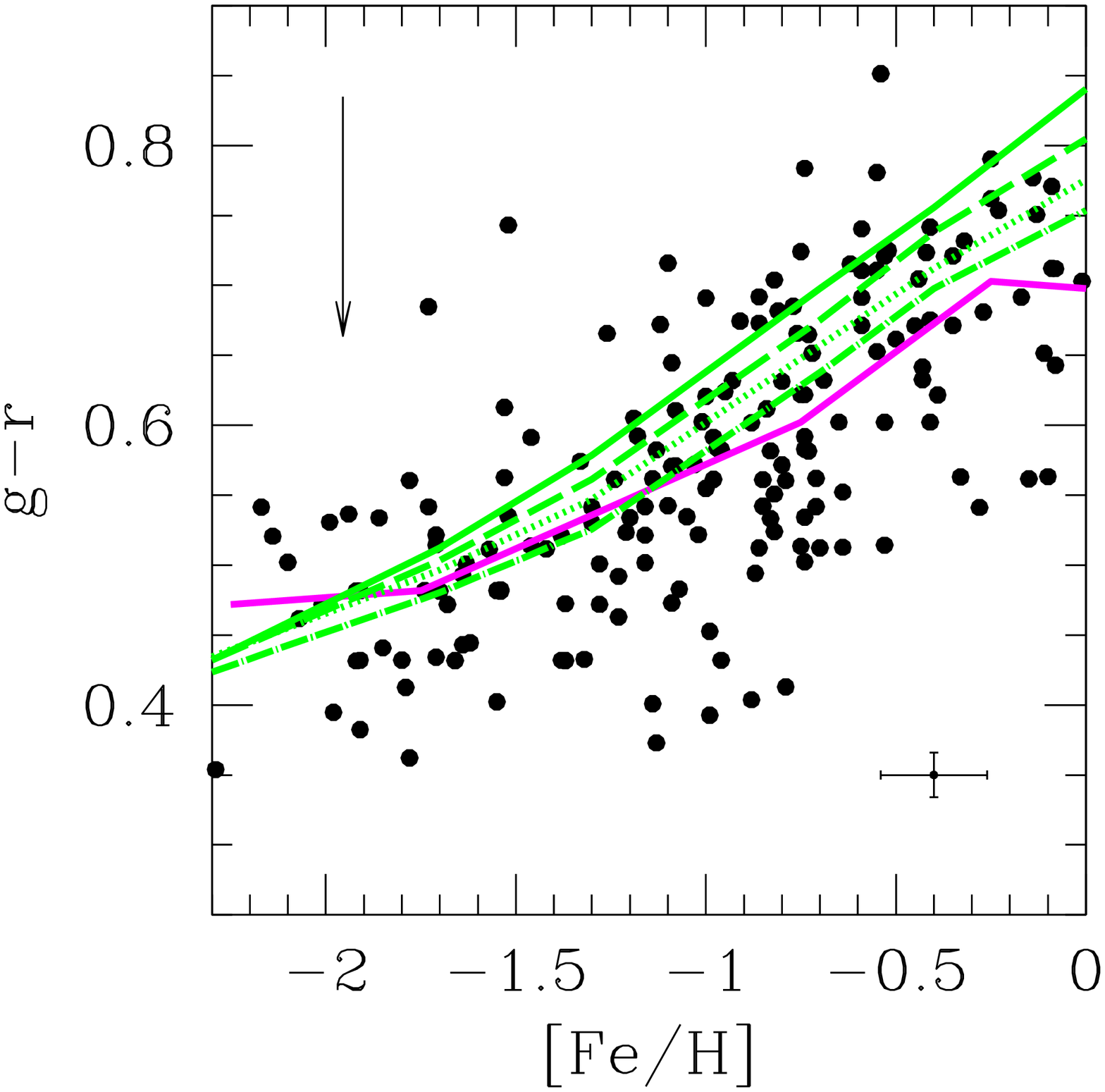}
\includegraphics[angle=0,width=0.69\columnwidth]{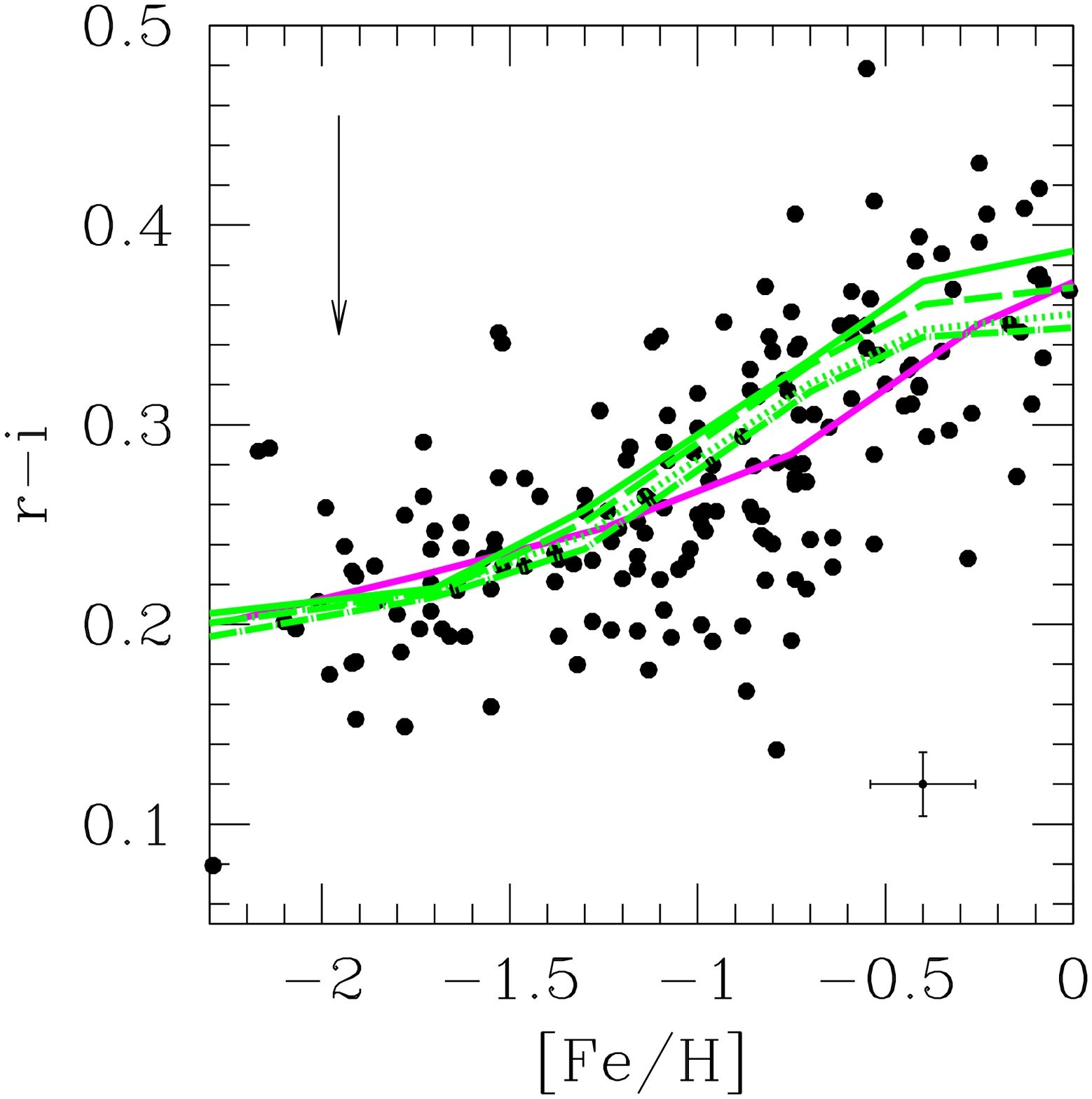}\\
\includegraphics[angle=0,width=0.69\columnwidth]{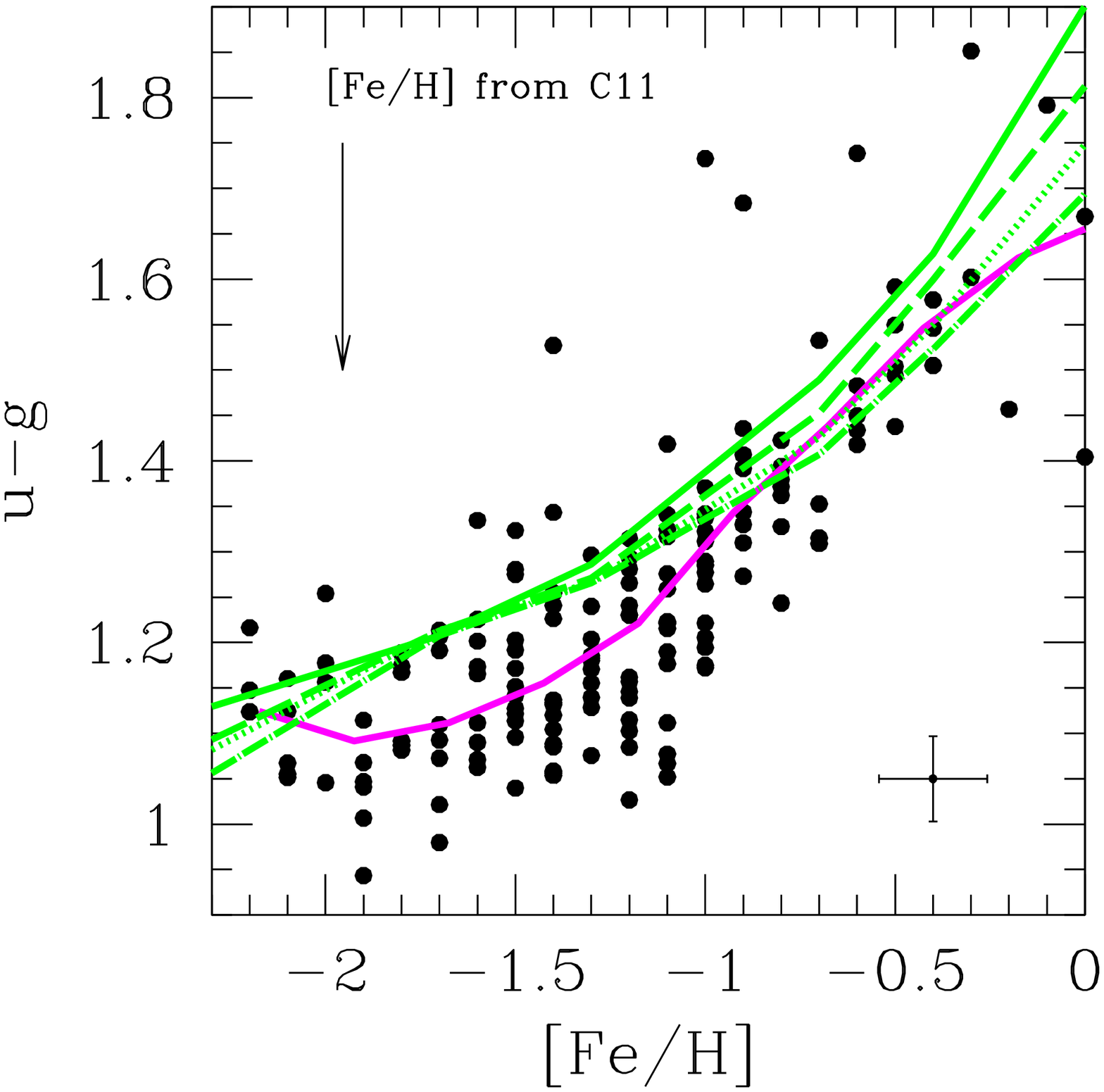}
\includegraphics[angle=0,width=0.69\columnwidth]{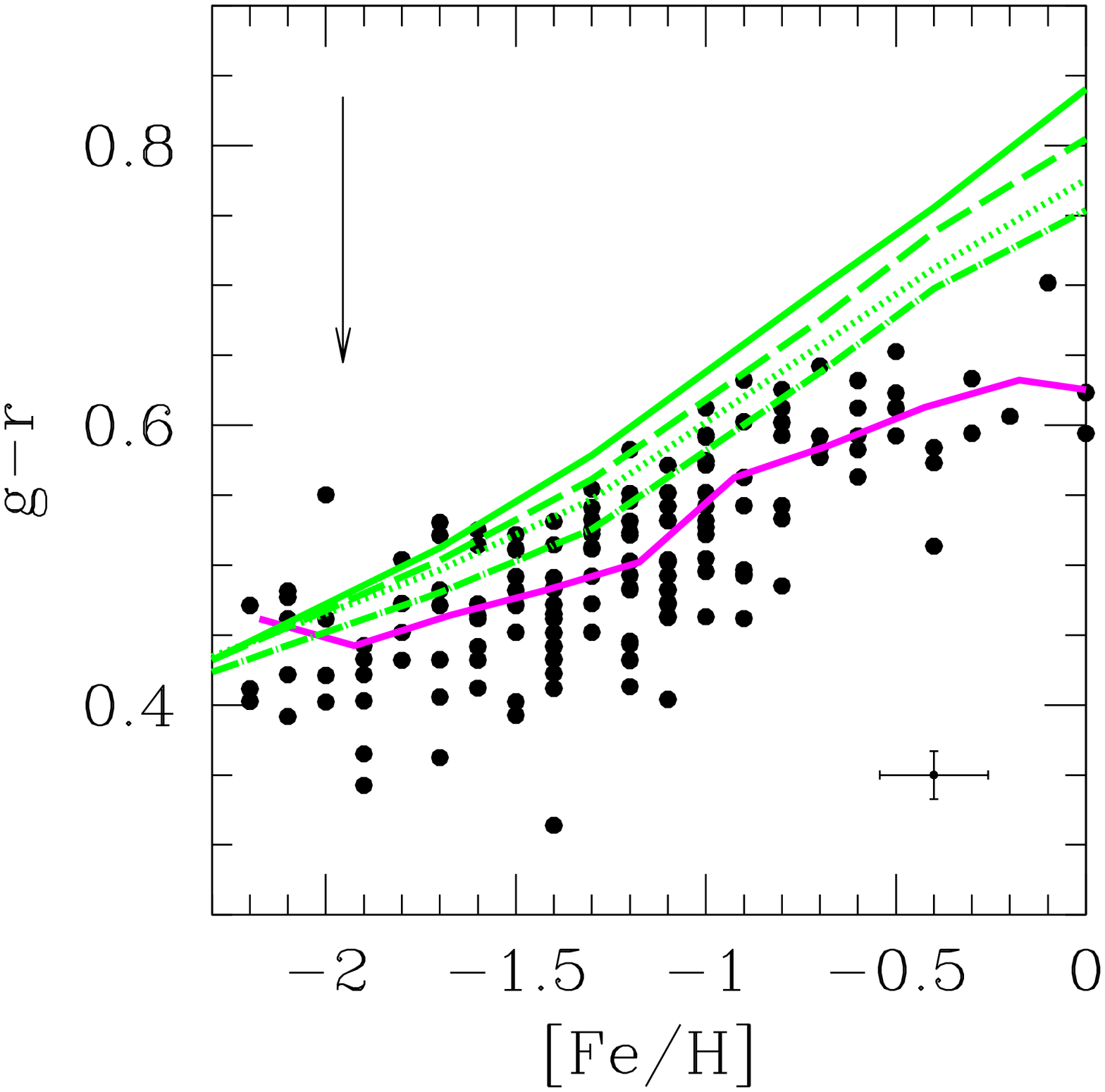}
\includegraphics[angle=0,width=0.69\columnwidth]{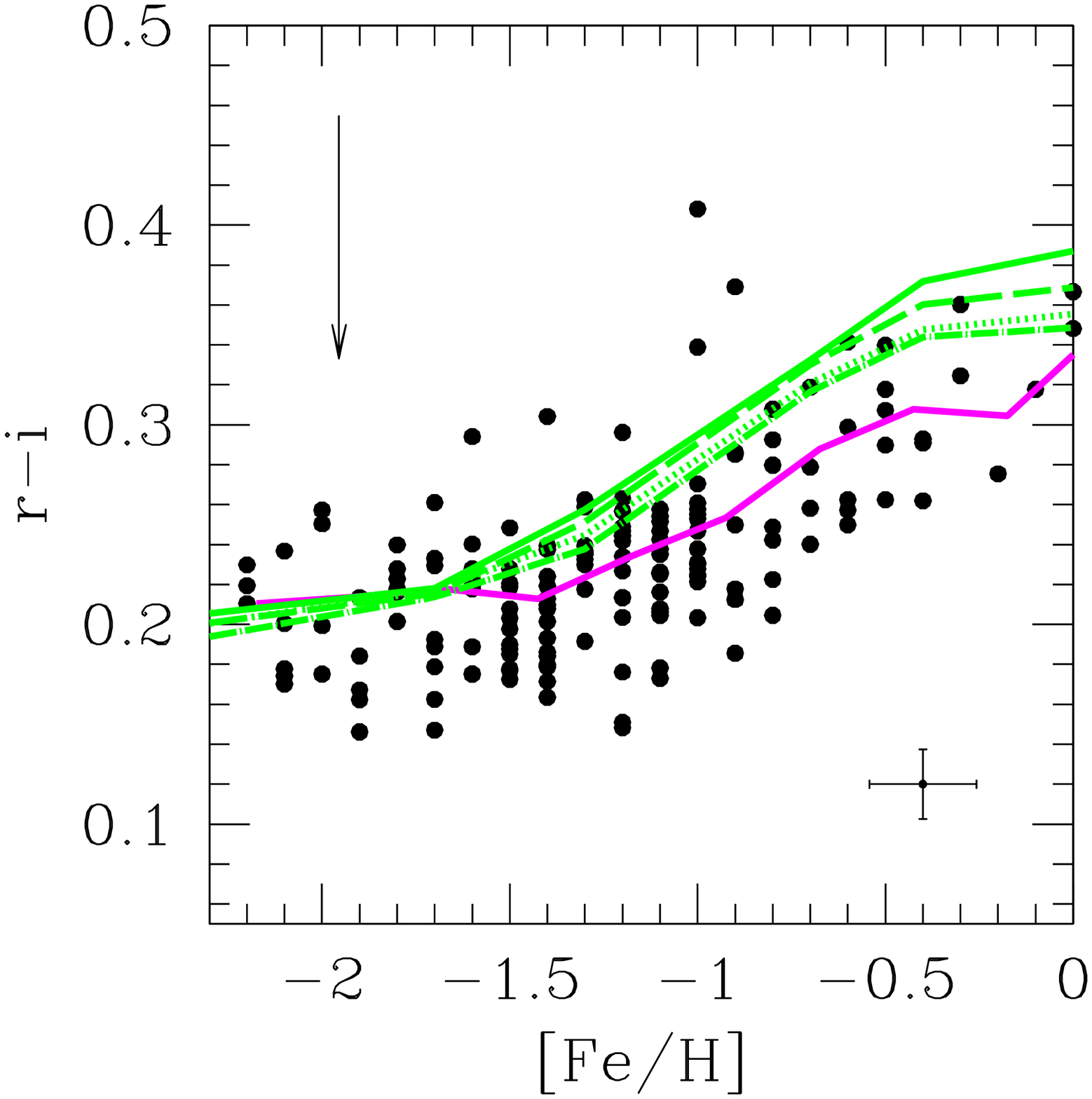}
\caption{Same as in Figure \ref{M31_J} but for colours in the SDSS
  bands. The errorbars in the lower-right corner indicate typical
  errors on the metallicity and the photometric measurements.}
\label{M31_sdss}
\end{figure*}

Figure \ref{M31_J} shows the Johnson-Cousins colours of the M31 GCs as a
function of metallicity. Given the high dispersion of the points we found that
the polynomial fit is not a good description of the data. To guide the eye, we
overplotted the median of the colours computed in intervals of
metallicity 
and compare it to the fit to the MW GCs. We
also show the model lines corresponding to the SSPs with ages ranging from 6 to
12.6\,Gyr (Kroupa Universal IMF). The Vega system used to compute the synthetic
magnitudes is the same as adopted for the MW data. The SDSS colours in the AB
system are shown in Figure \ref{M31_sdss}. Note that the scatter obtained for
the M31 data in these two figures is considerably larger than that observed in
the MW clusters. 
Moreover, as shown by the arrows, the M31 clusters display a considerable amount
  of reddening.  

Although the absolute colours of M31 can be subject to several
uncertainties, a robust result inferred from Figures 3 is that the
slope of the colour-metallicity relation is flatter in the M31
clusters. Given that the MW slope is consistent with that of the old
models, we argue that in M31 a significant  age variations with
metallicity is needed to account for the different slope. 
A flatter relation than that indicated from the models is also seen in Figure 4 with
SDSS colours. Old
stellar populations in the lower metallicity regime and young ages for the metal rich
clusters could explain such a trend.
 We obtain the worse fits to the data with the old models for the V$-$R and {\it
g$-$r} colours. Indeed, the M31 GC colours are in general bluer than their
Galactic counterparts, mainly in the high metallicity regime, as it can be
inferred from comparing to the MW fit. The reason for this discrepancy at nearly
solar metallicity may be linked to the fact that M31 clusters show higher
metallicities than their MW counterparts (e.g., \citealt{Beasley05} ).
In general the SDSS colours, even for intermediate metallicities,
are matched better with younger models. However, it is worth noticing that the
choice of the catalog may change our conclusion. In fact young models
are required to fit the trend in the {\it g$-$r} colour of
\citet{Fan08} dataset, whereas models of 6-8\,Gyr provide a fair match for all the colours of
the RBC and C11 catalogs.
From the comparison
with our model lines the best match to the data is obtained combining old models
at lower metallicities and younger models at the high metallicity regime.

In \citet{Peacock11} the synthetic {\it g$-$r} colour derived from the models
of \citet{Vazdekis10}  has been compared to the SDSS colours of M31 clusters
in a similar way as it is done here. Since the {\it g$-$r} colour falls in the
MILES range, it coincides with the predictions of the MIUSCAT models presented
here. In \citet{Peacock11} paper the \citet{Vazdekis10}  colours turned
out to be offset by $\approx$0.1 mag at solar metallicity with respect to M31 GC
colours. Although for the oldest model considered here we find similar
difference in colour, by using different colours and different photometric
systems we conclude that the data can be matched by considering younger models
than those in \citet{Peacock11}. Finally, as we obtain with the old models
very good fits to the MW clusters, which have more reliable data, and since
the M31 cluster colours in the same photometric bands differ from the ones of
the MW, we can conclude that we require younger models, at least in the high
metallicity regime, to fit the data.

Our results agree with 
evidences supporting that metal-rich GCs tend to be younger than the more
metal-poor clusters, at least in the MW (e.g.,
\citealt{Marin09, Deangeli05, BarmbyHuchra00,  Rosenberg99}). 
Moreover, it has been claimed that the GCs of M31 tend to be on average
younger than their Galactic counterparts \citep{Beasley05}, though this result is yet to be
conclusive. 
Spectroscopic ages for the high-metallicity M31 clusters have been
derived in C11.  
In the low-metallicity regime the authors could not get accurate age
determination, as these clusters are contaminated by blue HB stars,
hence we can not test the age variation with metallicity.
For the high-metallicity clusters in our sample ([Z/H]$\gtrsim -1$) the C11
determinations give a mean age of 11.3 Gyr, that apparently
contradicts the young nature of the M31 clusters.
Note however that these authors employed the Balmer
absorption line indices as age indicators and it is well known that
this method can give very uncertain absolute ages, even larger than
the age of the Universe \citep{Vazdekis01, Schiavon02, Mendel07, Poole10}.
Moreover, the models of \citet{Schiavon07} used in C11 
give broad-band colours in excellent agreement with our values.
For instance, a solar metallicity model with age of 11.2 Gyr  has a
B$-$V colours of 0.978 mag, very similar to B$-$V$=$0.977 mag that we
obtain for the MILES
models.
Both photometric predictions are much larger than the M31 colours
(Figure \ref{M31_J}), confirming the discrepancy between the spectroscopic
and photometric age measurements.

Furthermore, given
the high uncertainties involved in the extinction, even our results
should be taken with caution. 
Indeed, the parameter that varies mostly among the
  three catalogs is the reddening.
  The C11 estimates tend to be biased towards
  larger values than the other two catalogs, leading to bluer de-reddened
  clusters (see for instance the g$-$r colour in Figure
  \ref{M31_sdss}).  In C11 the quoted uncertainties for the $E(B-V)$
  is  0.1 mag, hence of the order of colour variations 
  due to different ages.  Therefore, it is not possible to draw
  robust conclusions.

\section{Colour evolution of Luminous Red Galaxies}

\begin{figure}
\includegraphics[angle=0,width=0.8\columnwidth]{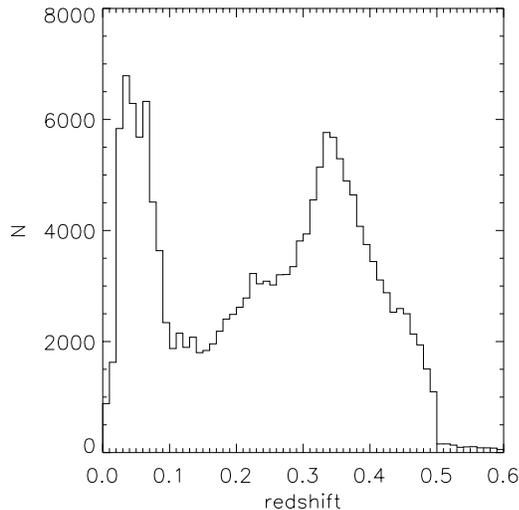}
\caption{Redshift distribution for the LRG sample, which is composed of
  168737 galaxies.}
\label{zdistr_lrg}
\end{figure}

In this section we consider the Luminous Red Galaxy (LRG) sample of
\citet{Eisenstein01}, for which typically the definition of early-type galaxy
(ETG) also applies. The LRG sample has been extracted from the SDSS
spectroscopic galaxy survey (Data Release 7, \citealt{dr7}), by selecting
galaxies with the GALAXY\_RED flag in the SDSS database. LRGs constitute a
uniform sample of objects with the reddest rest-frame colours, selected with
cuts in the ({\it g$-$r, r$-$i, r}) colour-colour-magnitude cube (see
\citealt{Eisenstein01}). To obtain the same galaxy population at different
redshifts, the LRG sample has been selected using two different selection
criteria below and above z=0.4, as the Balmer break moves from the {\it g} band
to the {\it r} band at this redshift. To increase the redshift range, galaxies
from the 2df SDSS LRG and Quasar (2SLAQ) survey \citep{Cannon06} have been
also included ($\approx$15000 objects) and they contribute to the high redshift
tail of the redshift distribution (see Figure \ref{zdistr_lrg}). By restricting
the sample to galaxies with reliable redshift determination (zWarning$=$0) we
end up with $\approx$170000 galaxies. 
The SDSS database provides a variety of
measured magnitudes for each detected object. Throughout this paper, we use
dereddened model magnitudes because they provide an unbiased colour estimate in
the absence of colour gradients \citep{Stoughton02}. 

Previous works \citep{Eisenstein01, Wake06} have highlighted the
difficulties of stellar population models in reproducing the colour evolution in
observed frame of LRG at intermediate redshift ($0.1<z<0.7$), with the
prediction of too red colours in the {\it g$-$r} and too blue in the {\it
r$-$i} for the lowest redshift LRGs, while the opposite effect is observed for
the highest redshift LRGs. Similar disagreements have been reported for red
sequence colours in galaxy clusters at similar redshifts \citep{Wake05}.

\begin{figure*}
\begin{center}
\includegraphics[angle=0,width=0.4\textwidth]{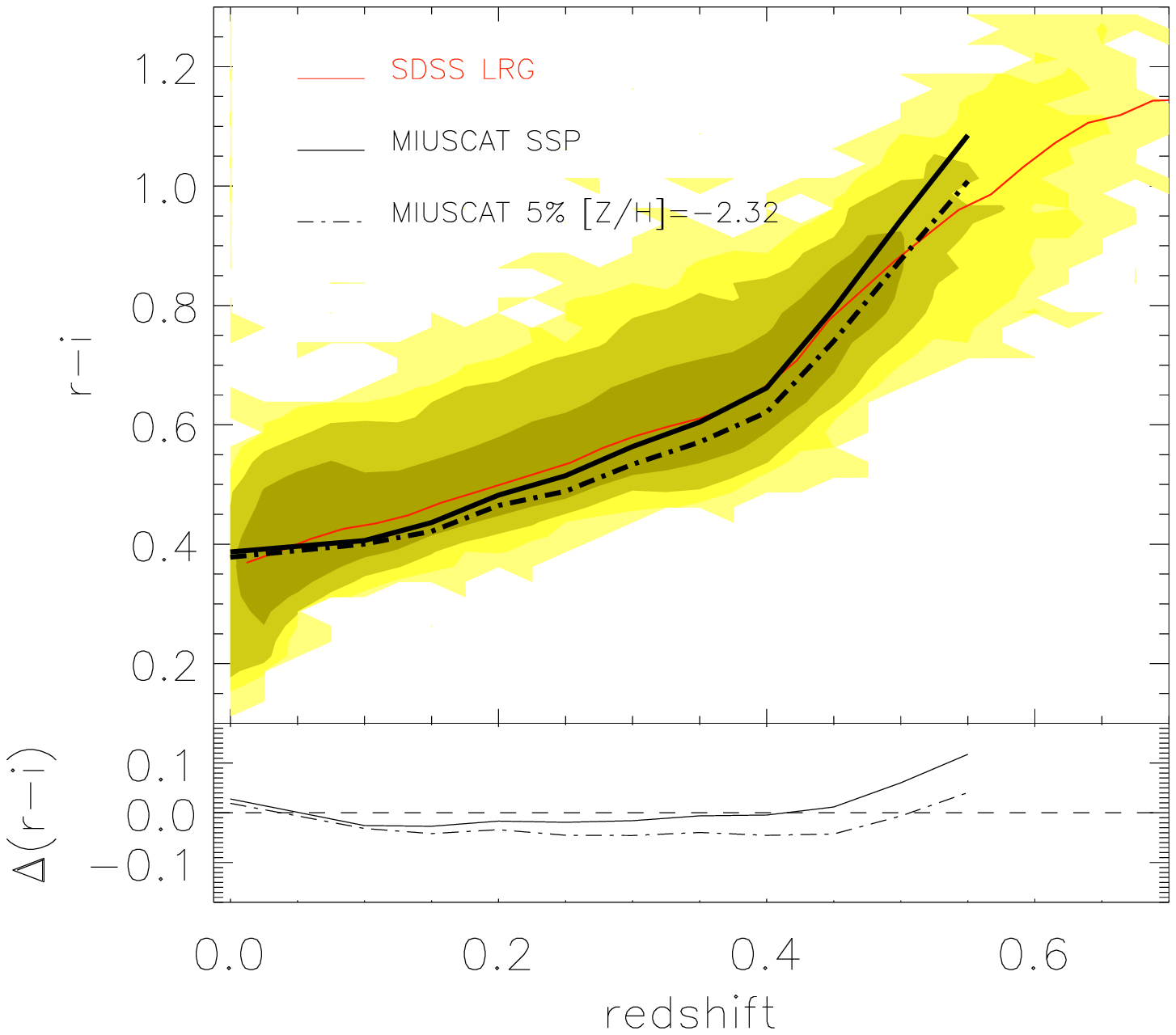}
\includegraphics[angle=0,width=0.4\textwidth]{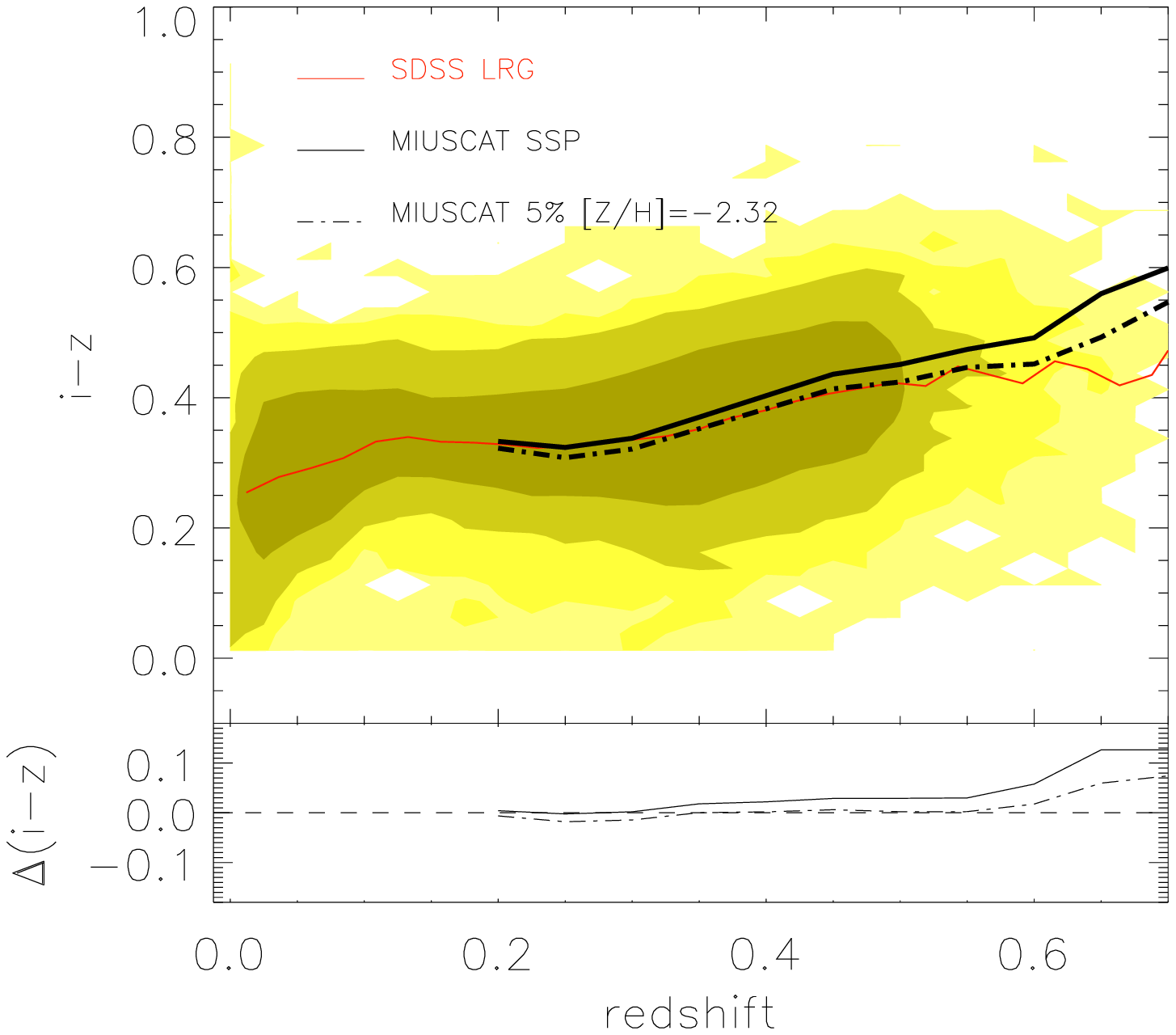}\\
\includegraphics[angle=0,width=0.4\textwidth]{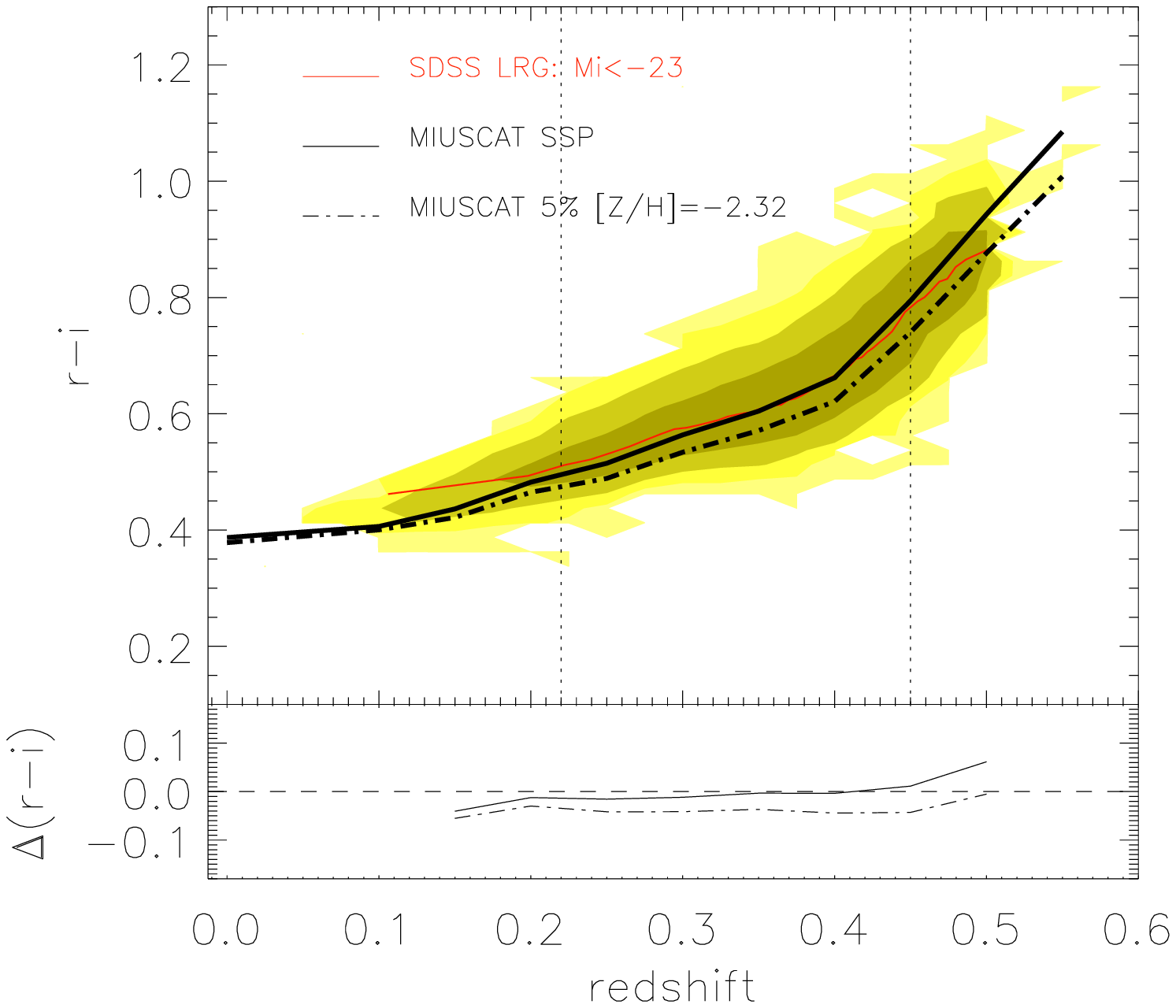}
\includegraphics[angle=0,width=0.4\textwidth]{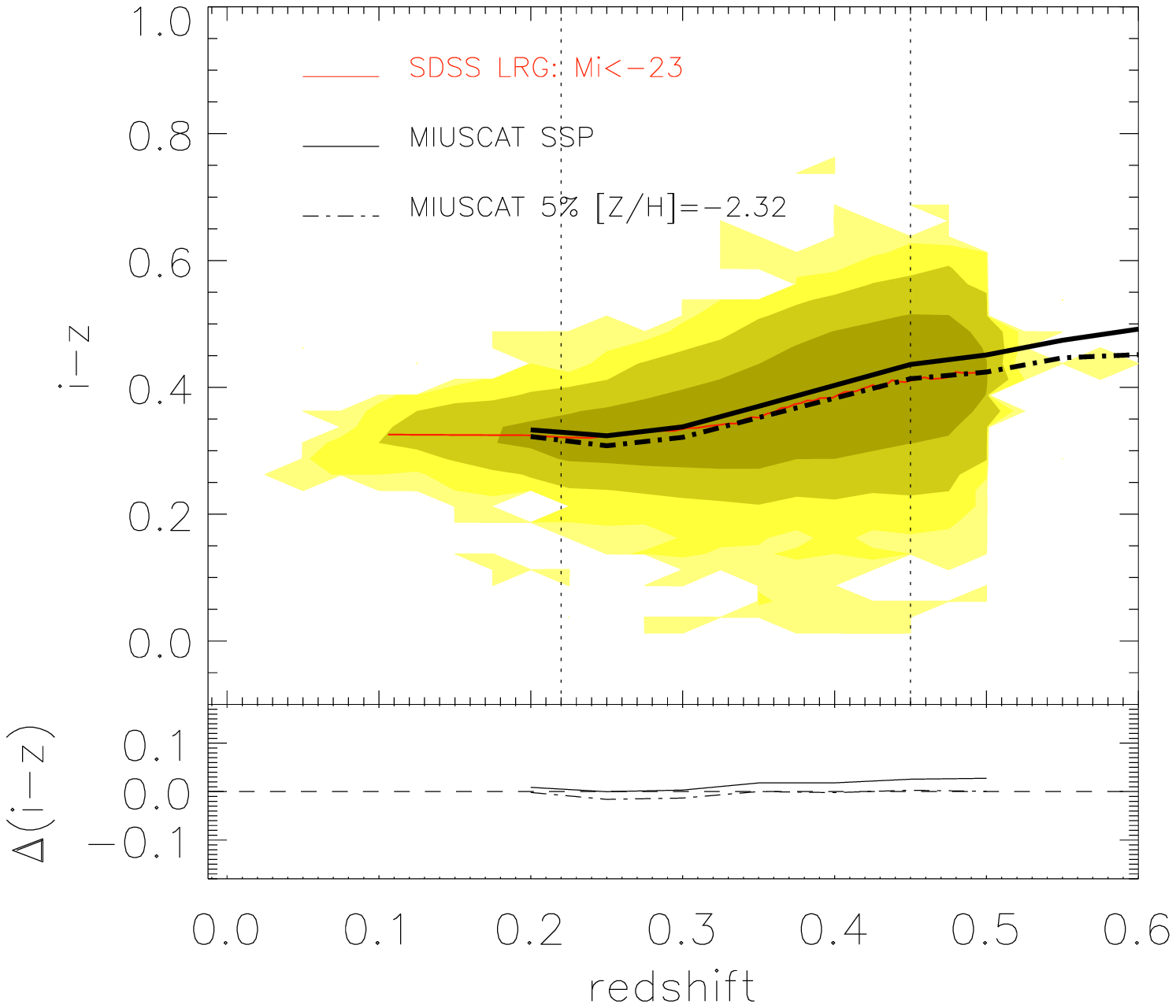}\\
\caption{
Colour evolution of the LRG sample in the  {\it r$-$i} (left-hand panel)
and {\it i$-$z} (right-hand panel) bands. 
The contours
enclose 20\%, 40\%, 60\% and 95\% of the galaxies. The red thin line shows the
median colour at different redshift bins equally spaced in intervals
of 0.025. The black thick line shows a
passive evolution model with solar metallicity and redshift of formation of 5
(corresponding to an age of 12.6 Gyr at z$=$0) for a Kroupa Universal
IMF. The dot-dashed line indicates a model that includes a 5\% contribution in mass from a metal-poor ([Z/H]=-2.32)
stellar population. The
residuals between the synthetic and observed colours are shown in
the bottom panels using the
same scale.
The bottom panels show the colour evolution for a subsample of LRGs
having  absolute magnitude: $M_{i}<$ -23 mag. The dotted vertical lines
indicate the redshift range where these bright galaxies are 
 a coeval population according to \citet{Toj11b}. 
 } 
\label{col1_z}
\end{center}
\end{figure*}

In Figure \ref{col1_z} we compare the predictions from the MIUSCAT models to the
colour evolution of the LRG sample. According to the fiducial model for the
formation of LRG, in the simplest case we assume that they formed in a uniform
burst at high redshift (z$=$5) and evolve passively thereon. Solar metallicity
and Kroupa Universal IMF are assumed. The observed colours are reproduced by
shifting the model spectra to the corresponding redshift and taking into account
the ageing effect. Synthetic colours in the SDSS filters have been computed in
the AB system \citep{Oke83}. For consistency, the observed colours from the
SDSS database have been transformed to AB system by applying the relative
shifts: $u_{AB}=u_{SDSS}-0.04$ mag; $z_{AB}=z_{SDSS}+0.02$
mag\footnote{www.sdss.org/dr7}.  
The model predictions are shown for the redshift
interval where the MIUSCAT spectra cover the {\it r$-$i} and {\it i$-$z}
colours. The {\it g} filter falls outside the measurable range at z$=$0.05, thus
colours containing {\it u} and {\it g} are not shown. We see that the simplest
model is able to reproduce the observed distribution at low redshift in both
colours.  The model starts to deviate from the median colours at z$\sim$0.4,
producing too red colours. 

In Figure \ref{col1_z} it is represented also a second model, where a small
fraction of the mass (5\%) is contributed by a metal-poor population
([Z/H]$=$-2.32). It is expected \citep{Worthey05, Pagel97}  that galaxies
exhibit some spread in the metallicity distribution with a contribution also 
from metal-poor stars. The effect of including stars with such a low metallicity
is to bluen the colours, with the result of loosing the match with  the observed
distribution in the lowest redshift range, but improving it at high redshift. 
It is worth to say that the differences between this model and the median of
observed colour is only $\approx$ 0.02 mag. The uncertainties derived from the
zero-point calibration of the SDSS system (0.01 mag) and from the choice of the
IMF (0.01 mag of difference between Salpeter and Kroupa Universal) could also
account for this difference. Hence, it turns out to be difficult to discriminate
among these two possibilities. Nevertheless, in both cases the MIUSCAT models perform 
rather well, confirming the results of \citet{Maraston09}, according to which
models constructed with empirical stellar libraries provide a good fit to the
LRG colour distribution.

Although the LRG sample has been selected to follow a passive
evolution model (see \citealt{Eisenstein01}),  this does not guarantee that all
the galaxies accomplishing the selection criteria follow this model.
In \citet{Toj11} it has been shown that a non negligible
fraction of LRGs does not evolve passively in a dynamical sense. They
can undergo mergers and the sample can be contaminated with galaxies
that belong to the LRG sample for a short period of time.
However, the brightest tail of the sample turns out to be coeval 
 (i.e. with almost no merging) 
for galaxies in the redshift range: 0.22-0.45. For this reason, in the bottom
panels of Figure \ref{col1_z} we compare our passive evolution
models with these bright galaxies ($M_{i} <$ -23
mag).  
For this subsample, the high-redshift galaxies are truly the progenitors of
the low-redshift ones.
In the redshift range where the galaxies appear  
coeval, our passive model performs remarkably well.

\section{Nearby Early-Type Galaxies}

\begin{figure}
\includegraphics[angle=0,width=0.8\columnwidth]{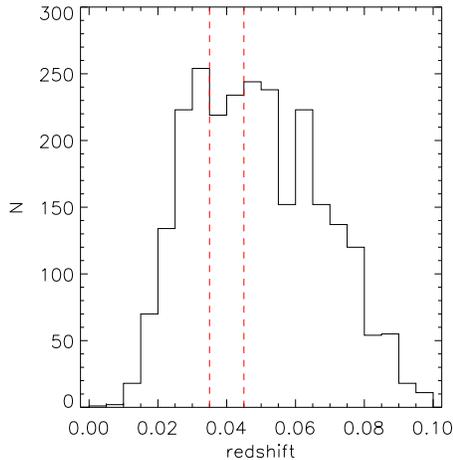}
\caption{
Redshift distribution for the ETG sample of \citet{Nair10}. The red lines
indicates our sample at z$\simeq$0.04, which includes 370 objects.
  }
\label{zdistr}
\end{figure}

\begin{figure*}
\includegraphics[angle=0,width=0.8\columnwidth]{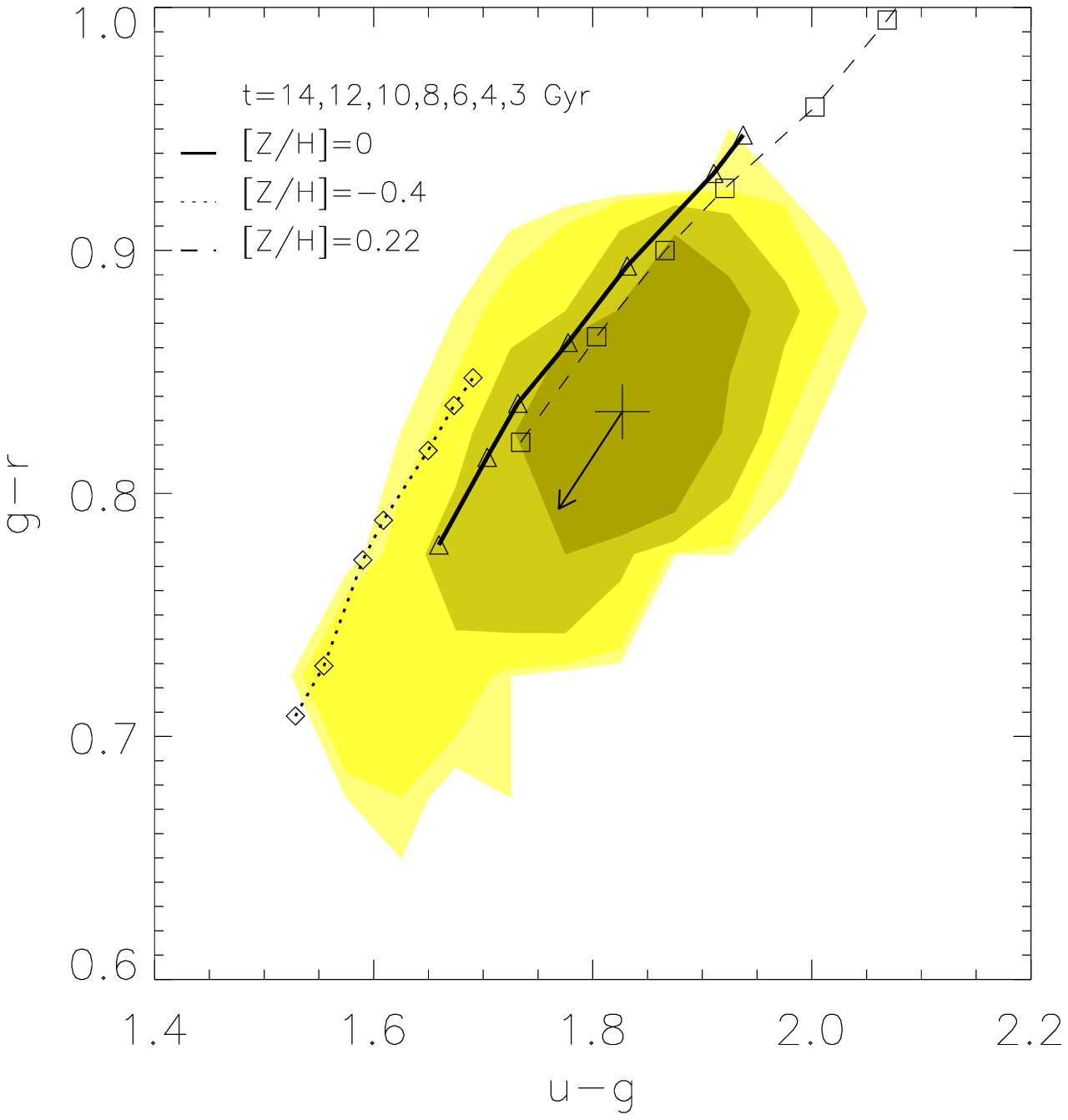}
\includegraphics[angle=0,width=0.8\columnwidth]{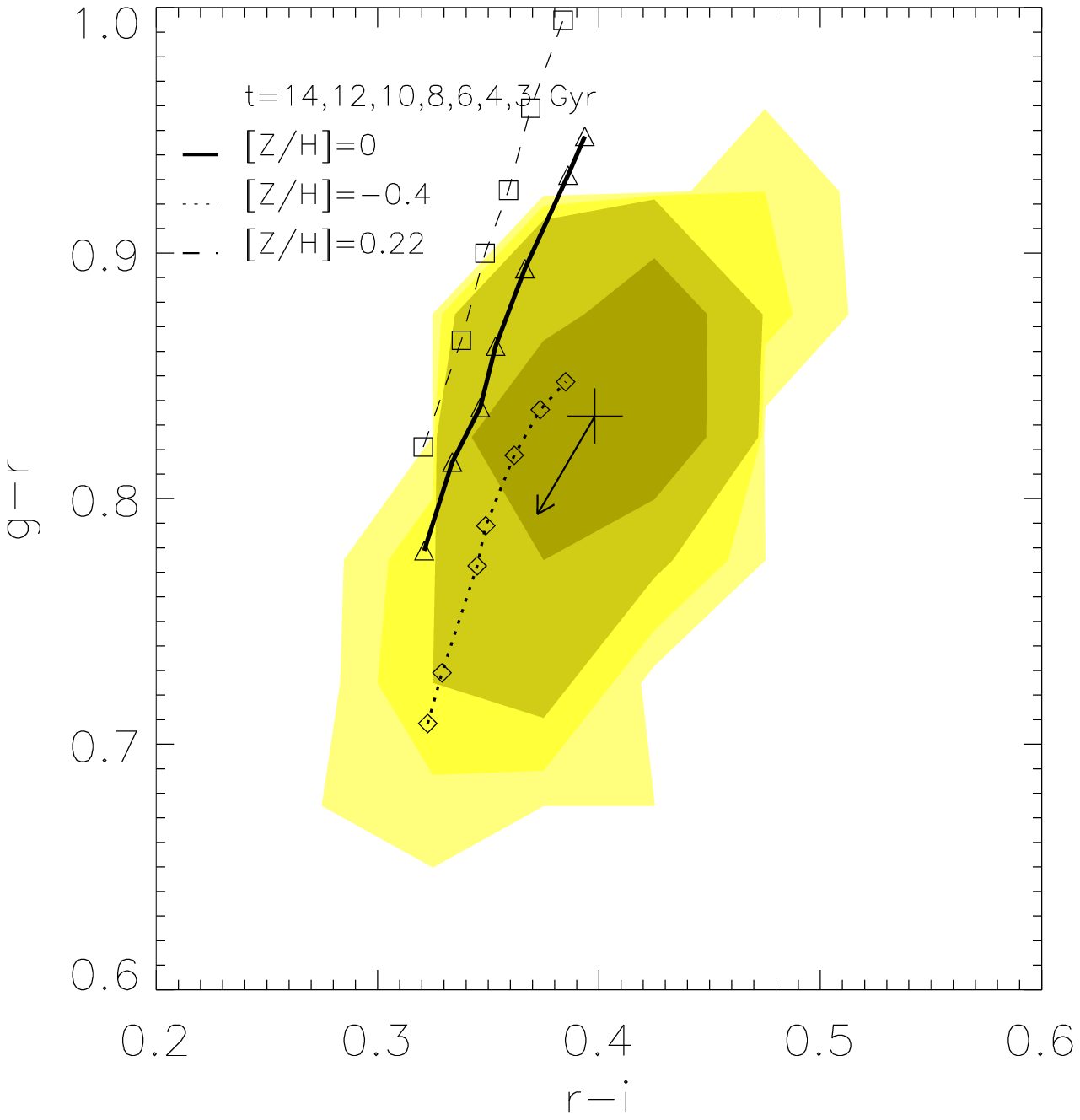}
\caption{
Observed colour distributions in the SDSS bands for the ETG sample at
z$\simeq$0.04. The contours enclose 20\%, 40\%, 60\% and 95\% of the galaxies
and the cross indicates the median colours ($u-g=1.83$, $g-r=0.83$,
$r-i=0.40$ mag). The arrow indicates the colour
de-reddening when an internal extinction of $A_V=0.2$ mag is assumed. Synthetic
colours derived from the MIUSCAT SEDs for a Kroupa Universal IMF are
overplotted. The different lines join models with the same metallicities for
ages ranging from 3 to 14\,Gyr.
}
\label{etg_ssp}
\end{figure*}

A further insight into the capabilities of the models can be obtained by
comparing them to a sample of nearby galaxies, for which we can use three
different colours as constraints. For this purpose we select a homogeneous
sample in the Local Universe with a large number of SDSS galaxies with visual
morphological classification \citep{Nair10}. We select a subsample of early-type
galaxies (morphological type $T\le-4$) with reliable redshift measurements. We further restrict our sample to a narrow
redshift slice, $0.035<z<0.045$, to end up with 370 targets. This choice of
redshift is motivated by two main reasons. First, at this redshift the number of
galaxies peaks (see Figure \ref{zdistr}). Second, given the limited  spectral
coverage of the MIUSCAT models in the blue end, the correction factors we need
to apply to measure the SDSS {\it u} magnitude become too large for redshifts
above this value, as the wavelengths at which the {\it u} filter response peaks
start to fall outside the spectral range of the models (see Paper I).
Before comparing
the colours derived from the MIUSCAT SEDs, once redshifted to z$=$0.04, 
we have tested that the selected redshift interval is small enough
that colour differences due only to variation in redshift within such
interval are negligible ($<$0.01 mag). 

As for the photometric estimates,
we rely on the model magnitudes from the DR7 catalog. It is worth noticing that
all the colours involved in this study belong to the optical spectral range,
which are known to be heavily affected by the age/metallicity degeneracy,
irrespective of the colour-colour diagnostic diagram in use. Therefore the
results obtained in this section are not intended to provide a well-constrained
picture for the stellar populations of these galaxies, which might require a
deeper analysis involving colours in other wavelengths and/or spectra. 

In Figure \ref{etg_ssp} we show the colour distribution of the ETG sample
compared to the MIUSCAT SSP models with ages ranging from 3 to 14\,Gyr for three
metallicity values: [Z/H]$=$-0.4, 0 and 0.22. We also include an arrow
indicating the effect of internal dust extinction on the observed colours. We
assume $A_V=0.2 $ mag, in agreement with the typical value derived for low
redshift LRG \citep{Toj11}. 
The de-reddening in the various SDSS bands has been
computed with the extinction law: 
$\tau_{\lambda}=\tau_V(\lambda/5500\AA)^{-0.7}$
 \citep{CF00}. We see that none of the SSP models is able to
match the observed colour distribution. Although the fiducial model for the
formation of ETGs predicts that they form in a uniform burst at high redshift
with high metallicity, the old models tend to produce too red colours in the
{\it g$-$r} vs {\it u$-$g} plot. On the contrary, the {\it r$-$i} colours turn
out to be too blue with respect to the real galaxies. Hence, this simple picture
does not seems to appropriate to describe the stellar populations of this galaxy
sample.  

The MIUSCAT models adopt the stellar isochrones of \citet{Girardi00}  (i.e.
Padova 2000), as described in Paper I. It has been claimed that these tracks
provide blue colours for old elliptical galaxies. \citet{BC03} claimed that
models computed with the isochrones of \citet{Bertelli94}  (i.e. Padova 1994)
are preferred because redder V-K colours are predicted, in better agreement with
the observations. For the optical spectral range the net effect of adopting
these two sets of stellar tracks is almost negligible. 
Using the Padova 1994 tracks, \citet{BC03} obtained redder
colours than using Padova 2000: 
{\it u$-$g} by $\approx$0.02 mag and {\it g$-$r}
and {\it r$-$i} by $\approx$0.01 mag. Therefore these results tend to worsen the
obtained mismatch for the bluer colours.

\begin{figure*}
\includegraphics[angle=0,width=0.8\columnwidth]{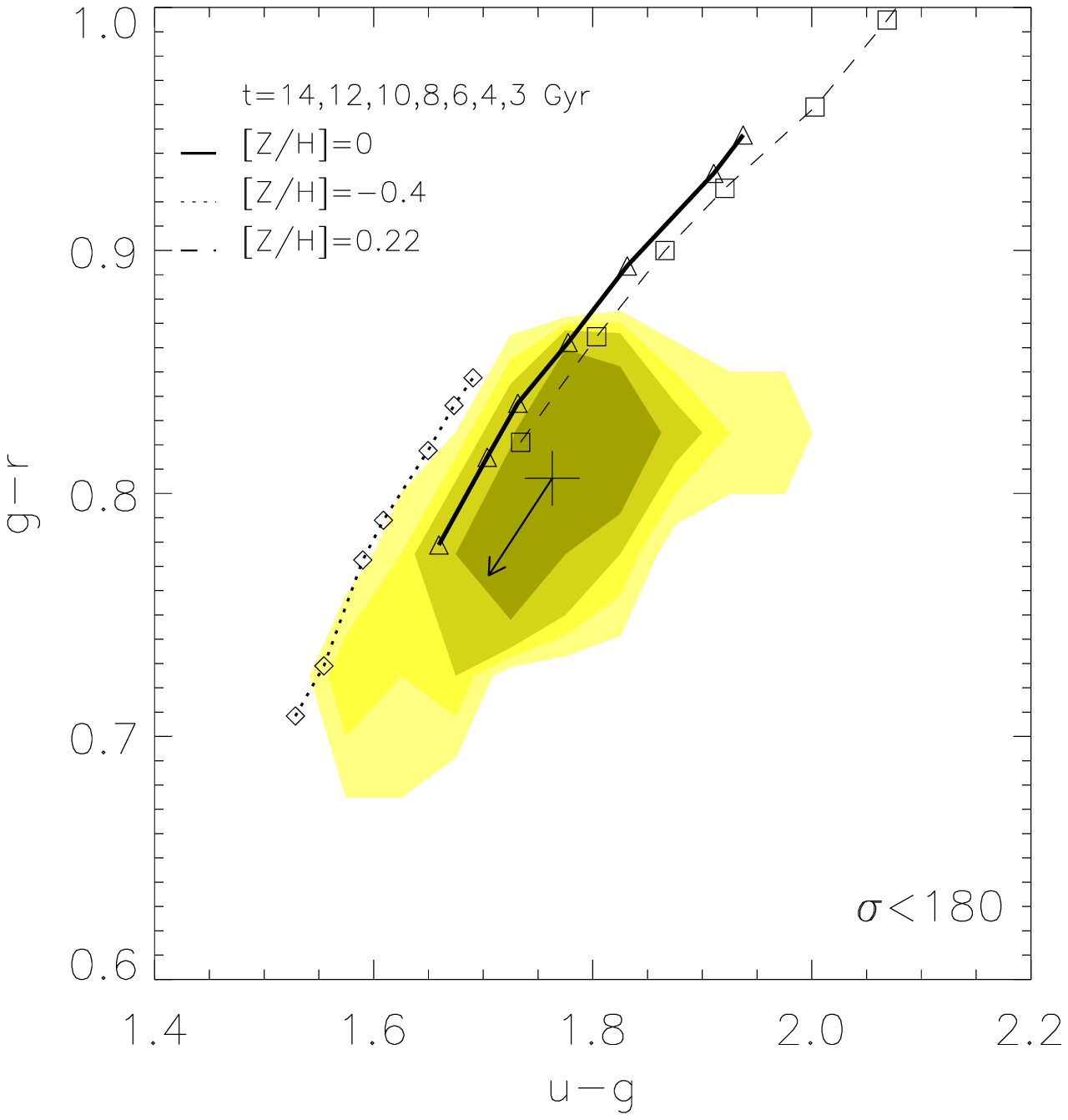}
\includegraphics[angle=0,width=0.8\columnwidth]{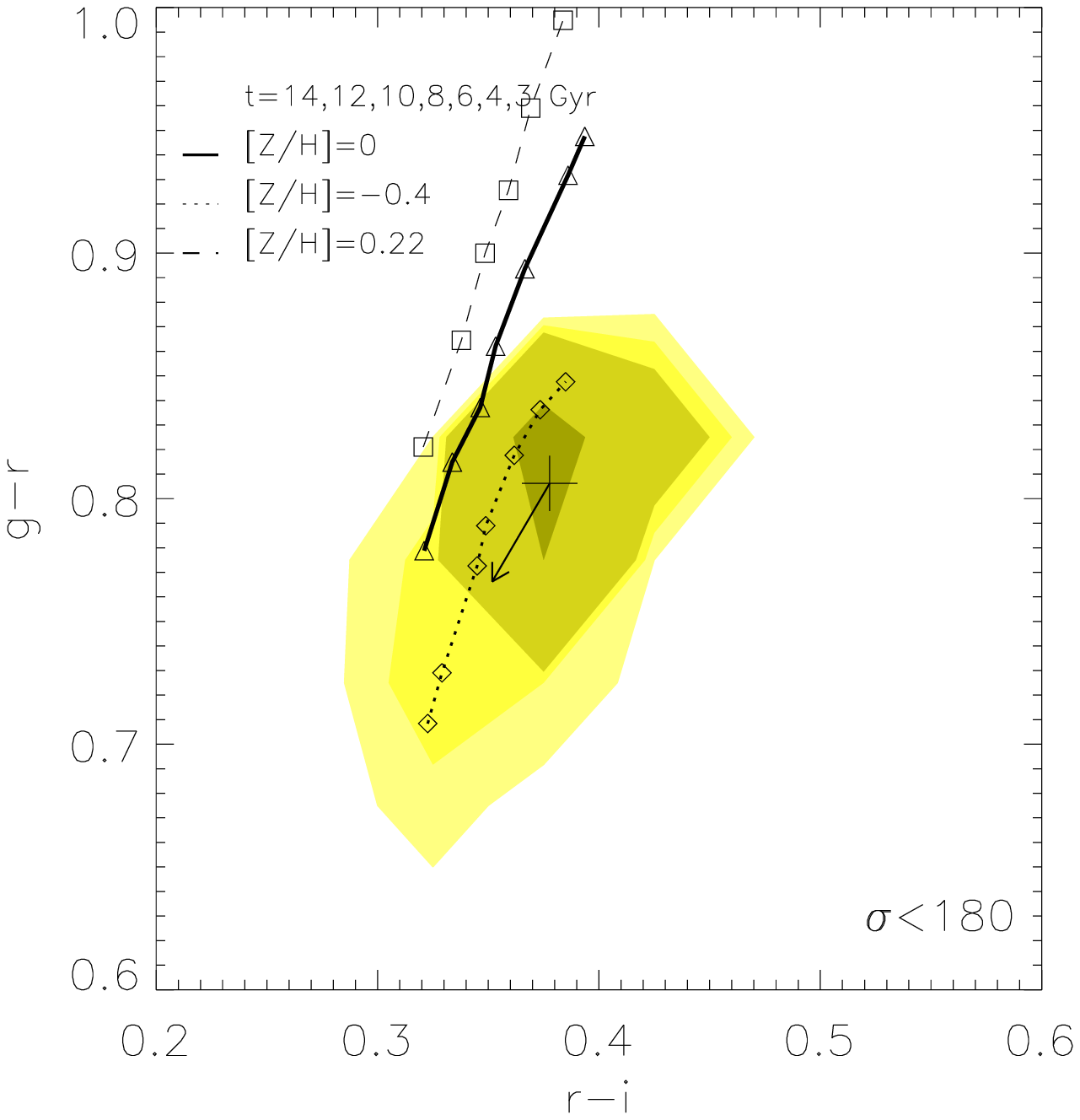}\\
\includegraphics[angle=0,width=0.8\columnwidth]{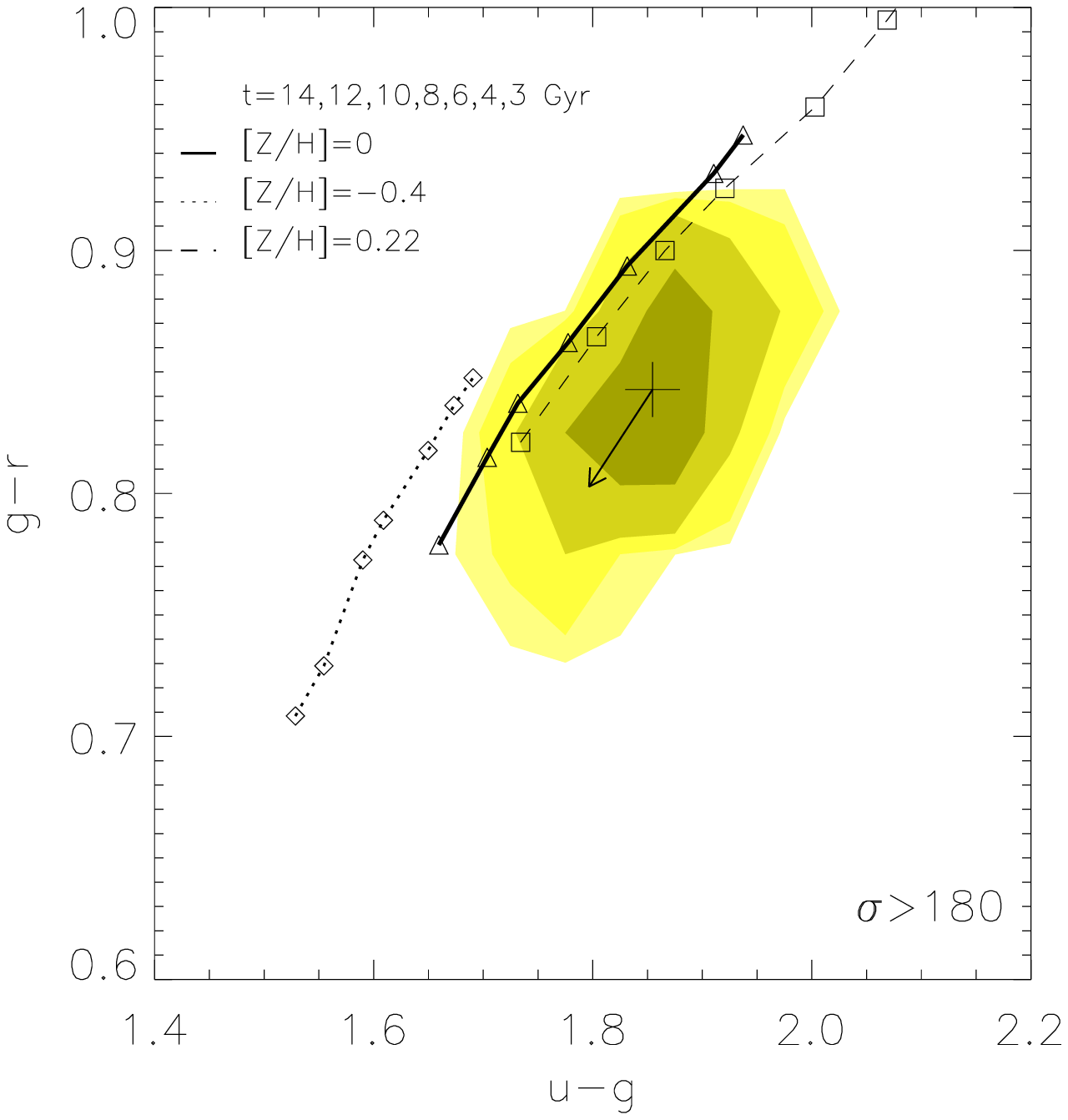}
\includegraphics[angle=0,width=0.8\columnwidth]{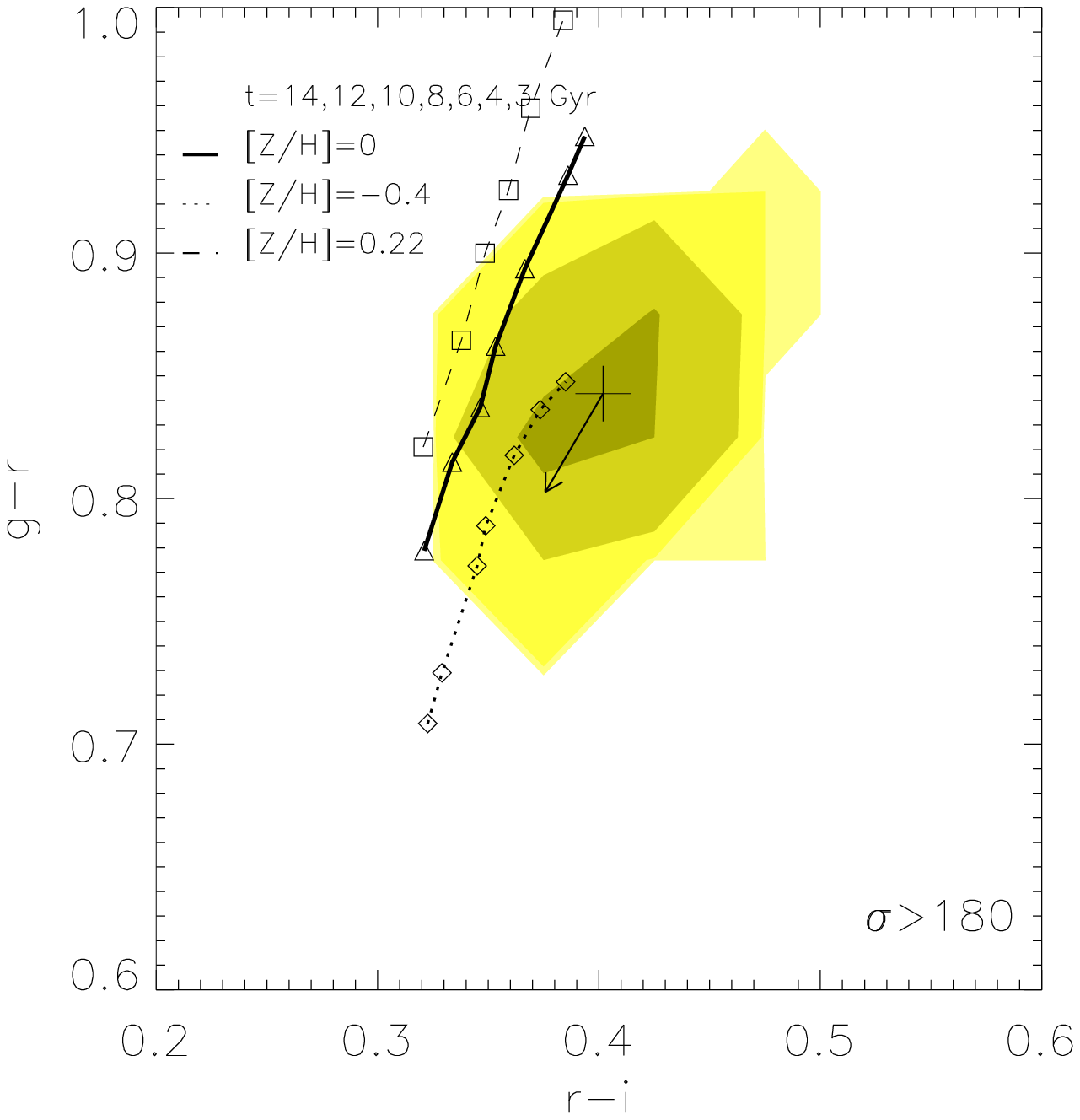}
\caption{
Upper panels: observed colour distribution for ETGs with velocity dispersion
below 180 \mbox{km s$^{-1}$}, compared to the same SSP models as in Figure
\ref{etg_ssp}. Bottom panels: colour distribution for ETGs with velocity
dispersion above 180 \mbox{km s$^{-1}$}. The median colors of the
observed distribution for low
and high velocity dispersion objects are, respectively: $u-g=1.76$, $g-r=0.81$,
$r-i=0.38$ mag and $u-g=1.87$, $g-r=0.85$,
$r-i=0.40$ mag.
  }
\label{etg_sigma}
\end{figure*}

To understand better the origin of this mismatch and to avoid possible sample
selection biases we have divided the ETG sample in intervals according to their
masses. In Figure \ref{etg_sigma} we show the colour distributions corresponding
to galaxies with velocity dispersions above and below 180 \mbox{km s$^{-1}$}.
This value allows us to separate galaxies with different properties, while
keeping almost equally populated these two galaxy mass ranges. ETGs with high
velocity dispersion are generally found to exhibit older ages and higher
metallicities than their low velocity dispersion counterparts \citep{Trager00,
Sanchez06B}, resulting in redder colours. We see that none of the SSP models
plotted in this Figure is able to match the distribution of massive ETGs.
Galaxies with lower velocity dispersion tend to be matched with models of
lower metallicity and age $\simeq$8\,Gyr, even though the predicted {\it u$-$g}
remain too blue ($\simeq$0.1 mag). Although a better agreement is obtained, the
models do not provide fully satisfactory fits. 

\begin{figure*}
\includegraphics[angle=0,width=0.8\columnwidth]{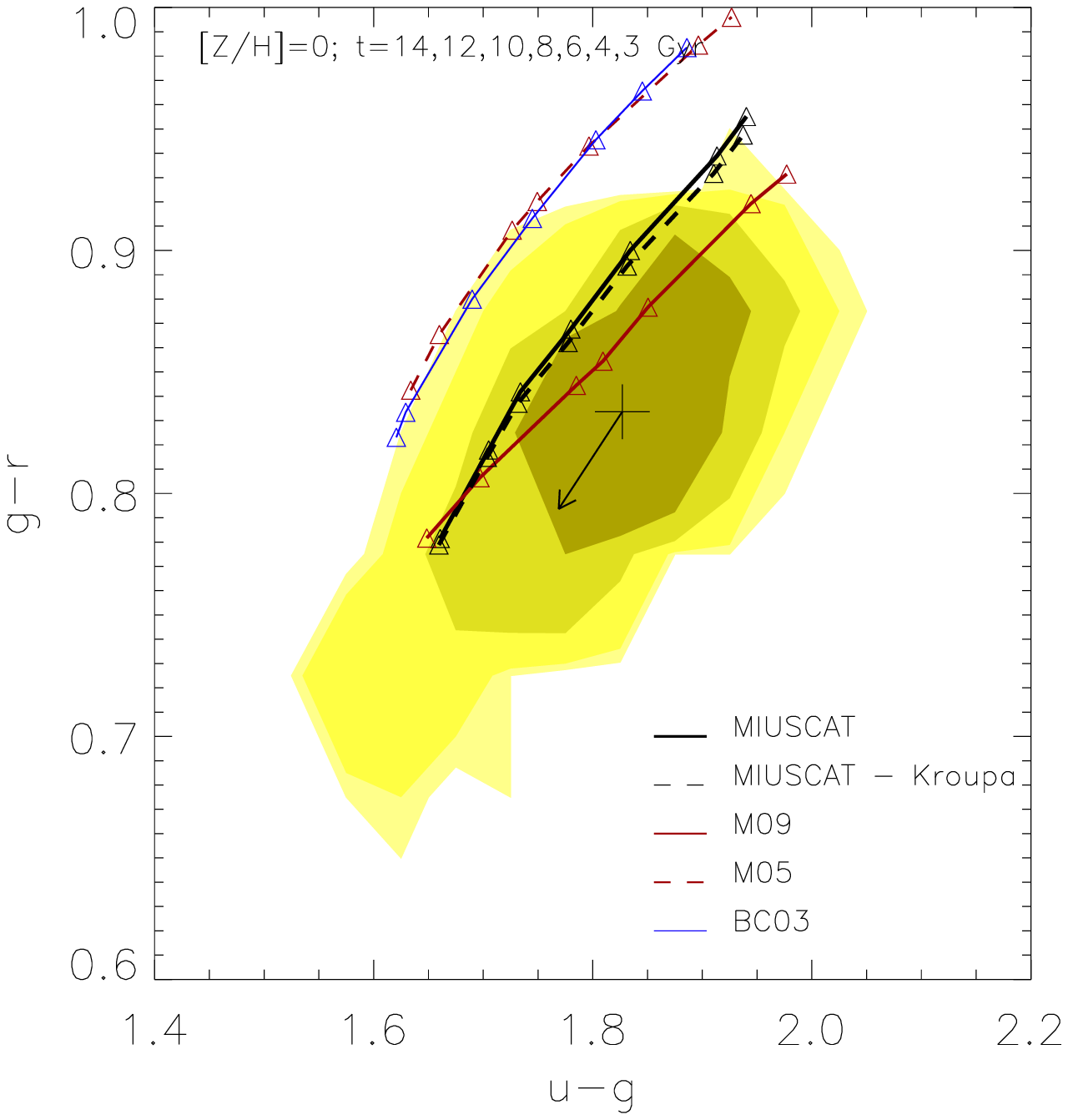}
\includegraphics[angle=0,width=0.8\columnwidth]{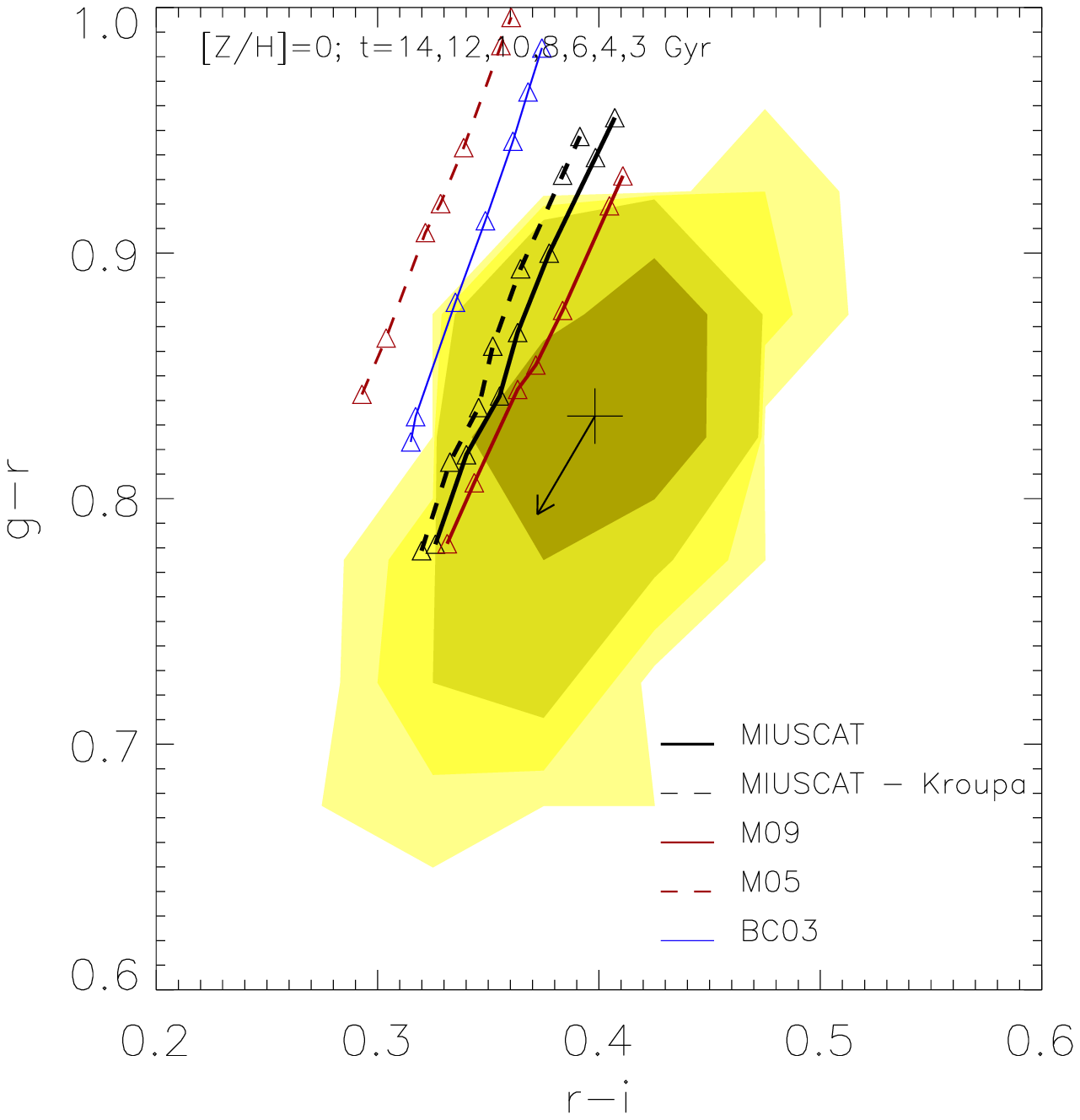}
\caption{
Same as in figure \ref{etg_ssp} for stellar populations models from different
authors. For this comparison only models for solar metallicity are shown. The
blue line corresponds to the BC03 models assuming a Salpeter IMF. Red dashed
line represents the M05 models based on theoretical stellar atmospheres, whereas
the solid red line shows the models from the same authors based on the Pickles
empirical stellar library \citep{Maraston09}. In these two cases a Salpeter
IMF is adopted by these authors. Finally the
black lines indicate the MIUSCAT model tracks adopting the Salpeter (solid line)
and Kroupa Universal (dashed line) IMF.
} 
\label{etg_mod}
\end{figure*}

The colour predictions computed by adopting SSP models from different authors are
shown in Figure \ref{etg_mod}. We considered the BC03 and \citet{Maraston05}
models, the two based on the BaSeL theoretical stellar library
\citep{Lejeune98} and the \citet{Maraston09} models that use the \citet{Pickles98}
empirical library. 
The BC03,  \citet{Maraston05} and \citet{Maraston09}  adopt a Salpeter IMF, whereas the MIUSCAT
models are shown for both Salpeter and Kroupa IMF, showing that the choice among
these commonly used IMFs has little effect on these colours. Although none of
the models can fairly reproduce the observed distribution, it is remarkable the
advantage of using models based on empirical-based libraries. These models
predict {\it g$-$r} bluer by 0.06 mag and {\it r$-$i} redder by 0.05 mag compared
to the models based on theoretical libraries. The \citet{Maraston09} models
predict slightly bluer {\it g$-$r} and redder {\it r$-$i} colours than the
MIUSCAT SEDs by $\approx$0.01 mag, but also redder {\it u$-$g} colour.

\subsection{IMF effects}
\label{sect_imf}

\begin{figure*}
\includegraphics[angle=0,width=0.8\columnwidth]{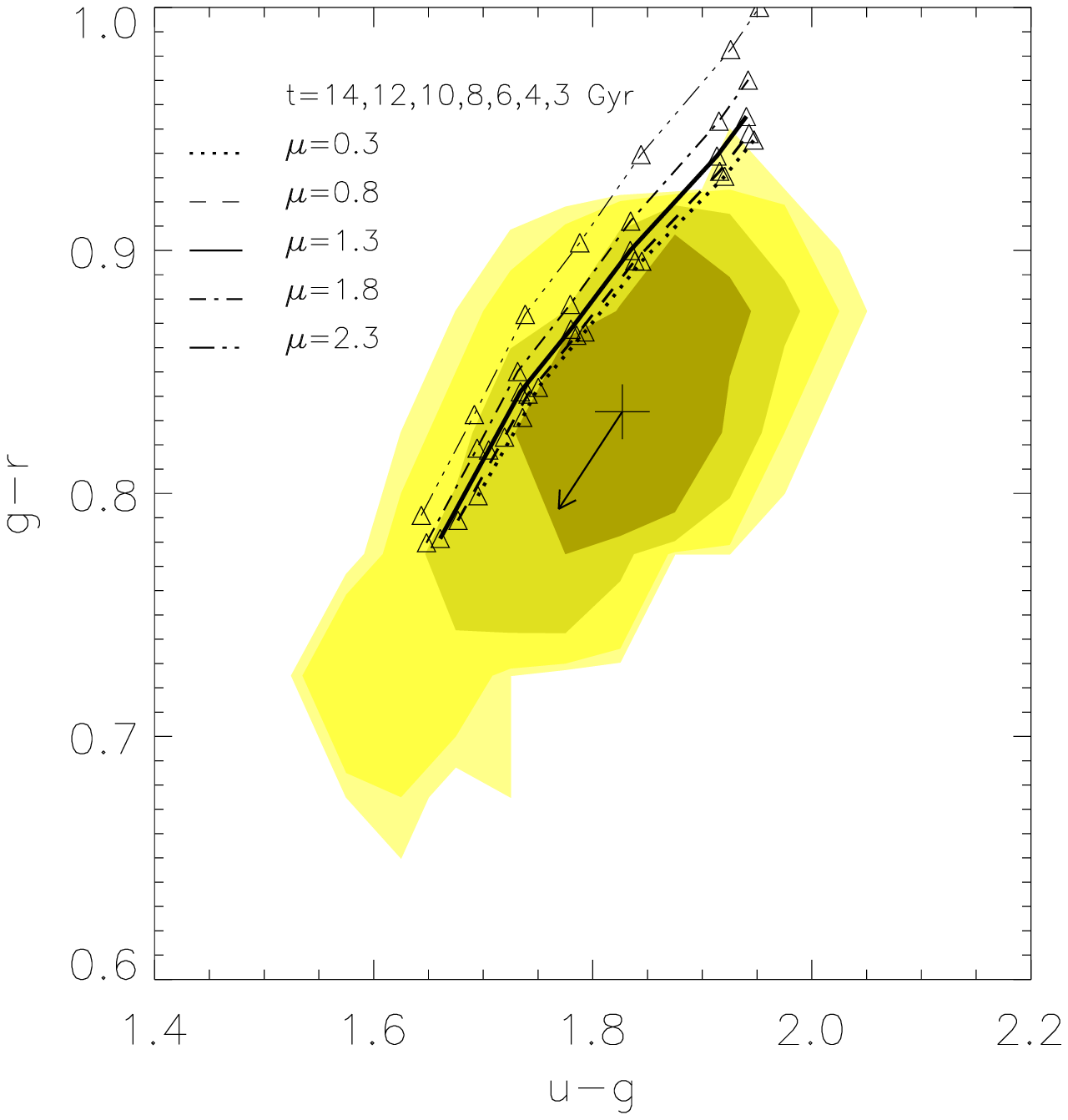}
\includegraphics[angle=0,width=0.8\columnwidth]{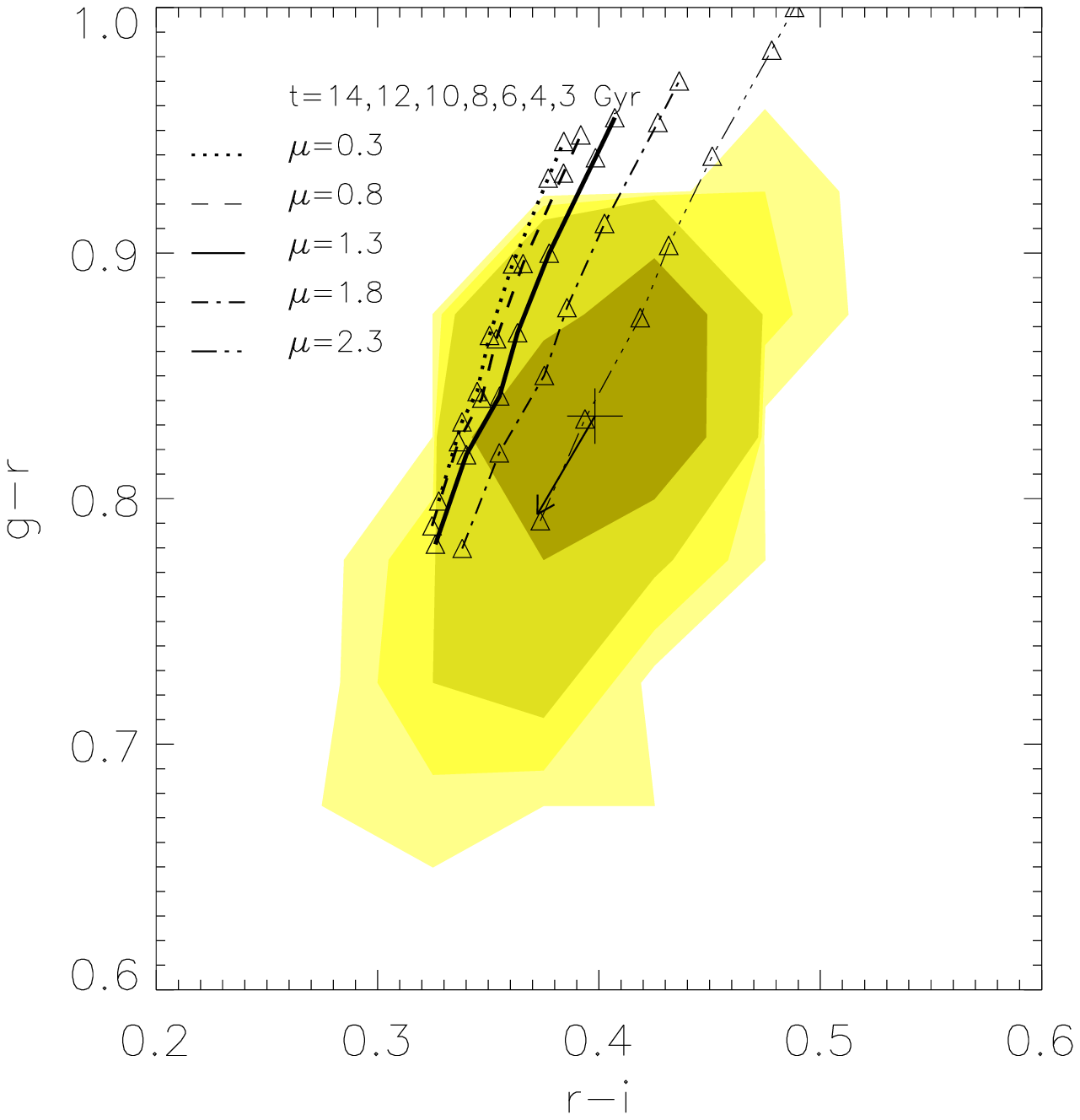}
\caption{Same as in Figure \ref{etg_ssp} for unimodal IMF with different slopes.} 
\label{etg_imf}
\end{figure*}

Here we explore how these colours vary when adopting different IMF shapes and
slopes. We consider the power-law unimodal IMF defined in Vazdekis et al. (1996)
for different slopes for the solar metallicity SSP models with the same ages
considered in Figure \ref{etg_ssp}. Figure \ref{etg_imf} shows a very modest variation of
the bluer bands. However varying the IMF turns out to have a significant effect
on the {\it r$-$i} colour. Indeed, by assuming a very steep IMF slope ($\mu=2.3$)
the colour distribution in the {\it g$-$r} vs {\it r$-$i} diagram can be matched,
though for relatively young ages (3-4\,Gyr).  On the other hand the {\it g$-$r} vs
{\it u$-$g} colour distribution is not reproduced for none of the represented
models.

\subsection{Two-burst model: inclusion of a metal-poor component}
\label{sect_mp}

\begin{figure*}
\includegraphics[angle=0,width=0.8\columnwidth]{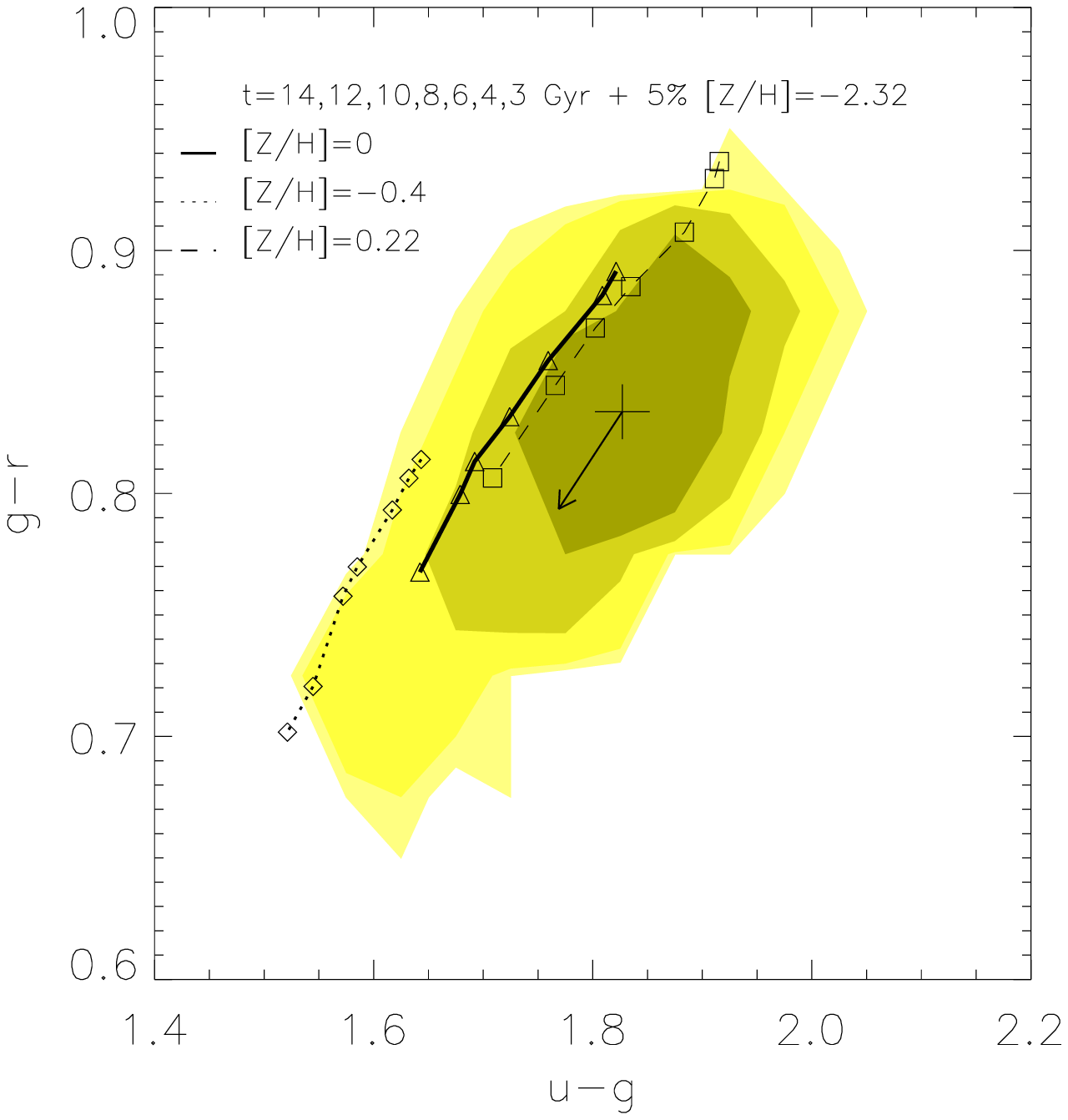}
\includegraphics[angle=0,width=0.8\columnwidth]{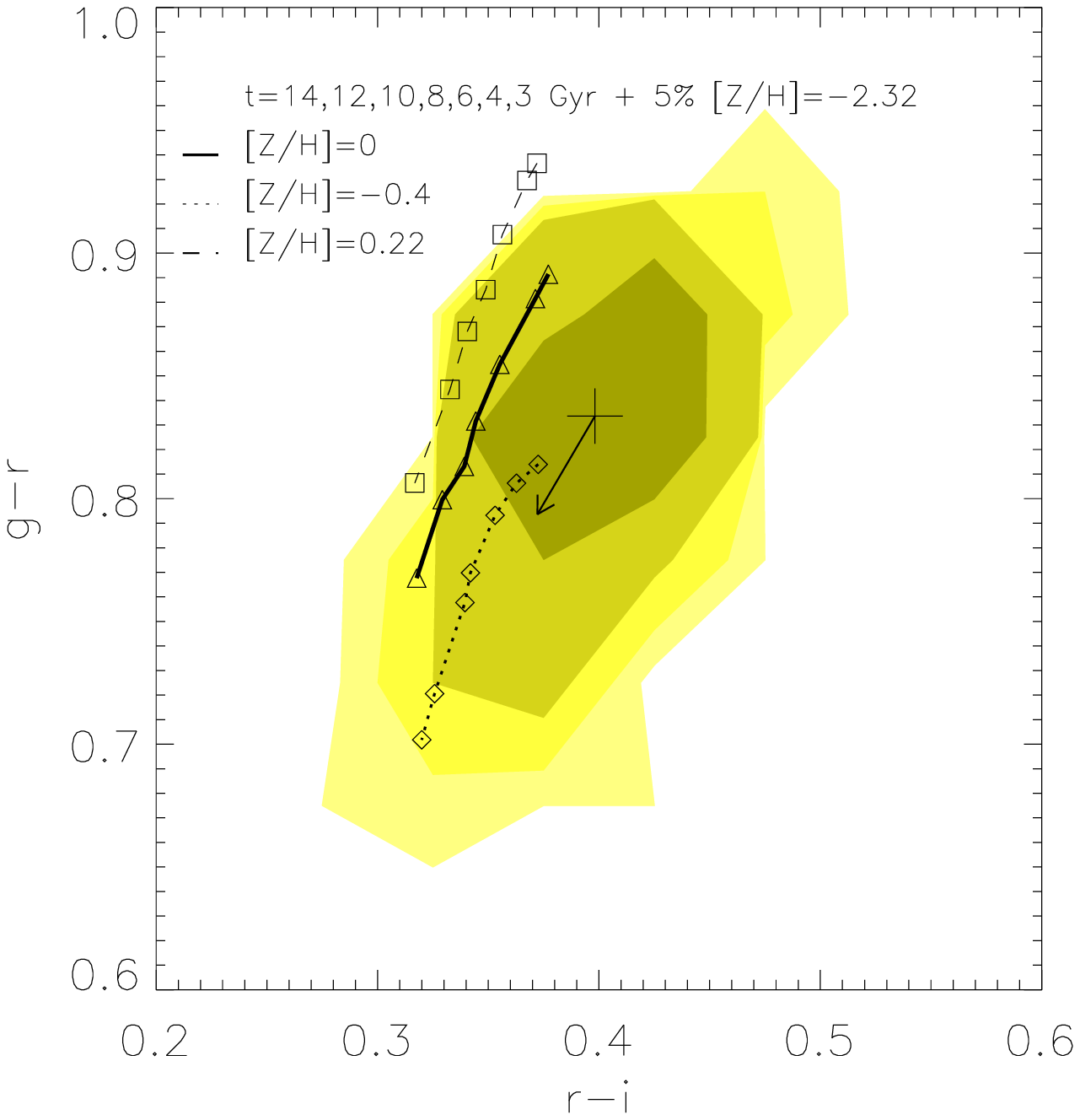}
\caption{
Same as in Figure \ref{etg_ssp} but for a two-burst model that includes a
metal-poor population ([Z/H]=-2.32) of 14\,Gyr, which contributes with a 5\% to
the total mass.
  } 
\label{etg_mp}
\end{figure*}

We relax here the assumption that ETGs are SSPs and use stellar populations
with two main bursts to model these galaxies. The first model tested is the same
used for the description of the LRG colour evolution shown in Section 5.  We
include a small contribution of 5\% to the total mass, with a primordial
chemical composition ([Z/H]=-2.32) and very old age (14\,Gyr). This should
resemble the first population of stars formed before being enriched by the
subsequent star formation. As shown also by \citet{Maraston09}, the contribution
of metal-poor stars is effective in bluening the {\it g$-$r}, without perturbing
too much the {\it r$-$i} colour. The resulting MIUSCAT two-burst colours are shown
in Figure \ref{etg_mp}. As expected, all the colours become bluer, improving the
match with the observed colour-colour distribution in the {\it g$-$r} vs {\it
u$-$g} diagram. The resulting {\it r$-$i} colour becomes slightly bluer, but since
the effect of the metal-poor component is stronger on the bluer bands the colour
distribution in {\it g$-$r} vs {\it r$-$i} improves as well, in agreement with
the finding of \citet{Maraston09}. However the inclusion of such metal-poor
component alone is not sufficient to fairly reproduce the observed colour-colour
distributions.  

\subsection{Two-burst model: inclusion of a young stellar population}
\label{sect_age}

\begin{figure*}
\includegraphics[angle=0,width=0.8\columnwidth]{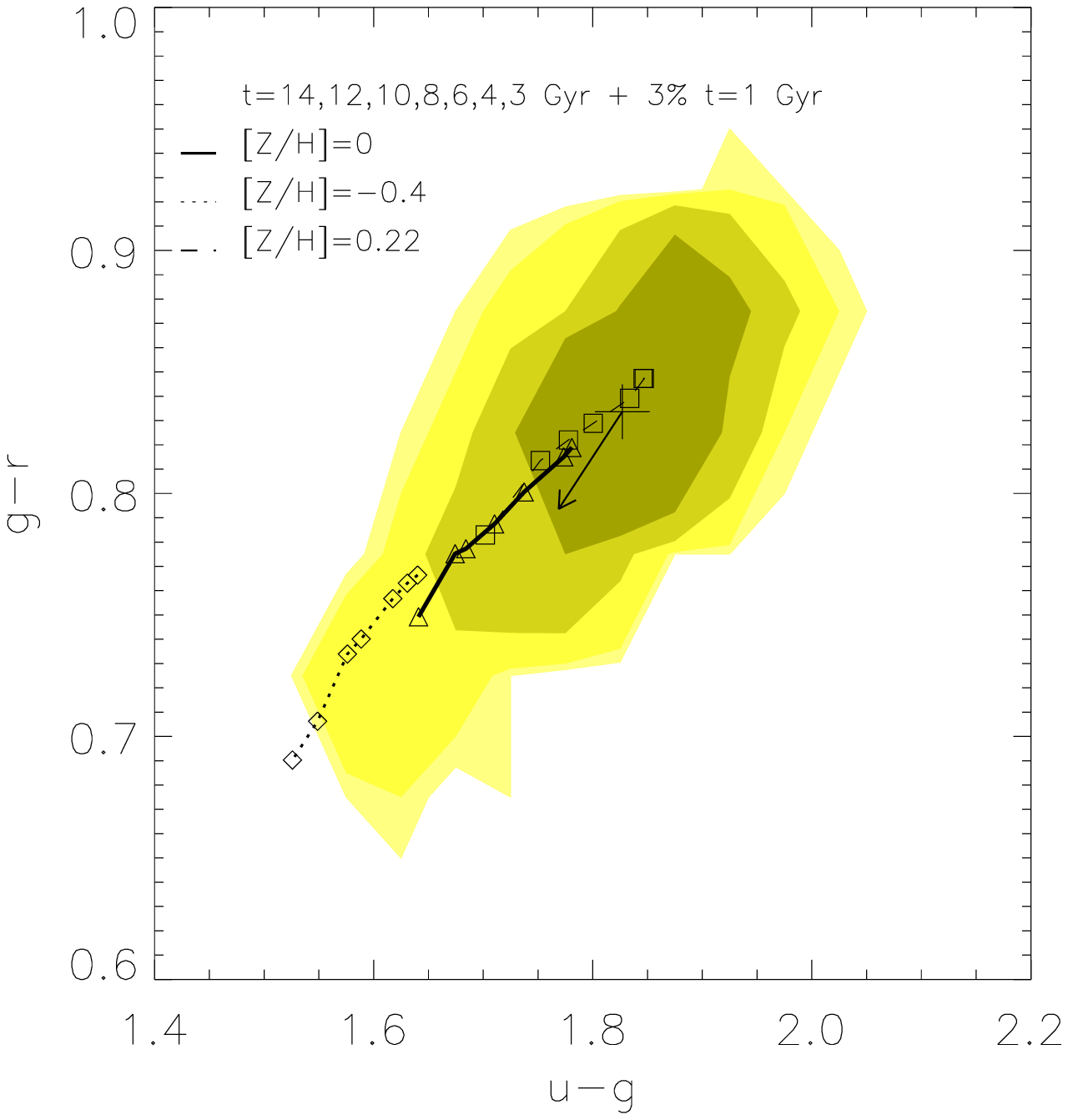}
\includegraphics[angle=0,width=0.8\columnwidth]{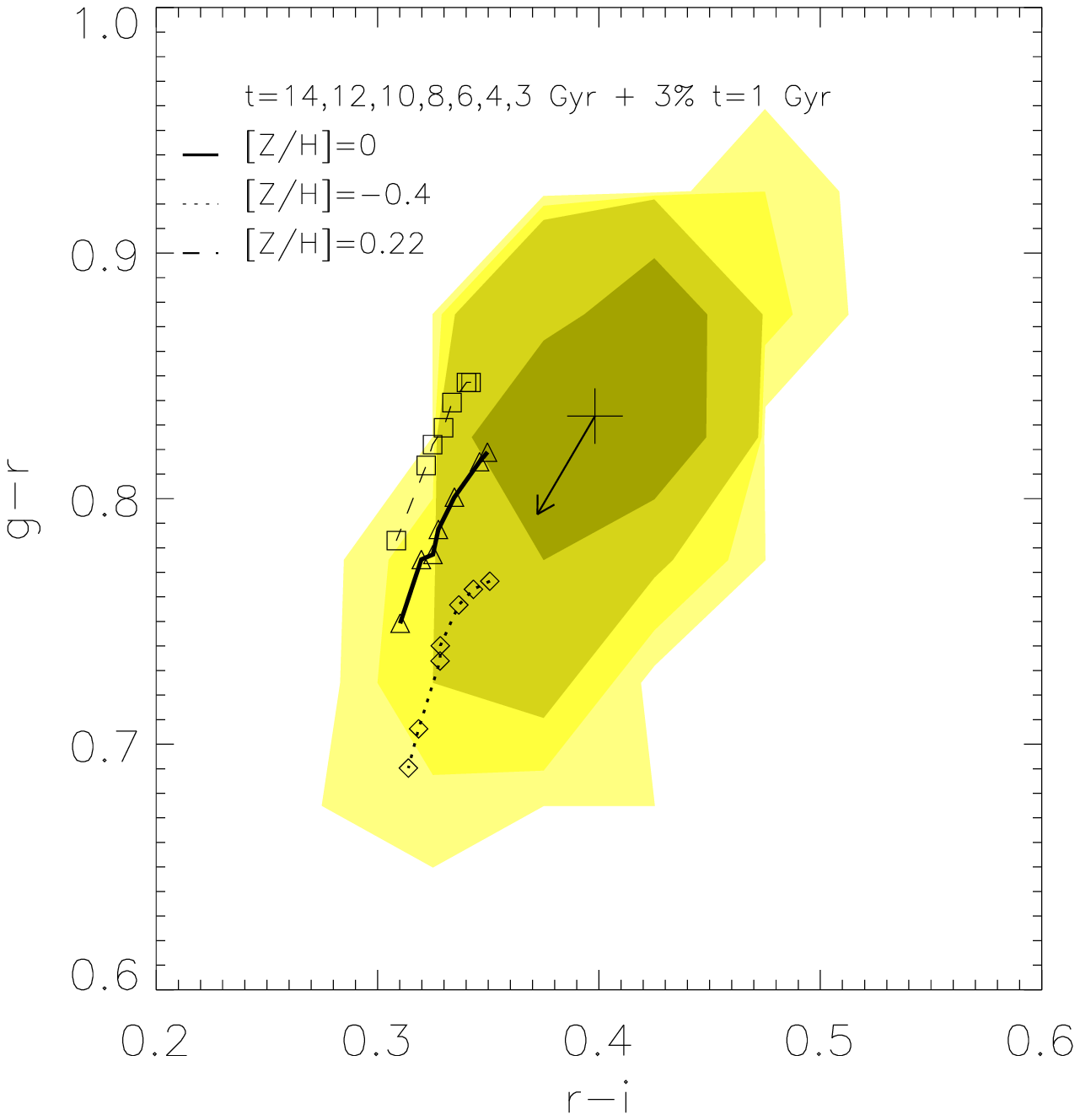}
\caption{Same as in Figure \ref{etg_mp} for a two-burst model including a 3\%
contribution in mass from a 1\,Gyr stellar population.} 
\label{etg_yp}
\end{figure*}

A similar effect is seen when a young component is included. In Figure
\ref{etg_yp} the plotted model includes a small contribution (3\% of the total
mass) of a young stellar population of 1\,Gyr. In this case the effects on all
the colours is more pronounced than in Section \ref{sect_mp}. This model can successfully
explain the colour-colour distribution in the {\it g$-$r} vs {\it u$-$g} diagram,
but not the {\it r$-$i} colour, since the predicted values are far too blue
($\approx$0.06 mag). Therefore, these models are not able to match the two
colour-colour diagrams simultaneously.

\begin{figure*}
\includegraphics[angle=0,width=0.8\columnwidth]{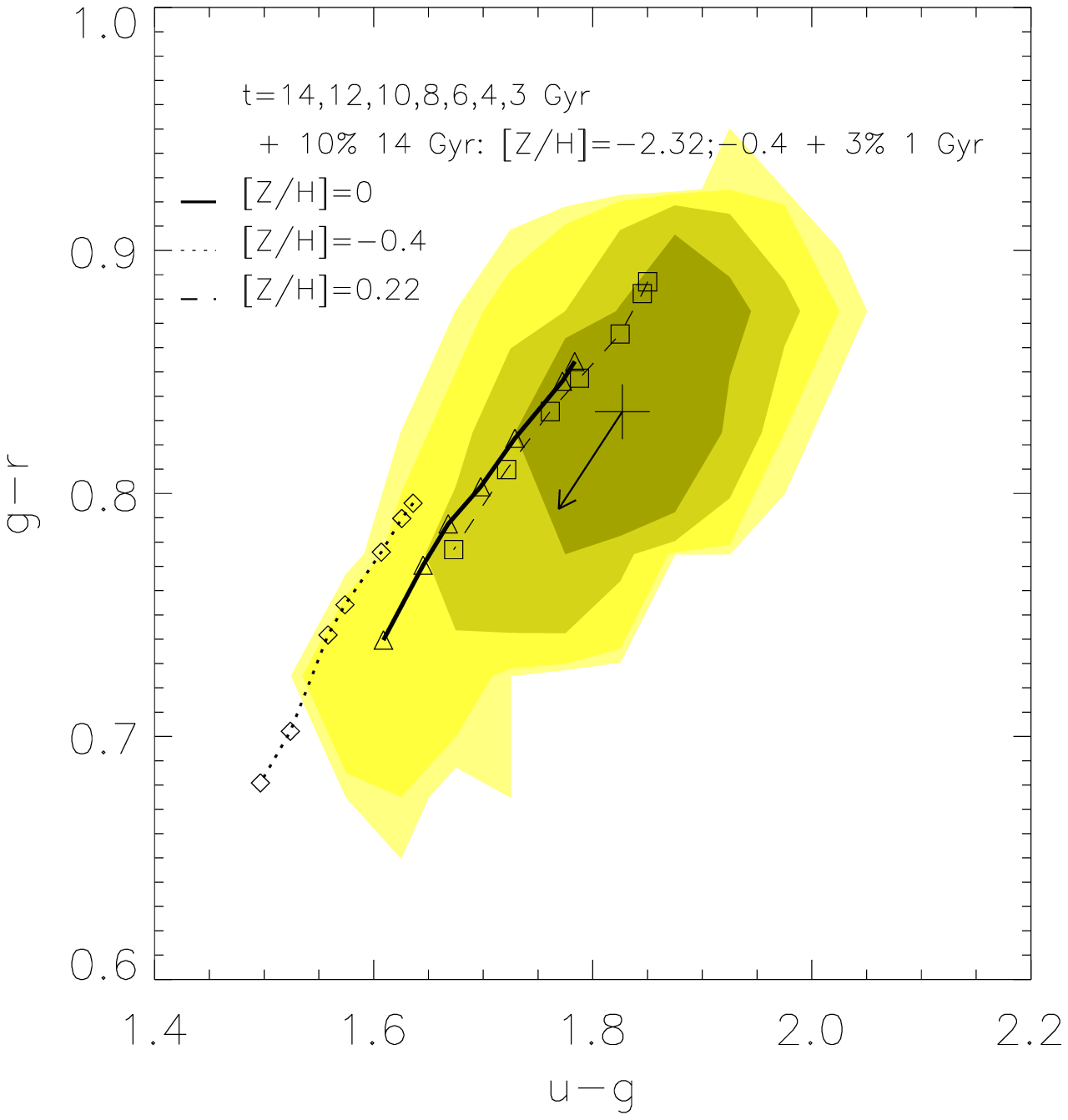}
\includegraphics[angle=0,width=0.8\columnwidth]{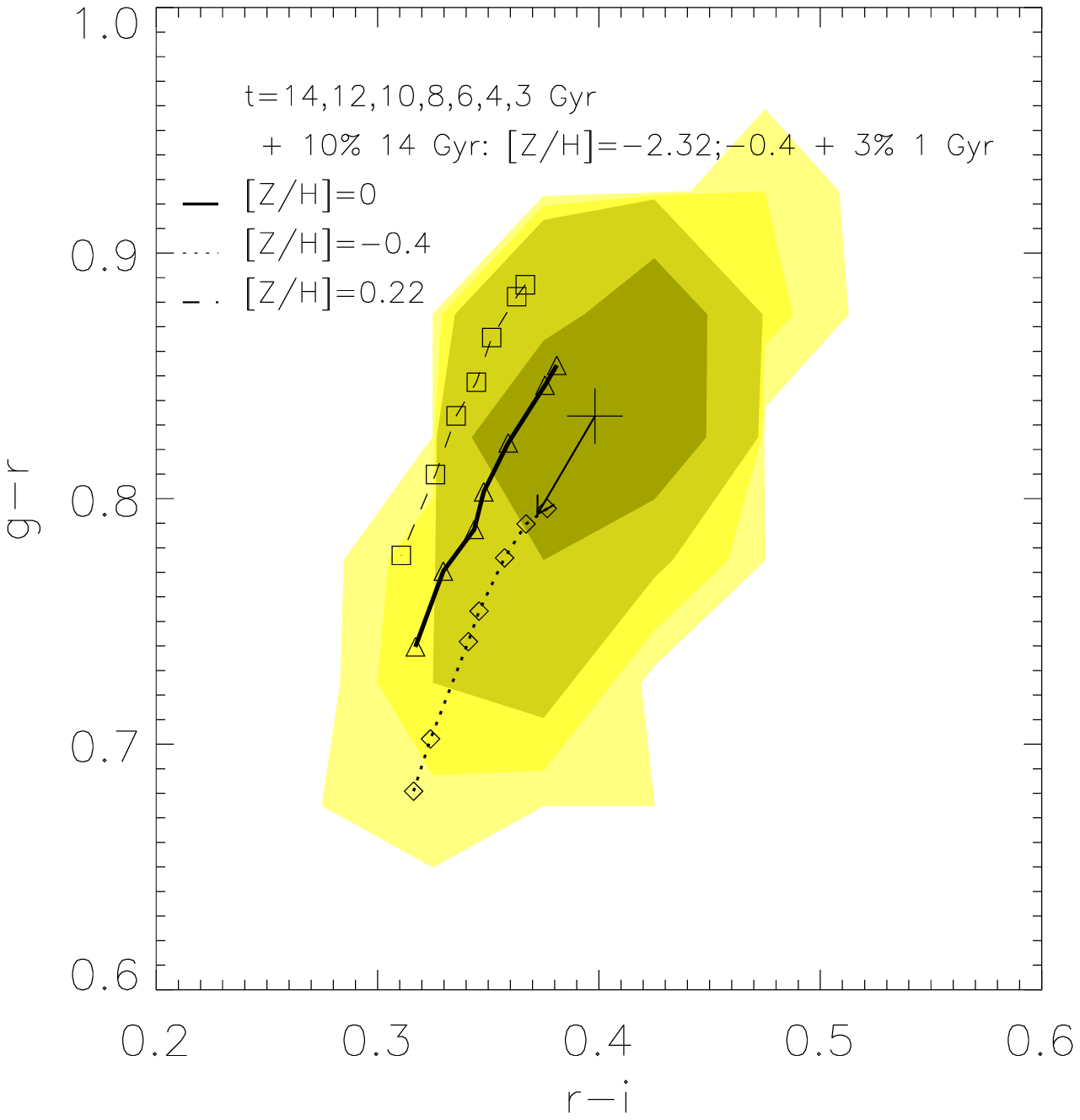}
\caption{
Same as in Figure \ref{etg_ssp} but for the SFH derived from the
chemo-evolutionary model described in Vazdekis et al (1996, 1997). We model it
by adding on the top of the main episode of star formation the contributions of
two stellar populations: 10 \% of the mass formed at 14\,Gyr with a top-heavy
IMF (slope $\mu$=0.8). Half of the mass formed in this burst is in metal-poor
component ([Z/H]=-2.32 dex) and half of it has subsolar composition, [Z/H]=-0.4
dex. Finally, there is a 3\% contribution to the total mass from a component of
1\,Gyr. The stars formed in the main episode and afterwards are skewed towards
lower masses according to a steeper IMF slope ($\mu$=2.3). We use here the
Bimodal IMF defined in \citet{Vazdekis96}. 
} 
\label{etg_topbh}
\end{figure*}

\begin{figure*}
\includegraphics[angle=0,width=0.8\columnwidth]{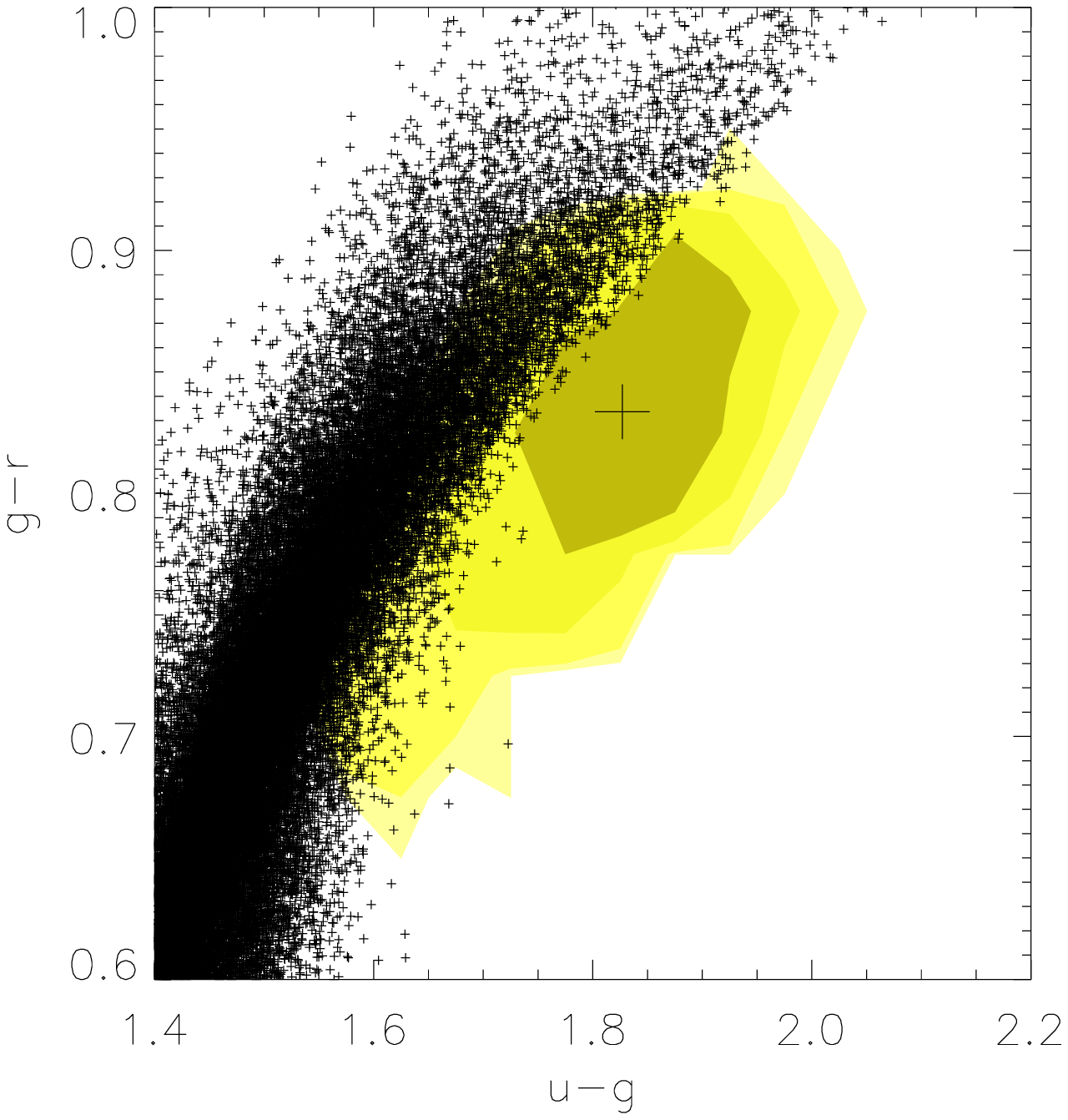}
\includegraphics[angle=0,width=0.8\columnwidth]{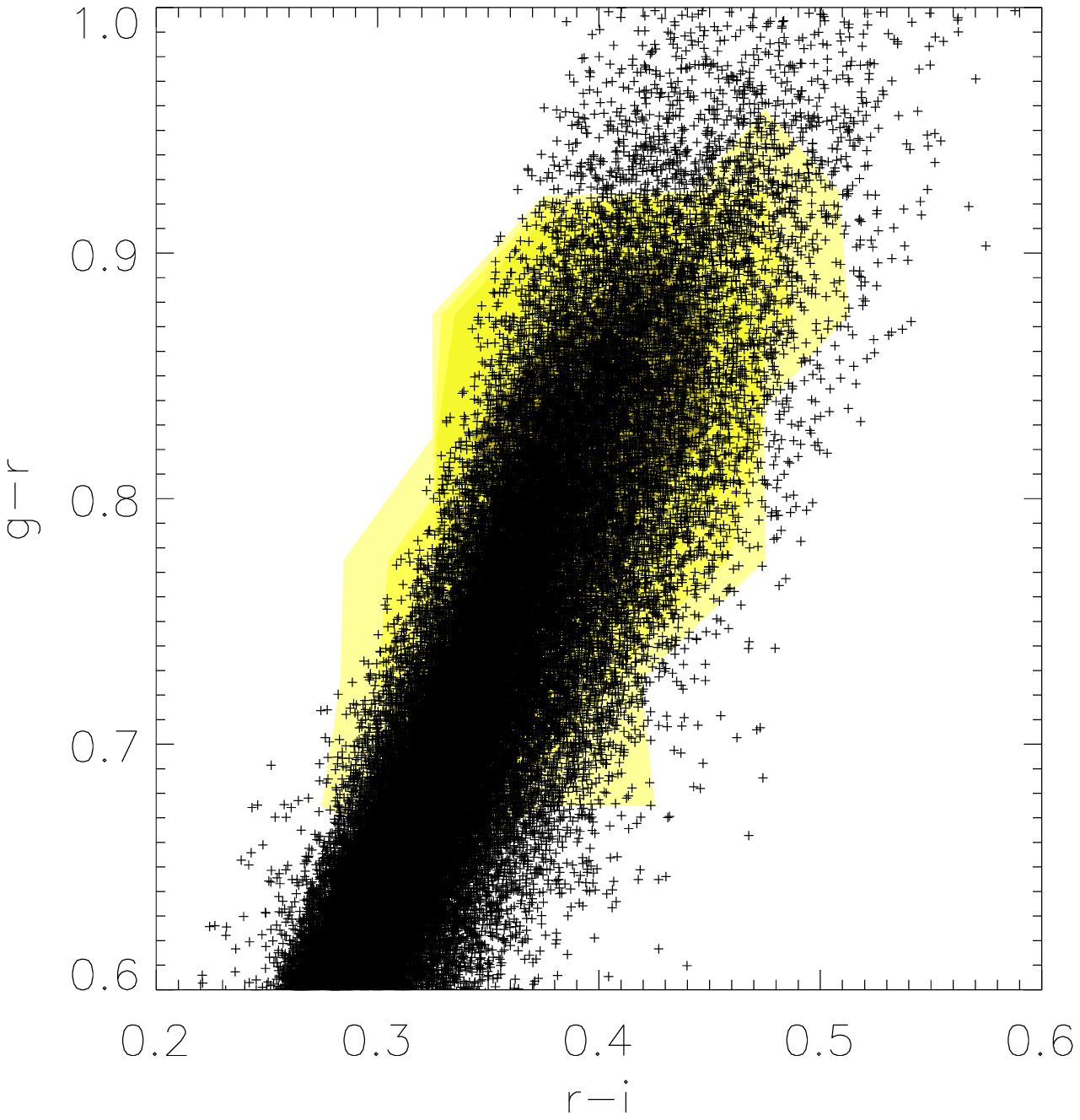}
\caption{The observed colour-colour distributions are compared with
  simulated  points derived from the three-bursts Montecarlo
  SFHs described in Section \ref{sect_mc}.  } 
\label{etg_mc}
\end{figure*}


\subsection{Three-burst model: chemo-evolutionary model with varying IMF slope}
\label{sect_chemo}

Here we switch for the first time to the combination of more than two stellar
populations. In Sections \ref{sect_mp} and ~\ref{sect_age} we have seen that the
combination of an old metal-rich stellar population with a small fraction of
either a metal-poor or a young component has significant effects on the
resulting colour-colour plots. Furthermore in Section~\ref{sect_imf} we tested
the effects on the colours from non standard IMFs. In fact, very recently
\citet{Vandokkum11} showed spectroscopic evidences pointing towards steeper IMFs
in massive galaxies, in agreement with previous studies (e.g., \citealt{Faber80,
Carter86, Vazdekis96, Cenarro03}). Therefore it is worth to test the colours
that result from combining all these stellar populations. An example of such
scenario is the one derived in \citet{Vazdekis97} from the best fits to both
broad-band colours (from $U$ to $K$) and the Lick line-strength indices, of a
number of prototype massive galaxies, employing a full chemo-evolutionary
version of our models \citep{Vazdekis96}. 

In this scenario the IMF varies with time, being skewed towards high-mass stars
during an initial short period of time ($<$0.5\,Gyr) that
quickly enriches the interstellar medium and, afterwards, the IMF becomes even
steeper than the Salpeter one. This is reproduced  with a multiple bursts model by
adding on top of the main episode of star formation, that forms 87\% of the
mass,  two populations. In the first population 10 \% of the mass formed at
14\,Gyr with a top-heavy IMF (slope $\mu$=0.8). Half of the mass formed in this
burst is associated with a metal-poor component, [Z/H]=-2.32 dex (as in
Section~\ref{sect_mp}), and the other half has subsolar composition
([Z/H]=-0.4). Another 3\% of the mass is formed at the age of 1\,Gyr (as in
\ref{sect_age}). The stars formed in the main episode and at late times are
skewed towards low-mass  ($\mu$=2.3) (as in Section~\ref{sect_imf}). A bimodal
IMF is assumed for all these models. The adoption of the IMF with slope 2.3 in
the main burst is consistent with recent evidences for a steep IMF in massive
galaxies \citep{Vandokkum11}.
\begin{figure} 
\includegraphics[angle=0,width=1.\columnwidth]{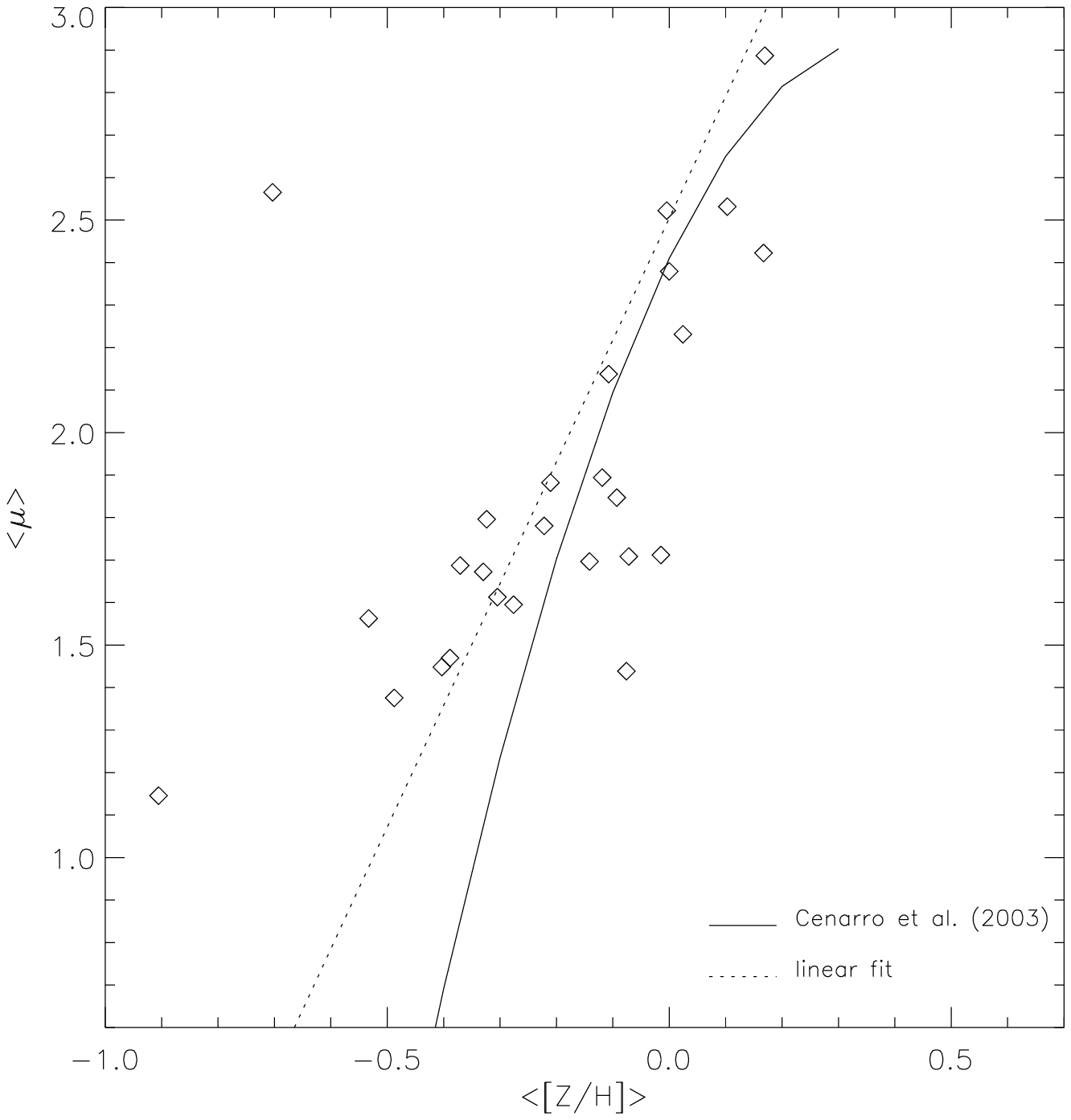}
\caption{Mean IMF slope versus mean metallicity for the best solution
  SFHs from the Montecarlo simulations described in Section \ref{sect_mc}. The solid  black line shows the
  relation found in \citet{Cenarro03}, whereas the dotted line is the
  linear fit to the points, obtained with a least squares fit 
  in the X direction.
   }
\label{mu_met} 
\end{figure}

This model is shown in Fig. 14. For solar metallicity it is  quite successful at matching the bulk of the
galaxy distribution, although some problems persist in the {\it r$-$i} colour,
which remains too blue. Although quite unlikely, part of this discrepancy can
also be attributed to internal extinction, leading too red colours that we
cannot fit with dust-free models. 

\subsection{Three-burst models: Montecarlo simulations }
\label{sect_mc}

In order to explore all the possible combinations of SFHs that can match the two
colour-colour distributions we have run a set of Montecarlo simulations, where
we allow all the stellar population parameters to vary: age, metallicity and IMF
slope.  We only consider seven representative IMF slopes: 0.3, 0.8, 1.3,
  1.8, 2.3, 2.8 and 3.3. 
To keep the case as general as possible we use three bursts, where the
fraction of mass involved in each of the burst is also a free parameter. 
Figure~\ref{etg_mc} shows the results for 100000 simulations.  On one hand we
see that the cloud of models does not properly match the observed galaxy
distribution in the {\it g$-$r} vs {\it u$-$g} plot. On the other hand, the
cloud of models matches the region enclosed by data in the  {\it g$-$r} vs {\it r$-$i}
plot. 

To understand which kind of SSP mixtures most likely resemble the data we
selected the closest 1700 points to the median of the observed colours in each
of the distributions. Among these combinations we only choose those that are
common to these two distributions to end up with 26 models, whose SFHs are shown
in the Appendix. We have ordered all these models following an increasing
contribution of young stellar populations, i.e., with ages smaller than 5\,Gyr.
We see that a large number of models require very important contributions from
SSPs with old ages, high metallicity (solar or higher) and IMF slopes steeper
than Salpeter (i.e. $\mu > 1.3$). All these models require smaller, but not
negligible, contributions from young or/and metal-poor stellar populations.
Finally SSPs with slopes lower than the Salpeter value seem to be required for a
number of models. A mean solution for these models can therefore be summarized
as mainly  composed of metal-rich and dwarf-dominated stellar populations with
much smaller contributions from young or metal-poor components. Interestingly
the scenario discussed in Section~\ref{sect_chemo} would be a good
representative of this mean trend of solutions. In fact we nearly recover it in
the models shown in the  first and second  rows of panels of
Figure~\ref{sfh}. 
Whereas many
studies support the presence of old ages and metal-rich stellar populations in
massive galaxies (see for example the review of \citealt{Renzini06}), the
requirement of non standard IMFs is a matter of debate as recently pointed out
by \citet{Vandokkum11}. 
Among our solutions, the steeper IMF slopes are recovered for the highest
metallicity objects. 
In Figure~\ref{mu_met} we show the mass weighted mean IMF slope as a
function of the  mass weighted mean metallicity for the 26 solutions.
The existence of a IMF slope - metallicity relation has already pointed out
by \citet{Cenarro03}. Interestingly, the linear fit to the simulated
points lies close to the relation derived by the above authors by
means of near-IR IMF sensitive spectral features.
It is worth to note that, using completely independent diagnostic
techniques, it is inferred a similar IMF-metallicity relationship in
the sense that larger metallicities correspond to larger effective IMF
slopes.

There are a number of models that depart significantly from the mean
trend depicted above (see last
rows of Figure~\ref{sfh}). These models require important contributions from
stellar populations with ages smaller than $\sim$5\,Gyr. However we also see
that most of these models also require steeper IMF slopes than the models that
are clearly dominated by old stellar populations. This can be explained by the
fact that the colours redden with increasing age, metallicity or IMF slope as
shown in Paper~I. Such degeneracy between these three parameters for the optical
colours does not allow us to provide a unique solution. To constrain the
solutions it is required to use spectroscopic data or/and colours in other
spectral ranges (e.g., \citealt{Vazdekis97}).

However it is worth recalling that none of the SFHs explored in the simulations
provide satisfactory fits to these two observed colour-colour distributions.
Indeed, the departure of the models from the median of the observed colours is
non negligible, with the $u-g$ colour showing the largest residuals. It is
beyond the scope of this paper to recover the SFHs of the SDSS early-type galaxy
sample, but the fact that in these complex models we allow to vary all, the age,
metallicity and IMF slope of the SSPs, without reaching a fully satisfactory
solution, supports the need for a fourth parameter to be taken into account in
the models. We consider such a possibility in the next section.

\subsection{Effect of $\alpha$-enhancement}
\begin{figure*}
\includegraphics[angle=0,width=0.8\columnwidth]{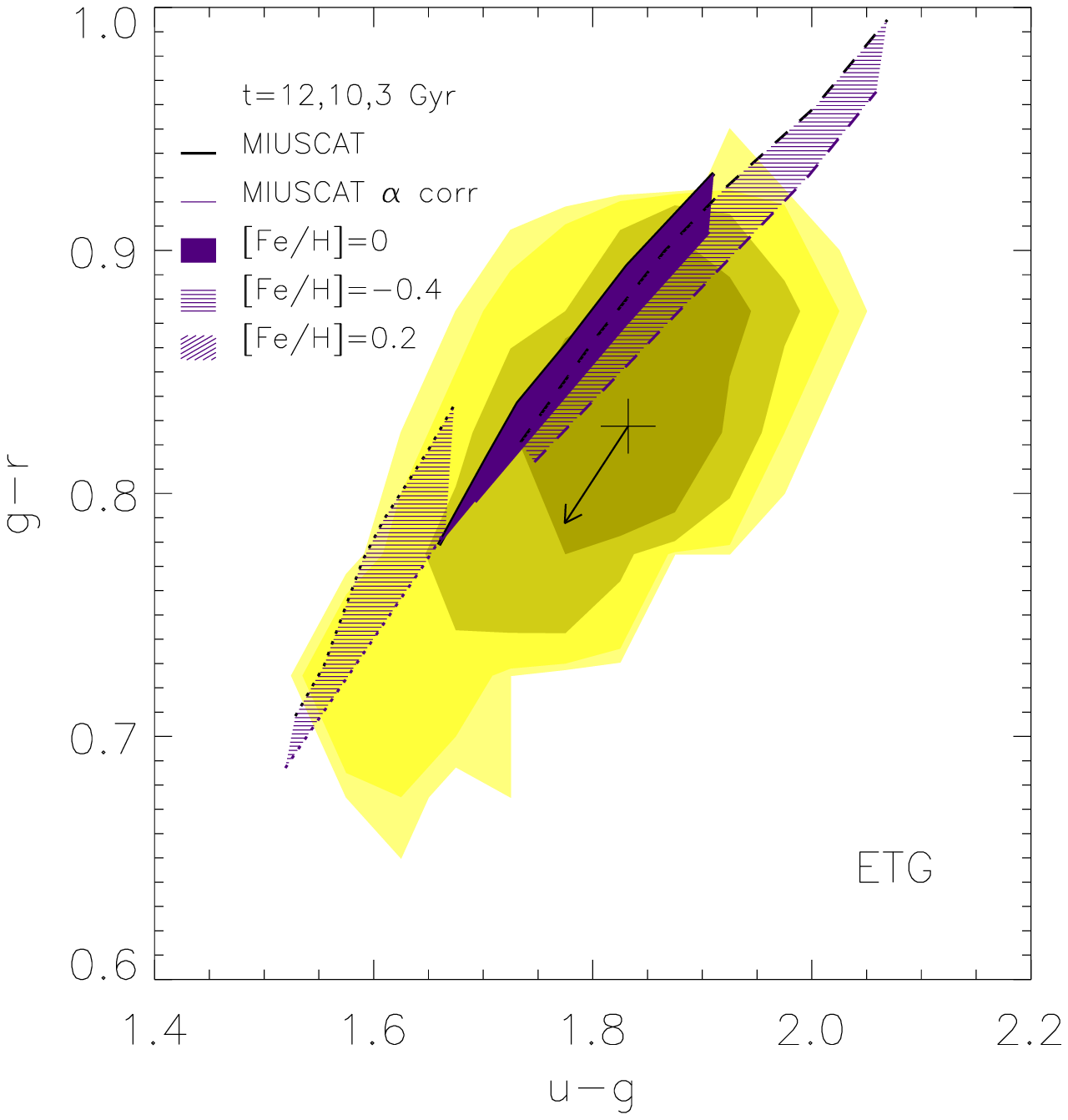}
\includegraphics[angle=0,width=0.8\columnwidth]{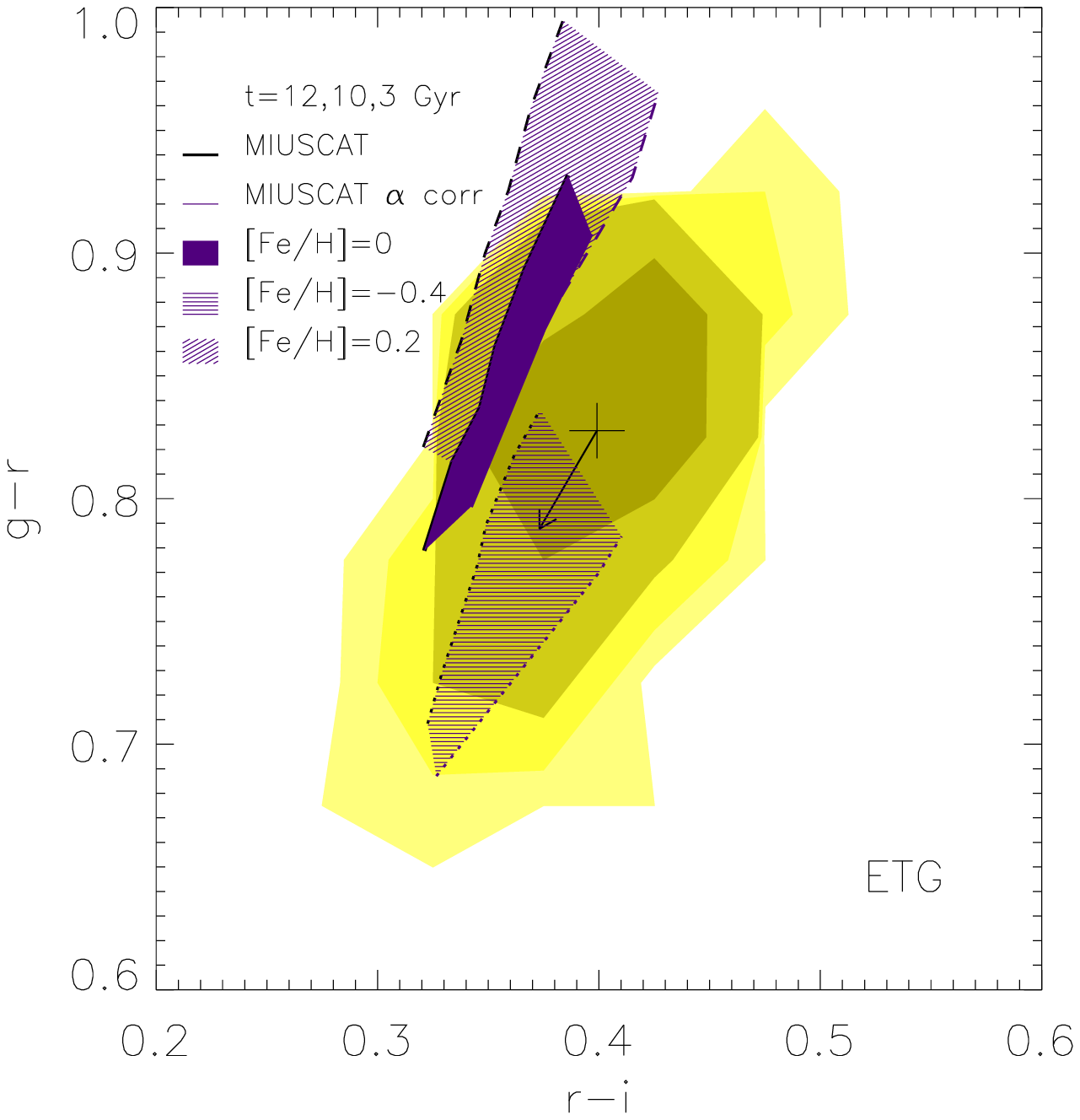}
\caption{
The observed colour-colour distributions are compared to the MIUSCAT SSP models
once corrected from the $\alpha$-enhancement effect. There are three shaded
regions corresponding to each metallicity. For each of these regions the upper
envelope indicates the MIUSCAT SSP models (same as in Figure \ref{etg_ssp}),
whereas the lower envelope shows the resulting models once corrected from the
$\alpha$-enhancement effects with the aid of the models of \citet{Coelho07}. For
this purpose we estimate the colour difference obtained from comparing the
scaled-solar and $\alpha$-enhanced ([$\alpha$/Fe]=0.4 dex) models from these
authors. 
}
\label{etg_alpha}
\end{figure*}
As explained in Paper~I the MIUSCAT SSP SEDs employed here are computed on the
basis of scaled-solar isochrones and stellar spectra with nearly scaled-solar
abundance ratios for the high metallicity regime that is required to fit massive
galaxies. However, it is well known that the spectra of these galaxies show an
enhancement in [Mg/Fe]. To assess the effects of $\alpha$-enhanced element
partitions on the SSP colours we correct them, in a differential way, with the
aid of the models of \citet{Coelho07}. These fully theoretical models are
constructed on the basis of a high resolution library of synthetic spectra
\citep{Coelho05}, where all the $\alpha$ elements, including magnesium, are
enhanced with respect to iron with the same proportion, i.e., [$\alpha$/Fe]=0.4.

We correct our MIUSCAT SSP colours with the difference in colour obtained from
the scaled-solar and $\alpha$-enhanced models of \citet{Coelho07}, keeping
constant the iron abundance. The corrected SSP colours are shown for
different metallicities in Figure \ref{etg_alpha}. Interestingly, the effect  of
$\alpha$-enhancement tends to bluen the {\it u$-$g} and {\it g$-$r} colours and
to redden {\it r$-$i}, as required by the data. Note that although this effect
is relatively modest it certainly goes in the right direction for improving our
fits to the most massive galaxies of our sample. Preliminary tests performed with our own version of the models
constructed with $\alpha$-enhanced mixtures based on the MILES library, making
use of the [Mg/Fe] determinations of \citet{Milone11}, lead to even larger
differences with respect to the scaled-solar predictions in the same directions
as with the models of \citet{Coelho07}.


\section{DISCUSSION AND CONCLUSIONS}

In the present work we have explored the constraints on the MIUSCAT models
provided by photometric data of globular clusters and quiescent galaxies. In
Paper~I we have presented the new models and described the improvements. Here,
we have focused on the comparison of the model predictions with broad-band
photometry.


The integrated colours of MW globular clusters can be remarkably reproduced by
the MIUSCAT models in all the visible bands.  Only colours involving the U
filter can not be used to constrain the models, as the uncertainty relative to
the filter response used to reproduce the U magnitude is too high. The ability
of our models for matching the colours of the MW stellar clusters in the other
bands is remarkable. In particular, the B$-$V colour has always challenged
models based on theoretical stellar libraries, which fail at matching the range
covered by the $V$ filter. Here we have shown that the problem can be solved by
the use of empirical stellar libraries (see also \citealt{Maraston11}).


The comparison with M31 globular clusters is less straightforward but it allows
us to test the models even in the region of solar metallicity. The comparison
with our models suggests that the M31 globular clusters are on average younger
than usually assumed for the MW globular clusters. Although an old model (12.6
Gyr) can fit the metal-poor clusters, we need ages in the range 6-10\,Gyr to
match the colours of the clusters with high metallicities. 

The spectral coverage of the MIUSCAT library allows us to test here the models
in several bands and, by relaxing the assumption that all the M31 GC are old, we
can obtain a good fit to the data  if we assume that metal-rich M31 clusters 
are younger than metal-poor ones. The fact that in MW clusters we do not need
young models may be ascribed to the lack of metal-rich clusters.
Therefore our results disagree with previous works \citep{Peacock11}
claiming that the {\it g$-$r}  colours of M31 globular clusters could not be
matched accurately with our models.
Nevertheless, the large amount of extinction  in these clusters
  makes their colours very uncertain and not suitable for calibration purposes.


When the models are compared to the colour evolution of LRGs, under the
assumption of a passive evolution model, the observed colours are fairly
reproduced. Both, a simple model where stars formed in a single burst at high
redshift, and a model that includes a small component of metal-poor stars, can
be used to fit the observed distribution with increasing redshift. 
However, in the high redshift interval (z $\gtrsim$ 0.4) where the observed
colours probe the rest-frame {\it g$-$r}, the models predict redder colours than
the median distribution.


Although the colour evolution of the LRGs can be fairly described within the
fiducial model of purely passive stellar evolution, the combination of three
colours, {\it u$-$g}, {\it g$-$r}, {\it r$-$i}, poses a challenge. We find that
the colour-colour diagrams based on these colours allow us to discriminate among
different models. Given the wavelength coverage of the MIUSCAT models, we
restrict our study to a limited redshift range, at z$\simeq$0.04, where we can
measure all these three colours. Our study shows that single-burst models do not
provide a good description for the local population of quiescent galaxies.
Irrespective of their metallicity, the old SSPs provide too red {\it u$-$g} and
{\it g$-$r} colours and too blue {\it r$-$i} colour. We also find that the SSP
model predictions from different authors do not match these colour-colour
distributions. Such disagreement has also been shown by
\citet{Conroy10}. We note
that the use of only two colours, {\it u$-$g} and {\it g$-$r} does not allow to
highlight the problem in the {\it r$-$i} colour, which goes in the opposite
direction. Nonetheless, an important result drawn from such a comparison is that
models based on empirical stellar libraries (MIUSCAT; \citealt{Maraston09})
predict colours much closer to the observed distribution than models based on
theoretical spectra \citep{BC03, Maraston05}.  

We rule out that the inability of the MIUSCAT SSP models for providing fully
satisfactory fits to the colours of LRGs is due to flux-calibration issues.  As
shown in Paper~I, the MIUSCAT spectra provide colours with photometric 
uncertainties $\lesssim$ 0.02 mag. Furthermore, the colours of our models are in
good agreement with globular cluster data.

We have tested the hypothesis that quiescent galaxies formed their stars in a
strong burst at high redshift. Our study based on the {\it u$-$g} vs. {\it
g$-$r} and {\it r$-$i} vs. {\it g$-$r} colour-colour diagrams suggest that the
constituent stellar populations of local ETGs are not necessarily fully coeval
and old, and might require small contributions from either young or/and
metal-poor stellar populations. We have explored several models with such
complex stellar populations. We find that these smaller contributions are very
effective at bluening the colours and matching the distribution in the {\it
u$-$g} and {\it g$-$r} colours. Note however that the resulting {\it r$-$i}
colour is too blue.

The chemo-evolutionary model shown in \citet{Vazdekis96} provides one of the best
fits to these two colour-colour diagrams. Apart from the inclusion of a
metal-poor population at early-times and a small burst at 1\,Gyr, this model
adopts a variable IMF scenario. In this model the first generation of stars is
skewed toward high-mass (top-heavy IMF), while at later times stars formed with
a steeper IMF slope. The adoption of such a steep IMF is the main reason for the
improvement also in the {\it r$-$i} colour. Although this model is just one
among several possible solutions, it clearly shows the need for a more complex
SFH than the single burst model to explain the formation of ETGs. Moreover, the
finding that quiescent galaxies have experienced a small amount of recent star
formation is not new and has been revealed by several observational evidences
\citep{Trager00, Bressan06, Kav07, Sarzi08, Toj11}  and predicted by
hierarchical models of galaxy formation \citep{Delucia06, Ricciardelli10}. However, it is worth noting that although the colours
obtained from these complex models are approaching the observed values, the
agreement is not satisfactory.  

An experiment done by simulating three-burst Montecarlo SFHs confirmed the
impossibility of the three model parameters (age, metallicity and IMF slope) to
fairly match the color plane {\it g$-$r} vs {\it u$-$g}. 
We are aware of the degeneracies affecting  the optical colours, which do not allow us to properly
constrain the solutions. It is beyond the scope of this work to fully
constrain the SFHs, for which spectroscopic data or colors in other
spectral range would be required, but the inability of these
simulations to reproduce the observed colours
suggests the need of another parameter to model the observations. We
have identified that the inclusion of $\alpha$-enhancement improves the fits.  A
limitation of the models based on empirical libraries is that they rely on stars
from the solar neighborhood, which, most likely, has experienced a chemical
enrichment that differs from that of massive elliptical galaxies. Indeed, it is
now well established that massive early-type galaxies are enhanced in Magnesium
with respect to Iron, showing $[Mg/Fe] \simeq 0.2$ \citep{Worthey92, Trager00,
Yamada06}. We have estimated the effect of $\alpha$-enhanced models on these
colours by applying differential corrections derived from the models of         
\citet{Coelho07} predictions on the MIUSCAT colours.  The overall effect is in
the direction of improving the fits to the observations, as the colours in the
bluer bands become bluer whereas the {\it r$-$i} tends to
redden. Therefore, 
the combination of both effects, a more complex SFH and the inclusion in
the models of abundance ratios in agreement with the observations should provide
better fits. 

We conclude that the impact of $\alpha$-enhancement on the colours
of massive galaxies can not be neglected to properly reproduce the
observations. Our forthcoming models taking into account the
$\alpha$-enhanced mixture of MILES stars will clarify this
issue.

\section*{Acknowledgements}

We are very grateful to Paula Coelho for providing us with her stellar
population models. We thank the referee for helpful suggestions that
improved this paper.
 AJC and JFB
are {\it Ram\'on y Cajal} Fellows of the Spanish Ministry of Science and
Innovation. 
This work has been supported by the Programa Nacional de
Astronom\'{\i}a y Astrof\'{\i}sica of the Spanish Ministry of Science and
Innovation under grants AYA2010-21322-C03-01 and  AYA2010-21322-C03-02
and by the Generalitat Valenciana under grant PROMETEO-2009-103.

\appendix
\section{Results from Montecarlo simulations}

In this section we show the SFHs of the best solutions found in the  Montecarlo
simulations described in Section~\ref{sect_mc}.  We selected the 1700 closest
points to the median of the observed colours in each of the colour-colour plots.
Then we only considered the solutions  that satisfy both distributions, ending
up with 26 models, whose SFHs  are shown in Figure~\ref{sfh}. The SFHs are
ordered according to an increasing fraction of mass in bursts younger than
5\,Gyr (the first panels show the solutions with the smaller fractions). 

The resulting SFHs can be divided into two broad categories. In the first group
we can identify SFHs dominated by an old, metal-rich population, enriched with
dwarf stars (high IMF slopes). On top of the old populations, we find small
contributions from young populations. In some cases we do
not see any significant young component. In these cases, blue {\it g$-$r} and
{\it u$-$g} colours result from the contribution of a metal-poor population (see
also section \ref{sect_mp}). 

The other group of solutions includes SFHs with an important component of young
populations ($\le$ 5\, Gyr). In some cases, the fraction of mass formed in the
youngest burst can be as large as 80\% (see SFH in the last row). As already
shown in Section \ref{sect_age}, the age is the dominant parameter at decreasing
both the {\it g$-$r} and {\it u$-$g} colours and, thus, it provides a good match
of the {\it g$-$r} vs {\it u$-$g} plot. The SFHs with the largest component of
young stellar populations also show the steepest IMF slopes, which are needed in
order to redden  the {\it r$-$i} colour.  This confirms the results found in
sections 6.1, 6.2 and 6.3, where we have illustrated how the IMF is mainly
affecting the {\it r$-$i} while age and metallicity  variations are needed to
match the {\it g$-$r} vs {\it u$-$g} colour distribution. 

Given the age/metallicity/IMF degeneracies affecting the optical colours, the
two kind of SFHs produce similar colours and, hence, lead  to the same level of
agreement with the observed colour-colour distributions. There are many
observational evidences, from spectroscopic and photometric studies favouring
the first group of solutions for the formation of massive galaxies (e.g.
\citealt{Renzini06}). Indeed, galaxies that formed a considerable amount of stars
very recently, as in the second group, would show strong $H_{\beta}$ absorption
in the spectra that are not observed in the local population of massive
ellipticals (e.g. \citealt{Trager00}).

Interestingly, within the first group of models we can identify a number of
solutions that resemble the ones found in section \ref{sect_chemo}. A small
burst at early times occurring with a flat IMF and metal-poor stars is followed
by a strong burst at nearly solar metallicity and steeper IMF and then by a
small burst at later times (see first and second rows).

\begin{figure*} 
\includegraphics[angle=0,width=1.5\columnwidth]{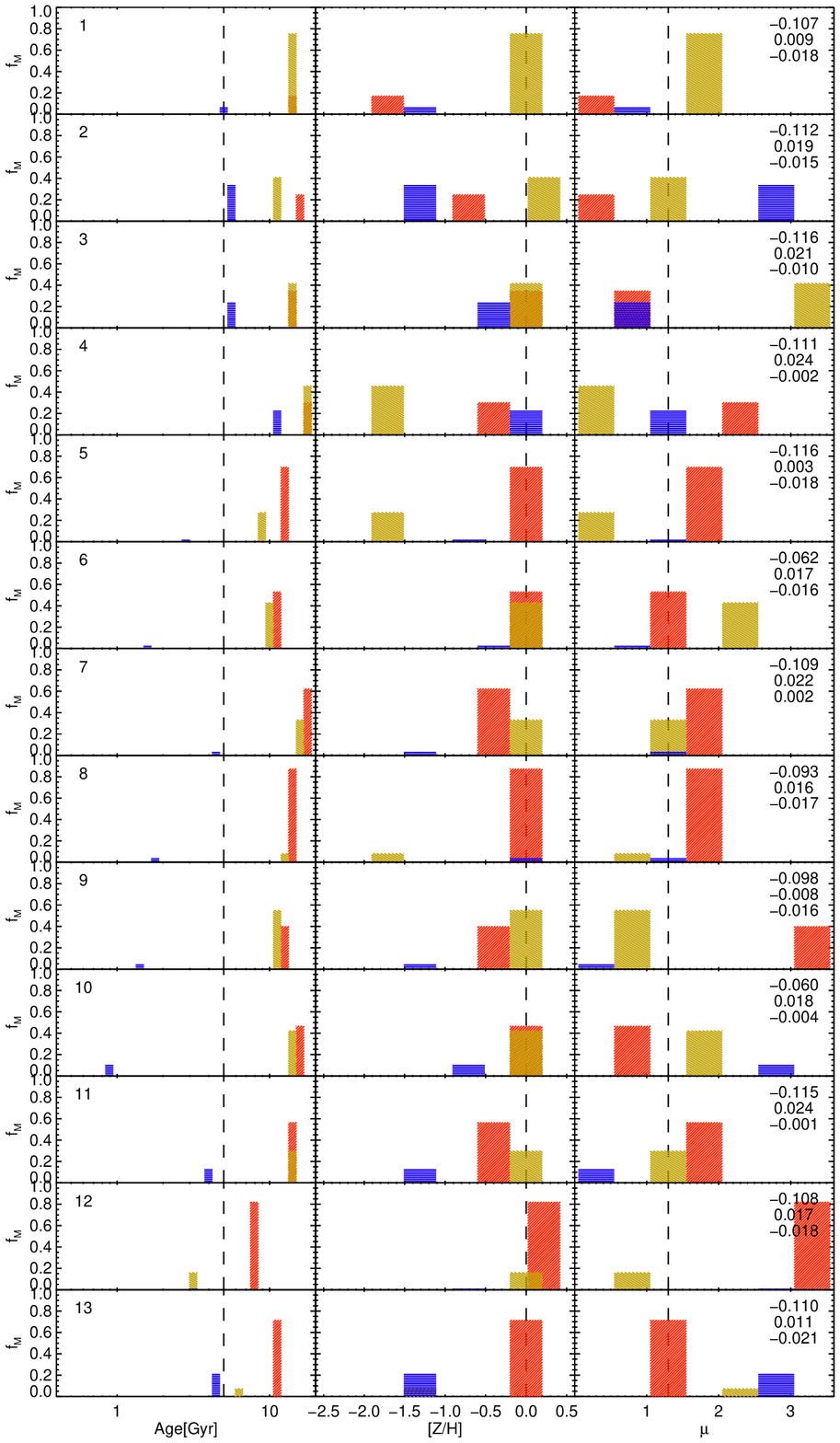}
\caption{Best solution SFHs from the Montecarlo simulations described in
Section~\ref{sect_mc}. The first column shows the SFH in terms of fraction of
mass formed at different ages. The second one shows the metallicity associated
to each of the bursts, and the third one the slope of the IMF, assumed to be
unimodal. In each panel the red, green and blue bars indicate the fractions
corresponding to the contributions of the old, intermediate and young burst,
respectively. The models are ordered according to the contribution of the young
stellar populations. The vertical dashed line in each panel indicate, from left
to right, an age of 5\,Gyr, solar metallicity and $\mu$=1.3 (Salpeter IMF),
respectively. We indicate in the last panel of each model the difference in
$u-g$, $g-r$ and $r-i$ colours with respect to the median of the observed
colours.}  
\label{sfh} 
\end{figure*}

\addtocounter{figure}{-1} 
\begin{figure*}
\includegraphics[angle=0,width=1.5\columnwidth]{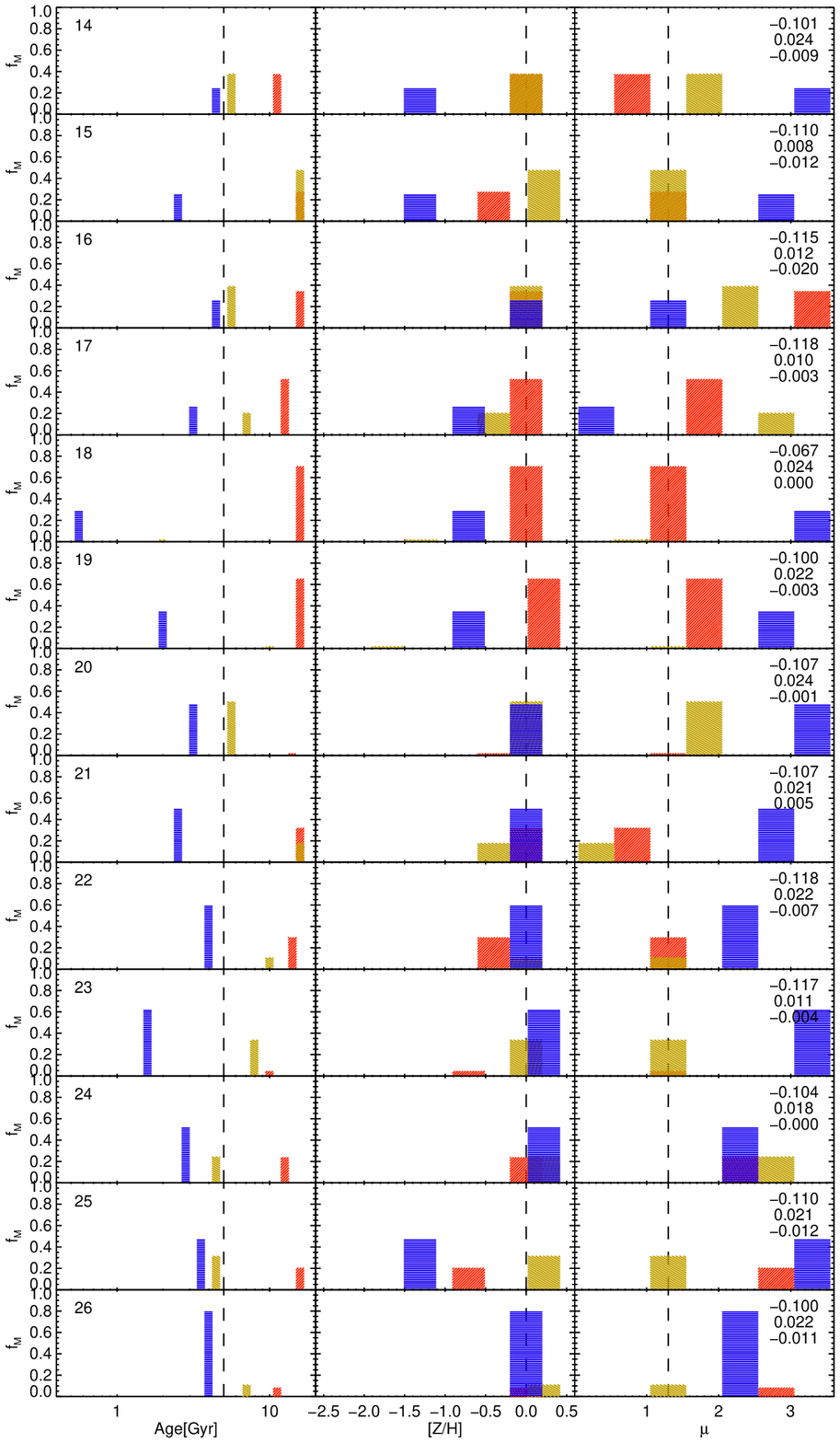}
\caption{Continued} 
\label{sfh}
\end{figure*}

\label{lastpage}

\end{document}